\documentclass[5p,times]{elsarticle}

\usepackage{subcaption}
\usepackage{mathtools}

\usepackage{enumitem}
\usepackage{xcolor,colortbl}

\usepackage[flushleft]{threeparttable} 
\usepackage{makecell}
\usepackage{booktabs}
\usepackage{arydshln}
\usepackage[
            ]{hyperref}

\renewcommand{\thesection}{\Roman{section}}

\usepackage{natbib}

\biboptions{sort&compress}

\journal{PRX Energy}

\newcommand{\beginsupplement}{%
	\setcounter{table}{0}
	\renewcommand{\thetable}{S\arabic{table}}%
	\setcounter{figure}{0}
	\renewcommand{\thefigure}{S\arabic{figure}}%
	\setcounter{section}{0}
	\renewcommand{\thesection}{S\arabic{section}}%
	\setcounter{equation}{0}
	\renewcommand{\theequation}{S\arabic{equation}}
	\setcounter{page}{1}
}

\definecolor{Gray}{gray}{0.85}

\newcolumntype{a}{>{\columncolor{Gray}}c}

\begin{document}

\begin{frontmatter}

\title{Cost and efficiency requirements for a successful electricity storage \\ in a highly renewable European energy system}

\author{Ebbe Kyhl G\o{}tske$^{a,b,*}$}
\author{Gorm Bruun Andresen$^{a,b}$}
\author{Marta Victoria$^{a,b,c}$}
            
\begin{abstract}

Future highly renewable energy systems might require substantial storage deployment. At the current stage, the technology portfolio of dominant storage options is limited to pumped-hydro storage and Li-Ion batteries. It is uncertain which storage design will be able to compete with these options. Considering Europe as a case study, we derive the cost and efficiency requirements of a generic storage technology, which we refer to as \textit{storage-X}, to be deployed in the cost-optimal system. This is performed while including existing pumped-hydro facilities and accounting for the competition from stationary Li-ion batteries, flexible generation technology, and flexible demand in a highly renewable sector-coupled energy system. Based on a sample space of 724 storage configurations, we show that energy capacity cost and discharge efficiency largely determine the optimal storage deployment, in agreement with previous studies. Here, we show that charge capacity cost is also important due to its impact on renewable curtailment. A significant deployment of \mbox{\text{storage-X}} in a cost-optimal system requires (a) discharge efficiency of at least 95\%, (b) discharge efficiency of at least 50\% together with low energy capacity cost (10~€/kWh), or (c) discharge efficiency of at least 25\% with very low energy capacity cost (2€/kWh). Comparing our findings with seven emerging technologies reveals that none of them fulfill these requirements. Thermal Energy Storage (TES) is, however, on the verge of qualifying due to its low energy capacity cost and concurrent low charge capacity cost. Exploring the space of storage designs reveals that system cost reduction from \mbox{\text{storage-X}} deployment can reach 9\% at its best, but this requires high round-trip efficiency (RTE$\geq$90\%) and low charge capacity cost (35~€/kW). 


\end{abstract}

\end{frontmatter}

\begin{minipage}{\textwidth}
	{\small \itshape 
		\noindent 
		$^{a}$ Department of Mechanical and Production Engineering, Aarhus University, Denmark\\
		$^{b}$ iCLIMATE Interdisciplinary Centre for Climate Change, Aarhus University, Denmark \\
		$^{c}$ Novo Nordisk Foundation CO$_2$ Research Center, Aarhus University, Denmark \\
		$^{*}$ Lead contact and corresponding author, Email: ekg@mpe.au.dk
	}
\end{minipage}


\clearpage

\section{INTRODUCTION}

To comply with global climate commitments \cite{UN2016} and greenhouse gases reduction targets, a massive deployment of renewable generators, comprised of wind turbines and solar panels, is anticipated \cite{victoria2022speed}. The integration of variable renewable generators is associated with some challenging aspects. The variable power output necessitates backup reserves, and increased transmission capacity is required to even out production over larger areas \cite{UECKERDT20151}. Extensive literature exists on the variability of solar and wind power generation \cite{ENGELAND2017600,BIANCHI2022560,RINGKJOB2020118377,SCHINDLER2020113016,Victoria_2020_role_of_photovoltaics}. Where diurnal cycles dominate solar generation, synoptic temporal fluctuations dominate wind generation. In Europe, both show a complementary seasonal cycle \cite{Graabak2016, KAPICA2021114692}; thus, an optimal seasonal mix of wind and solar exists \cite{HEIDE20102483,TAPETADO2021100729}.

To help relieve the aforementioned challenges, one option is to balance the fluctuations of the VRE production locally in time with the implementation of electricity storage. Here, electricity storage refers to the conversion from electrical energy to a storage energy carrier which is converted back to electricity when discharged at a later time step. Similar to the temporal variability of wind and solar, different time scales apply to electricity storage. For grid stability, certain technologies perform frequency or voltage regulation, intradaily smoothing of diurnal variability, or balancing of synoptic or seasonal variation \cite{AKTAS2021377,TONG2020101484,Victoria2019}. In this study, we limit ourselves to utility-scale electricity storage capable of providing balancing on a time frame longer than an hour.

Based on the already deployed capacity, pumped-hydro storage (PHS) constitutes the majority ($>$90\%) of electricity storage, globally \cite{CHENG2019386,EC_phs_status}. In Europe, PHS has a cumulative capacity of 55~GW power capacity \cite{iha2021} and 1.3~TWh energy capacity \cite{GETH20151212}. Electrochemical batteries account for only 1\% of today's storage capacity worldwide \cite{CHOI2021230419} (in Europe, residential batteries constituted 5.4~GWh storage capacity in 2021 \cite{solarpowereurope2022}), but contribute to a large extent to the short-range primary responses to continuous and sudden voltage and frequency instabilities \cite{SCHMIDT201981}. In addition, Lithium-ion (Li-ion), which currently accounts for 78\% of the battery systems in operation, has shown a large potential: Over the last decade, Li-ion battery packs have shown learning rates of 20\%, contributed substantially by the large growth in battery electric vehicles (BEV) \cite{Glenk2021}. 

In literature, power system models often use batteries to represent short-term storage and hydrogen (H$_2$) with electrolyzers and fuel cells as long-term storage \cite{Scholz2017,Cebulla2017}. In a low carbon-intense energy system, electricity storage is far from the only hydrogen use case. Hydrogen infrastructure is essential for the energy demands that are difficult to electrify, e.g. industrial processes or sections of the transport sector \cite{GILS2021140, FabianNeumann_hydrogen}. In systems with the coexistence of H$_2$ and Li-ion batteries, Victoria et~al. \cite{Victoria2019} find that significant electricity storage capacities are less likely to emerge with CO$_2$ emissions reductions lower than 80\% relative to 1990-levels, in agreement with other studies \cite{Scholz2017,Cebulla2017, HUNTER20212077}. Victoria et~al. \cite{Victoria2019} and Brown et~al. \cite{BROWN2018720} show that the presence of BEV replaces the need for high volumes of stationary battery storage, and, in general, that flexibility introduced by sector coupling further delays the moment in the CO$_2$ emissions reduction in which large storage capacities will be needed. In addition to this, Wang et~al. \cite{WANG20211187} showed that it is not certain that storage is always the cost-optimal strategy, compared to overplanting renewable capacities. This raises the question of under which circumstances electricity storage will be competitive.

Prior studies have evaluated the techno-economic potential and required characteristics of electricity storage to occur in renewable power systems \cite{Sepulveda2021,Dowling2020,ZIEGLER20192134,SCHILL20202059, TONG2020101484, Parzen}. With the scope of guiding storage technology development, Sepulveda et~al. \cite{Sepulveda2021} present the idea of a technology design space. The design space contains combinations of storage costs and efficiencies to evaluate the potential for long-duration energy storage systems in North-American power grids. They show that competitiveness with firm low-carbon emitting generators highly depends on the storage energy capacity cost and discharge efficiency, whereas charge and discharge capacity cost and charge efficiency are of secondary importance. Both Sepulveda et~al. \cite{Sepulveda2021} and Ziegler et~al. \cite{ZIEGLER20192134} find a threshold of \$20/kWh energy capacity cost for the storage to become favorable to the system. Studies by Dowling et~al. \cite{Dowling2020} and Tong et~al. \cite{TONG2020101484} both show that low-cost energy storage has a high potential of reducing the total cost of the power system. Parzen et~al. \cite{Parzen} considered the effect of including competition between multiple storage options in a European-wide power system cost-optimization model. As an evaluation metric, they used the optimal storage capacity as a market potential indicator. Here, the optimal storage capacity refers to the deployment of a given storage technology that entails the minimum total system cost. They showed that the system value of storage integration is not only determined by the cost but also by other performance measures such as the efficiency.\\

The previously reported results are subject to some methodological limitations:\\

\begin{itemize}
	\item They are obtained with single-node models that do not include the flow of electricity through transmission lines (i.e., copper-plate models). In that way, balancing through regional integration is neglected, and information on regional bottlenecks is lost. Moreover, regional differences within the energy mix are not accounted for, i.e., solar/wind share in every node, which also impacts the storage needed in the system. 
	\item \textit{and/or} The analyses do only consider the power system. Integrating other energy-consuming sectors such as the heating, transportation, and industry sectors have found to change the storage requirements \cite{Victoria2019,Lund_2016}. With sectoral integration, the electricity demand is substantially increased. A certain share of the additional demand from sectoral integration can be considered flexible, i.e., the production of an energy carrier can be shifted in time from the consumption. In addition, other energy-consuming sectors include other storage mechanisms, e.g., hot water tanks in the heating supply or BEV batteries which can provide additional demand flexibility. Thus, considering only the power system when determining the potential electricity storage capacity needs in a low-carbon emitting system could lead to a substantial misestimation. 
	\item \textit{and/or} Storage investment costs and efficiencies are assumed fixed and reflect current or predicted future cost and efficiency levels. Therefore, the results obtained in the given studies are constrained by the assumptions made of each specific technology.
\end{itemize}

With PHS and batteries as dominant storage options, and the integration of flexible demand options from sector coupling, it is not clear which requirements are needed for an additional storage technology to enter the cost-optimal system. Literature has shown that, in renewable power systems capacity expansion models, low energy capacity cost and high discharge efficiency are good prerequisites. The former incentivizes an expansion of the storage to overcome long periods with renewable droughts, while the latter represents a high utilization of the stored energy with less energy loss. It is not known whether this priority is cost-optimal in a sector-coupled system, and which other strategies could also apply to a qualified technology. To build upon the existing body of literature and overcome the mentioned limitations, we raise the following research question: 
\begin{itemize}
	\item Which characteristics are needed for a successful additional electricity storage technology to enter the cost-optimal system design, considering the presence of other storage options such as PHS and batteries, in a sector-coupled interconnected renewable energy system?
\end{itemize}

To help enlighten this, we use a state-of-the-art energy system capacity and dispatch optimization model to derive the space of cost-competitive electricity storage technologies, on top of PHS and battery storage. In this study, we refer to the group of additional electricity storage technologies as "\mbox{\text{storage-X}}" and use the term "design space" to describe the space of cost-competitive \mbox{\text{storage-X}} options. The derivation is computed when already accounting for the competition from other backup reserves and interconnectivity, and the integration of the heating, transportation, and the industry sector. We do not limit ourselves to fixed investment costs or efficiency assumptions. Instead, we examine the required characteristics (charge and discharge power capacity cost, charge and discharge efficiency, energy capacity cost, and self-discharge) of a generic storage technology to feature in the system, similar to Ref. \cite{Sepulveda2021}. This study contributes to the existing literature by also including the effect of linking the power system with other energy-consuming sectors since this has shown a high impact on the storage needs, both in terms of volume and the characteristic \cite{Victoria2019,Lund_2016}. Furthermore, we resolve the system with a network that distributes the renewable resources and consumption over the geographical domain. In this way, we also account for the regional congestion in the transmission connecting each of the regions which is an important factor in the assessment of the storage capacity allocation. 

Section \ref{sec:emergin_techs} presents a short review of the current state of seven emerging storage technologies that we consider as \mbox{\text{storage-X}} candidates. The model framework to derive the \mbox{\text{storage-X}} requirements to appear in the optimum system is presented in Section \ref{sec:materials_methods}. Our results, which cover the design requirements for \mbox{\text{storage-X}} and a delineation of the corresponding impact on the system design, are presented in Section \ref{sec:storage_x}. Section \ref{sec:discussion} encapsulates our findings in a discussion on whether current technologies can comply with the derived requirements, followed by the main conclusion of our work in Section \ref{sec:conclusion}. 

The code and data used for producing the figures are published and is openly available with an open license at Github: \href{https://github.com/ebbekyhl/storageX.git}{ebbekyhl/storageX.git}. 

\section{CURRENT STATE OF STORAGE-X CANDIDATES}\label{sec:emergin_techs}

\begin{figure}[!h]
	\centering
	\includegraphics[width=0.5\textwidth]{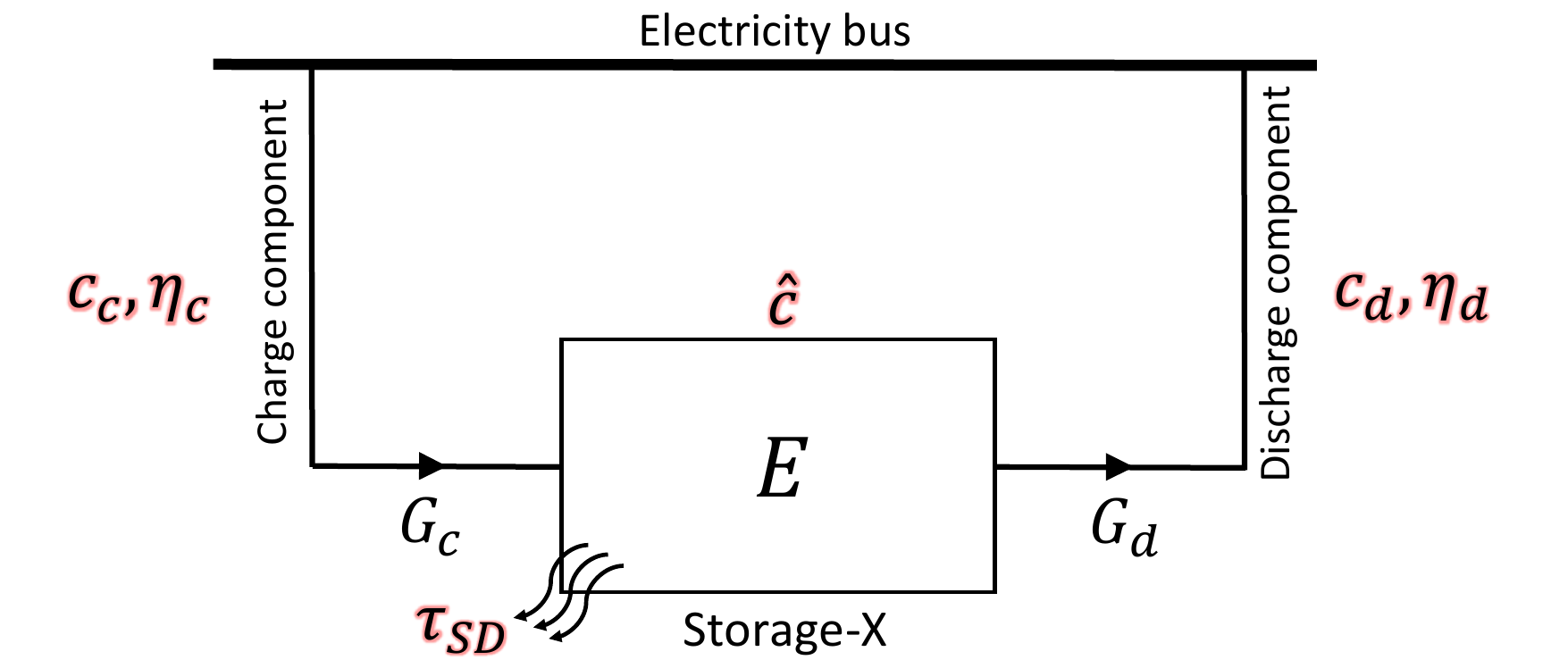}
	\caption{\textbf{Storage-X parameters}. Highlighted parameters (charge and discharge power capacity cost $c_c$ and $c_d$, charge and discharge efficiency $\eta_c$ and $\eta_d$, energy capacity cost $\hat{c}$, self-discharge time due to standing losses $\tau_{SD}$) define the storage design, whereas the remainder (charge and discharge power capacity $G_c$ and $G_d$, and energy capacity $E$) represent the volume of storage.}
	\label{fig:storage_x_illustration}
\end{figure}

The additional utility-scale electricity storage technology appearing in a decarbonized energy system could potentially be provided by a wide range of different storage technologies. A wide palette of additional storage technologies is emerging \cite{Gautam2022}. Here, we review the current state of seven \mbox{\text{storage-X}} candidates, based on their proven deployment scale together with expected energy efficiency and capacity costs. Before this, we establish the parameters used for this comparison in Fig.\ref{fig:storage_x_illustration}. The generic storage itself is characterized by a storage energy capacity cost $\hat{c}$ (€/kWh) and a self-discharge time $\tau_{SD}$ (days), assuming an exponential decay of the state of charge:

\begin{equation}\label{eq:standing_loss}
	\bar{e}_t = \bar{e}_{t_0} \exp \left(-\frac{t}{24\tau_{SD}} \right)
\end{equation}

where $\bar{e}_t$ is the state of charge (SOC), i.e., the fraction of the storage energy capacity contained at time $t$ (hours), and $\bar{e}_{t_0}$ is the initial SOC. We use the subscript ``SD'' to signify the distinction from another characteristic time constant of the storage, namely the duration (i.e., the ratio between the optimal storage energy capacity and the discharge power capacity).

The storage technology is furthermore characterized by charge and discharge power capacity costs, $c_c$ (€/kW) and $c_d$ (€/kW), and charge and discharge efficiencies, $\eta_c$ (\%) and $\eta_d$ (\%). Here, power capacity costs are given per unit of electricity. 

The combination of the six parameters constitutes the storage configuration (i.e., the design) of \mbox{\text{storage-X}}. Installed energy capacity and power capacities, $E$, $G_c$, and $G_d$, define the volume of storage.

Here, we consider a portfolio of seven emerging storage technologies, covering both thermo-mechanical and chemical energy storage. For the thermo-mechanical storage technologies, we refer to the thorough review by Gautam et~al. \cite{Gautam2022} from which we acquire the costs and efficiency assumptions for 2025. For the remaining technologies, we use data from Refs. \cite{Sepulveda2021,DEA2022_fuel,DEA2022_gen,BUDISCHAK201360}. The reported data is converted into the six descriptive \mbox{\text{storage-X}} parameters described above. See \ref{App:Disagg_storage_assumptions} for these calculations. The following subsections shortly describe the fundamental mechanisms behind every storage concept. A summary of investment costs and efficiencies is presented in Table \ref{tab:existing_storage_assumptions_short}.

\subsection{Adiabatic compressed air energy storage}
The adiabatic compressed air energy storage (aCAES) is a thermomechanical energy storage. In an aCAES, electrical energy is utilized to power the compression of atmospheric air to high-pressure air. Subsequently, the pressurized air is stored, either in an underground cavern or an overground pressurized tank. In the case of overground tanks, the system can be modularized, allowing the size of the plant to be scaled according to specific needs. Thermal energy generated in the compression is stored in parallel, making it an adiabatic compressed air energy storage. When discharged, the compressed air is released and heated by the thermal energy storage, to run a gas turbine which produces electricity. In that way, the heat recovery avoids the need for additional heat (usually obtained with the combustion of fossil gas) injection in a diabatic CAES, and it increases the overall energy round-trip efficiency of the storage system. Diabatic CAES has been proven on utility-scale (321~MW plant in Huntorf, Germany built in 1978 \cite{power_technology_Huntorf} and 110~MW plant in McIntosh, USA built in 1990 \cite{power_technology_McIntosh}). The first large-scale aCAES plant (100~MW and 400~MWh) was reported to be commissioned in Zhangjiakou, China, in 2022 \cite{acaes_china}. Prior to this, aCAES was yet to be demonstrated on a commercial scale \cite{BARBOUR20211914}.

\subsection{Redox-flow battery}
The redox flow battery (RFB) is an electrochemical energy storage. Vanadium redox flow battery is mentioned as the most promising of the available redox flow batteries for large-scale energy storage \cite{VINCO2021103180}. The RFB is different from a Li-ion battery since it separates the electrolyte from the cell stacks in two exterior tanks. One of the tanks contains the positive electrolyte solution while the second tank contains the negative solution. The vanadium electrolyte solution flows without phase change in a circuit driven by pumps, in which the electrolyte tanks are connected to the anode (negative side) and the cathode (positive side) of the cell stack, without any mixing. Electricity is converted into chemical energy and vice versa through a reversible reduction and oxidation (redox) reaction. This conversion occurs at high efficiency (round-trip energy efficiency of 65-80\% according to Ref. \cite{Sepulveda2021}. Vinco et~al. \cite{VINCO2021103180} reports a range of 69\%-91\%.) compared to other energy storage technologies. Due to this design, the energy capacity is decoupled from the power capacity, which qualifies it for long-duration energy storage \cite{BARRERAS2022}. Recently, China commissioned a 100 MW/400 MWh vanadium RFB to the grid in Dalian, which at the time of writing is the largest facility worldwide \cite{pvmagazine_rfb}.

\subsection{Molten-salt energy storage}
Molten salt energy storage (MSES) is a thermomechanical energy storage that stores electricity as sensible heat (i.e., temperature increase without phase change). Electrical energy is converted into heat through resistive heaters which occurs at almost 100\% energy efficiency. The liquid salt is contained either in a singular tank with a thermocline or in a two-tank (hot and cold) configuration. The stored heat is converted back to electricity typically through a steam-based Rankine cycle. The discharge efficiency is, thus, limited by the achievable temperature of the storage. Albeit a higher investment cost compared to the single tank, the two-tank system is more common due to its better performance. The two-tank system is capable of sustaining its output temperature throughout the process of discharging. For the single tank, the output temperature decreases when the thermocline reaches the top of the tank, reducing the efficiency of the discharging stage \cite{ANGELINI2014694}. Today, molten-salt facilities are mostly associated with concentrated solar power (CSP) plants, while large-scale MSES electricity storage is yet to be demonstrated.

\subsection{Solid thermal energy storage}
Solid thermal energy storage (TES) is a thermomechanical energy storage and is in the same category as MSES. It is similarly charged with electrical resistive heaters to increase the temperature (i.e., sensible storage) of a container with stacked rock material, firebricks, or other low-cost solid material that is thermally and chemically stable at high temperatures. The thermal energy is converted back to electricity through a Brayton cycle. Such setups are also known in literature as Carnot batteries \cite{HOLY2021103283}. Here, data is acquired from Ref. \cite{Sepulveda2021} who assume firebricks as the storage medium. A demonstration plant exists in Germany (1.5~MW/130~MWh) \cite{Gautam2022}.

\subsection{Pumped thermal energy storage}
Pumped thermal energy storage (PTES) is a thermomechanical energy storage that utilizes the same mechanisms in a heat pump to achieve high charge efficiency. It consists of a dual tank system (one hot and one cold), each containing a packed bed of volcanic rocks. The system is charged with a heat pump cycle using air as its working fluid, which ensures a charge energy efficiency above 200\%. In the charging process, air from the cold tank is compressed and transferred to the hot tank. During this stage, one tank attains a high temperature ($\approx600^\circ$C) while the other one reaches a very low temperature ($\approx-30^\circ$C). In the discharge stage, the direction of the airflow is reversed, utilizing the high-temperature difference gained to run a Brayton cycle, generating electricity back to the grid \cite{Gautam2022}. At the time of writing, large-scale ($>$1~MW) demonstrations have been projected but not yet showcased.

\subsection{Liquified air energy storage}
Liquified air energy storage (LAES) can also be classified as a thermomechanical energy storage. In the charging stage, electricity is utilized to power the compression of ambient air until it reaches high pressure ($\approx$200~bar). The compressed gaseous air is then cooled (compression and subsequent (inter)cooling might occur over several stages), reaching cryogenic temperatures and a liquid state. The heat acquired from the cooling can be stored for the later discharge stage, increasing round-trip efficiency. When discharged, the air is exposed to atmospheric air and/or the stored heat, which entails an expansion and evaporation through a turbine generating electricity \cite{Gautam2022,DAMAK2020208}. Large-scale plants have been demonstrated, e.g., the Pilsworth plant in the UK (5~MW) \cite{power_technology_Pilsworth}.

\subsection{Hydrogen electricity storage}
Hydrogen (H$_2$), on top of its other wide-ranging applications, can act as a power-to-power storage option. In such a case, hydrogen is produced with electricity through water electrolysis, either with alkaline or proton exchange membrane (PEM) electrolyzers. The H$_2$ is then stored in either underground salt caverns or overground steel tanks. At a later stage, the H$_2$ can be converted back to electricity through fuel cells or in a combustion turbine. 
Here, we assume alkaline electrolyzers \cite{DEA2022_fuel}, overground steel tanks \cite{BUDISCHAK201360}, and PEM fuel cells \cite{DEA2022_gen}. We refer to this option as H$^X_2$ in which the superscript $X$ indicates that we treat it as a \mbox{\text{storage-X}} technology, meaning that it can only be converted back to electricity.\\

The acquired data for the seven above-mentioned technologies is presented in Table \ref{tab:existing_storage_assumptions_short}, sorted according to their energy capacity cost.

\begin{table}[!h]
	\caption{\textbf{Reported capacity cost, efficiency, and self-discharge time}. Charge and discharge power capacity cost $c_c$ and $c_d$, charge and discharge efficiency $\eta_c$ and $\eta_d$, energy capacity cost $\hat{c}$, self-discharge time due to standing losses $\tau_{SD}$, for seven emerging storage technologies. Values are obtained directly from or calculated based on data from Refs. \cite{Gautam2022,Sepulveda2021,DEA2020}. See \ref{App:Disagg_storage_assumptions} for these calculations. If not reported, we assume zero standing loss.}
	\label{tab:existing_storage_assumptions_short}
	\begin{tabular}{@{}lcaaaaaa@{}}
		\toprule
		& & $\hat{c}$ & $c_c$ & $c_d$ & $\eta_c$ & $\eta_d$ & $\tau_{SD}$ \\\midrule
		\footnotesize{\textcolor{gray}{unit}} & & \footnotesize{\textcolor{gray}{€/kWh}} & \footnotesize{\textcolor{gray}{€/kW}} & \footnotesize{\textcolor{gray}{€/kW}}  & \footnotesize{\textcolor{gray}{\%}} & \footnotesize{\textcolor{gray}{\%}} & \footnotesize{\textcolor{gray}{days}} \\
		\footnotesize{TES} \cite{Sepulveda2021,Gautam2022}	& & 8 & 38 & 864  & 98 & 38 & 100 \\
		\footnotesize{H$^X_2$} \cite{DEA2022_fuel,DEA2022_gen,BUDISCHAK201360} & & 11 & 450 & 1100 & 68 & 50 & $\infty$ \\
		\footnotesize{MSES} \cite{Gautam2022} 	& & 17 & 104 & 1040  & 99 & 43 & 100 \\
		\footnotesize{PTES} \cite{Gautam2022}  	& & 19 & 326 & 653 & 220 	& 25 & 100 \\
		\footnotesize{aCAES} \cite{Gautam2022} & & 26	& 314 & 629 & 92 & 65 & 100 \\
		\footnotesize{LAES} \cite{Gautam2022} & & 31	& 562 & 562 & 77 & 65 & 200 \\
		\footnotesize{RFB} \cite{Sepulveda2021} & & 115 & 176 & 176  & 85 & 85 & $\infty$ \\
		\bottomrule
	\end{tabular}
\end{table}

\begin{figure}[!h]
	\centering
	\includegraphics[width=0.5\textwidth]{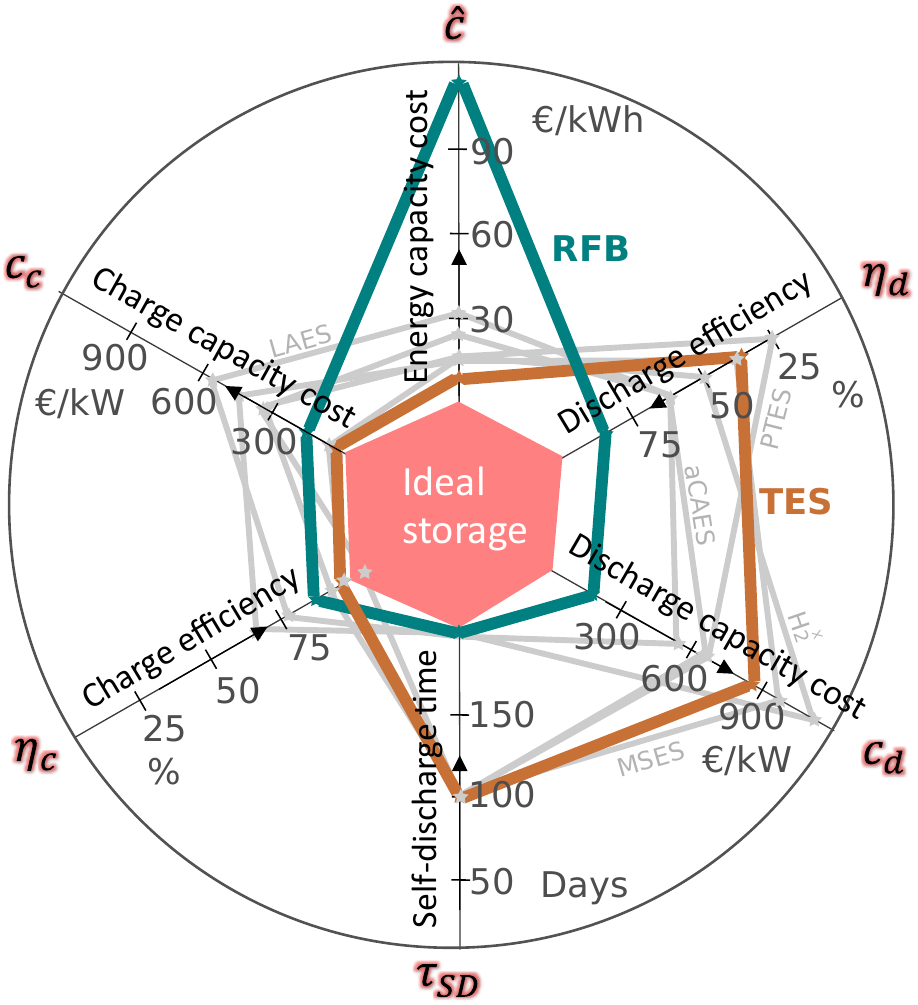}
	\caption{\textbf{Visual comparison of storage technologies}. The seven storage technologies, each given by a combination of six design parameters (charge $\eta_c$ and discharge $\eta_d$ efficiency, charge $c_c$ and discharge $c_d$ capacity cost, energy capacity cost $\hat{c}$ and self-discharge time $\tau_{SD}$) are here depicted. Highlighted are Thermal Energy Storage (TES) and Redox flow battery (RFB). On a given axis, when near the circular exterior, the technology is subject to a poor parameter (e.g., high capacity cost, low efficiency, or short self-discharge time). Oppositely, in the proximity of the center, the technology performs well on the given parameter. Due to the characteristics of TES and RFB, we see a distinct difference in the deviation from the ideal storage. Pumped-Thermal Energy Storage (PTES) crosses the boundary of the ideal storage since the charge energy efficiency is above 1.}
	\label{fig:existing_storage_sketch}
\end{figure}

 Each of the configurations is mapped in the radial plot in Fig. \ref{fig:existing_storage_sketch}. Here, the six axes represent the parameters describing \mbox{\text{storage-X}}. In a hypothetical case, the best-performing and most competitive storage would be characterized by zero capacity cost and 100 \% energy conversion efficiencies with zero standing losses. In this depiction, such an ideal storage configuration would be located in the center of the figure. Moving towards the exterior in one of the six axes is equivalent to either higher investment costs, lower energy efficiency, or higher standing losses. Here, we highlight the characteristics of a Thermal Energy Storage (TES) and a Redox-Flow Battery (RFB). The two storage technologies have distinct characteristics indicated by the parameters deviating from the ideal storage. RFB has a pronounced high energy capacity cost whereas TES deviates from the ideal storage on the discharge efficiency, discharge capacity cost, and self-discharge axes.

\section{METHODOLOGY}\label{sec:materials_methods}

\begin{figure*}
	\centering
	\includegraphics[width=0.8\textwidth]{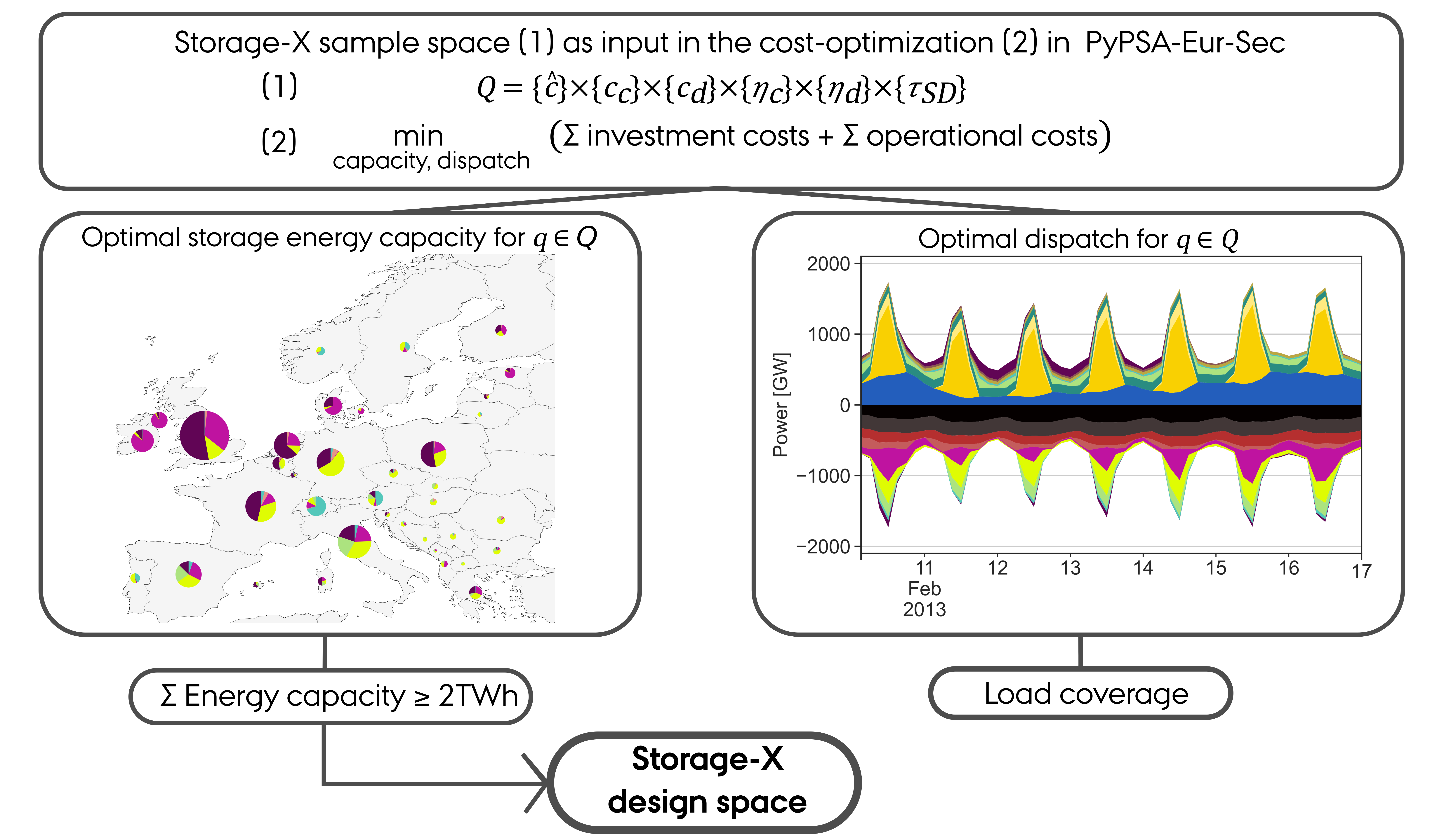}
	\caption{\textbf{Method workflow}. \mbox{\text{Storage-X}} sample space is used as input in the energy system model PyPSA-Eur-Sec to run as many scenarios as the size of the sample space. The capacity and dispatch are optimized in each of the scenarios. Subsequently, the \mbox{\text{storage-X}} deployment is evaluated based on the optimal energy capacity. If it fulfills a total energy capacity of $\geq$2~TWh, the given configuration qualifies for the \mbox{\text{storage-X}} design space. The load coverage (LC) is obtained from the dispatch as a proxy for the level of storage contribution to balance the electricity load and supply. The LC metric is used to compare the configurations but is not used as a selection criterion for the derivation of the design space.}
	\label{fig:method_workflow}
\end{figure*}

The emerging technologies presented in Tab. \ref{tab:existing_storage_assumptions_short} are examples of storage designs currently being developed. It is unclear whether their parameter combinations are sufficiently favorable for the technologies to appear in the cost-optimal energy system design. To shed light on this, we follow a workflow to derive the requirements for a general storage technology to be optimally deployed, on top of batteries and PHS, in a future highly renewable energy system. For this investigation, we consider the European energy system as a case study. To determine the cost-optimal system design, we use PyPSA-Eur-Sec, which is an open energy system optimization model, described in the following subsections. The workflow behind this investigation is visualized in Fig. \ref{fig:method_workflow}.

\subsection{Energy system model: PyPSA-Eur-Sec}

PyPSA-Eur-Sec is a linear model which performs a cost optimization subject to constraints (see \ref{sec:Appendix} for the mathematical formulation). In essence, it finds the optimal combination of generation technologies (solar, wind, gas, etc.), storage technologies (batteries, \mbox{\text{storage-X}}), and conversion technologies (heat pumps, electrolyzers) in every node to minimize the system costs while making sure that enough energy is available in every time to supply the demand (energy balance constraint). Electricity can be produced locally or imported from neighboring nodes through transmission lines, represented by the linearized power flow model assuming lossless links \cite{NEUMANN2022118859}. The variables that are optimized include the capacities (of different generating, storage, and conversion technologies) and how they are operated in every time step. Besides ensuring that the energy balance is fulfilled for every node and in every time, the optimization includes other constraints that e.g. limit the maximum global CO$_2$ emissions. The model measures the direct CO$_2$ emissions from electricity and heating production, industrial processes, and combustion of fossil oil, e.g., for e.g. internal-combustion engines in land transport. The model does not represent the full CO$_2$ footprint of each asset but only considers emissions throughout their operation.\\

\begin{figure}[!h]
	\centering
	\includegraphics[width=0.5\textwidth]{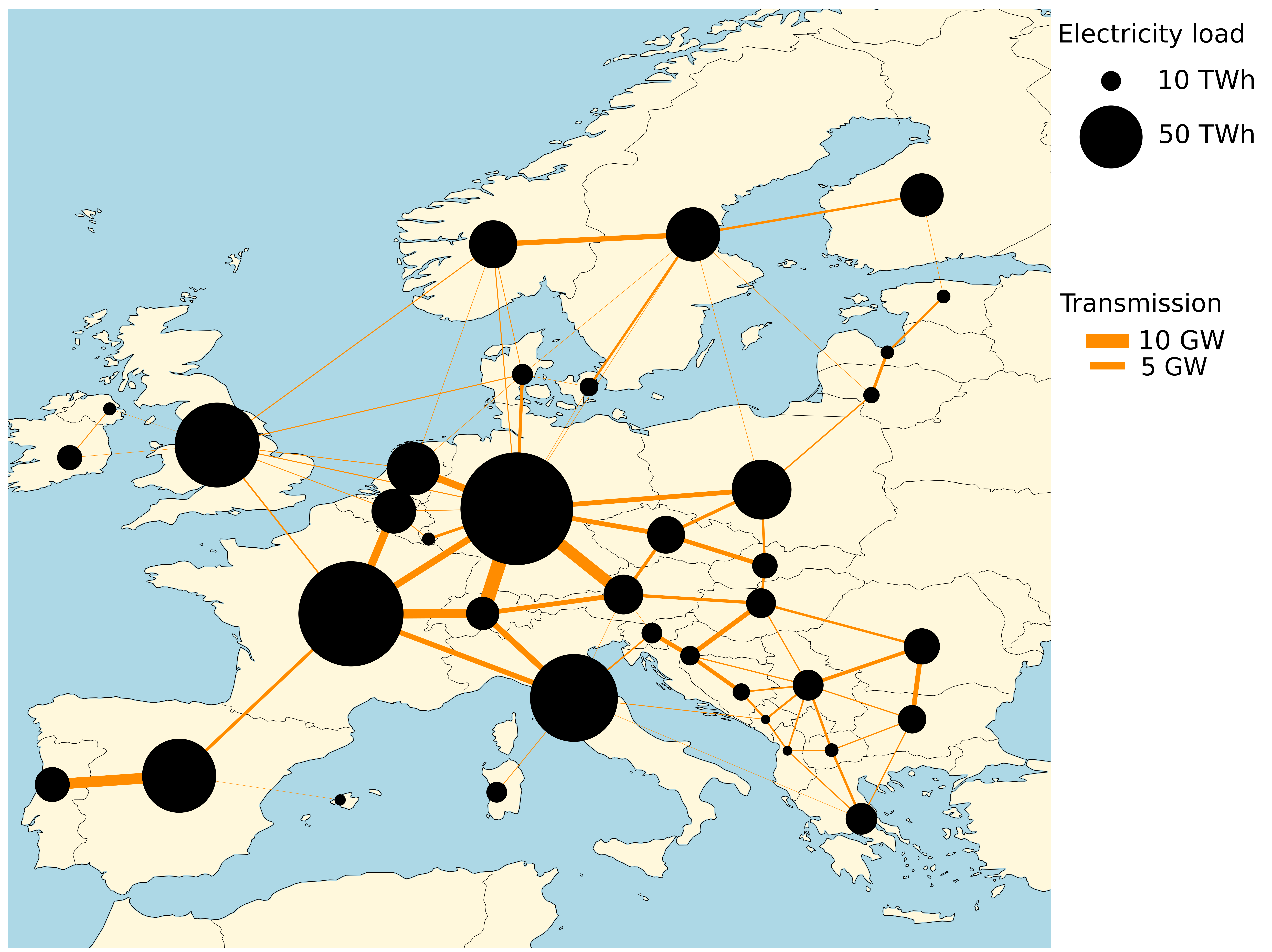}
	\caption{\textbf{Network topology}. The network consists of 37 nodes distributed over 33 European countries, all members of ENTSO-E. Italy, Spain, Great Britain and Denmark have multiple nodes since the countries cover different synchronous regions. The size of the nodes corresponds to the historical annual electricity consumption for 2013. The nodes are interconnected with AC and DC high-voltage lines corresponding to today's capacity including transmission lines projected in the TYNDP 2018 by ENTSO-E \cite{entsoe_tyndp}.}
	\label{fig:network_topology}
\end{figure}

In this study, the model is resolved with a network of 37 nodes spanning over 33 countries (see Fig. \ref{fig:network_topology}), all members of the European Network of Transmission System Operators for Electricity (ENTSO-E). Distribution grid investment cost is included but the distributional network is not modeled since we only focus on storage capacity allocation in the high-voltage grid. The optimization is computed for one year with a 3-hourly resolution. For this reason, only storage with duration (i.e., energy-to-power ratio) larger than or equal to 3 hours is considered for the capacity expansion since fluctuations within each time step are implicitly smoothened by the averaging. The internodal transmission capacity is exogenously included in the model and equals current transmission lines with the addition of lines under construction expected to be commissioned according to the Ten Year Network Development Plan (TYNDP 2018) by ENTSO-E \cite{entsoe_tyndp}. For clarity, we refer to Supplemental Tables S1 and S2, which contain the capacities of the included AC and DC interconnections, respectively.\\

The model assumes an ideal market with perfect competition between all included technologies and long-term market equilibrium, i.e., energy technology recovers exactly its full costs. Furthermore, the model assumes perfect foresight of energy supply and demand. This implies that the resources are distributed optimally over the year based on the year-ahead informed weather conditions and consumption patterns. For such a reason, any storage acting as a security backup against unforeseen energy droughts is not included. All technology costs and lifetime estimates are reported at Ref. \cite{technology_data}. Here, technology costs are acquired for 2030 to account for expected technology cost reduction while selecting a year relatively close to the present to reduce uncertainty in cost estimations. The investment costs are annualized with the reported lifetime estimates of each asset, assuming a discount rate of 7\%.

\begin{figure*}
	\centering
	\includegraphics[width=0.85\textwidth]{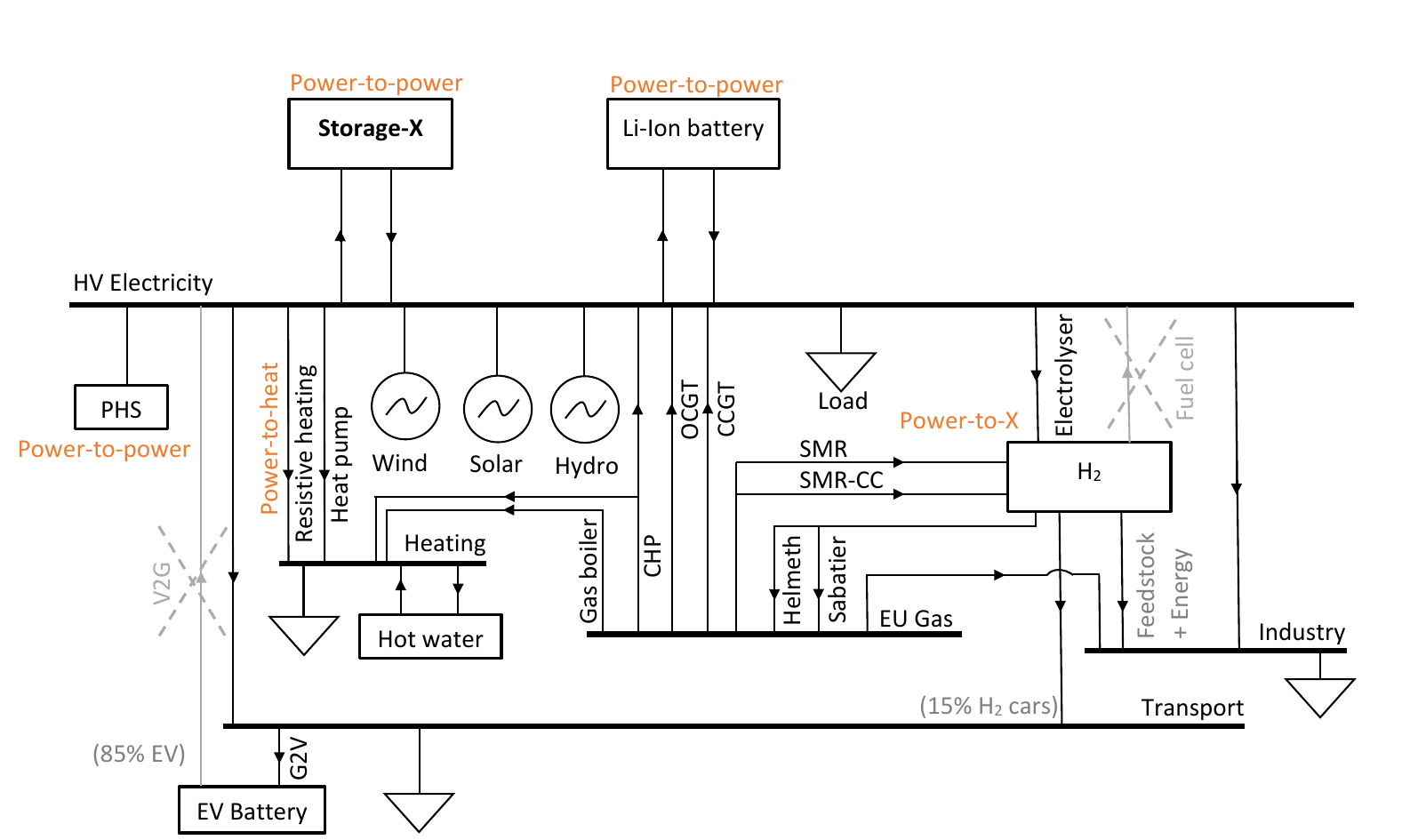}
	\caption{\textbf{Schematics of \mbox{\text{storage-X}} integration in the fully sector-coupled energy system}. \mbox{\text{Storage-X}} is linked to the high-voltage (HV) electricity bus. Hydrogen is only used as a power-to-X technology and, thus, does not have a direct link back to the HV electricity bus. Such setup (Power-H$_2$-Power) is intendedly resembled by the \mbox{\text{storage-X}}. Furthermore, smart charging, here named grid-to-vehicle (G2V), is included when adding land transport, allowing flexible charging of the EV batteries. Vehicle-to-grid (V2G), i.e., EV battery supplying power to the high-voltage electricity bus, is not allowed. In the fully sector-coupled system, solid biomass and biogas usage is also allowed which is not depicted in the diagram for simplicity.}
	\label{fig:method_energyflow}
\end{figure*}

The electricity load prior to optimization reflects historical national consumptions reported at ENTSO-E, collected from Open Power System Data \cite{openpowersystemdata}. This is based on 2013 and the data includes the electricity demand in the industry. The electricity load is assumed price-inelastic, which means that we do not include any demand response on the marginal electricity price. An additional electricity demand arises when including the heating and transport sectors while enforcing a CO$_2$ emissions limit. The CO$_2$ emissions limit causes a fraction of the heating demand to be electrified with the inclusion of heat pumps or resistive heaters, while the remainder can be covered by combined heat and power plants (CHP) or gas boilers. Land transport is included exogenously. Here, we assume that 85\% of the energy demand for land transport is delivered from BEV, while the remaining 15\% relies on hydrogen fuel cell vehicles (FCEV). The latter represent transport difficult to electrify such as heavy construction machinery or long-haul trucks for which a higher gravimetric energy density is more suitable. The industry sector includes the energy demand, e.g., for steel, iron, and aluminum production, as well as the production of chemicals such as ammonia and methanol, and the CO$_2$-intense cement production. The methane feed-in for some processes can be either fossil-based or synthetic, decided by the optimization. For other processes, carbon capture can be deployed, if cost-optimal. In addition, the model also includes energy demands in aviation (kerosene from synthetic or fossil oil) and shipping (H$_2$ and oil), as well as a European biomass supply (both solid and gaseous). For fuel production from synthetic and renewable gases, e.g., green hydrogen production with electrolyzers, additional power generation capacity is needed, compared to an electricity-only system. Fig. \ref{fig:method_energyflow} depicts a simplified scheme of how the generators, links, and storage are connected in the model. Each bus may have multiple loads and may be interlinked with multiple other buses which the presented schematic does not include for simplicity. For a detailed description of the sector representation, we refer to Ref. \cite{victoria2022speed}.\\

Here, we consider a future European energy system that is on the verge of achieving net-zero emissions. The annual net CO$_2$ emissions of the considered system are restricted to 5\% of the 1990-levels. Gross emissions can exceed this limit if they are compensated by negative-emissions technologies (direct air capture or carbon capture on point sources) which are deployed if cost-optimal. The optimization is performed \textit{overnight} which means that we consider one calendar year for the capacity deployment and energy dispatch without accounting for the pathway from today's system toward the resulting system. We investigate three system compositions (SC): 
\begin{enumerate}[
	leftmargin=1.5cm,
	label={SC\arabic*.},
	ref={SC\arabic*.}]
	\item \textbf{Electricity}:  The electricity demand is fixed to the historical levels while transforming the electricity supply to comply with the CO$_2$ emissions constraint, without the inclusion of other energy-consuming sectors.
	\item \textbf{Electricity + Heating + Land Transport}: In addition to transforming the electricity supply, the system includes the heating sector and the energy consumption in the land transport sector.
	\item \textbf{Fully sector-coupled}: Including the energy consumption in the heating, land transport, industry including feedstock, aviation, and shipping sector, and the supply of biomass.
\end{enumerate}

The main model assumptions specific to this study are summarized in Table \ref{tab:model_assumptions}.

\begin{table}[!h]
	\caption{\textbf{Model assumptions in this study}. Temporal and spatial resolution, technology costs assumption year, transmission volume level, and the net CO$_2$ emissions constraint used in the three system compositions (SC1 - SC3).}
	\label{tab:model_assumptions}
	\centering
	\begin{threeparttable}
		\begin{tabular*}{0.45\textwidth}{@{}lcl@{}}
			\toprule
			Assumption               			& &\textcolor{white}{text}\\ \midrule
			Type								& & Overnight optimization \\
			Time resolution      				& & 3 h \\
			Network resolution 					& & 37 nodes \\
			Weather year                        & & 2013 \\
			Technology costs					& & 2030 \\
			Transmission 				        & & Fixed to today + TYNDP \\
			Net CO$_2$ emissions			   	& & 5\% relative to 1990  \\
												& & \hspace{0.5cm} SC1: 74.1 MtCO$_2$\\
												& & \hspace{0.5cm} SC2: 148.2 MtCO$_2$\\
												& & \hspace{0.5cm} SC3: 229.9 MtCO$_2$\\
			\bottomrule
		\end{tabular*}
	\end{threeparttable}
\end{table}


\subsection{Generators}

The renewable generators available to the system are wind (off- and onshore), solar PV (utility and rooftop), and hydropower (reservoir and run-of-river). We distinguish between two voltage levels in the grid: High voltage (transmission grid) and low voltage (distribution grid). The renewable generators are connected to the high-voltage grid, except solar rooftop which is attached at low-voltage. For wind and solar, the installable capacity is limited by the estimated potentials based on the available land use from the Corine Land Cover (CLC) database \cite{corine} subtracting Natura 2000 protected areas \cite{natura2000}. The following limits are imposed, based on the study by Victoria et~al. \cite{victoria2022speed}: On- and offshore wind are limited to 20 \% of available land, utility-scale solar PV to 9 \%, whereas rooftop PV potential is estimated according to population density. Hydropower generation capacities, comprised of run-of-river and reservoir, are included exogenously since most of the potential is assumed to be exploited, and kept fixed at today's capacity (34.5 GW for run-of-river and 99.6 GW for reservoir) \cite{GOTZENS20191}). Time series of hydro inflow as well as wind and solar capacity factors are obtained with ATLITE \cite{Hofmann2021} and ERA5 \cite{era5}, using weather data for 2013. In every 3-hour time step, wind turbines and solar panels can provide electricity to the grid proportional to the concurrent capacity factor. In the case of exceeding the demand and/or congestion in the transmission, the excess energy is curtailed. The curtailed energy in each time step is calculated as:

\begin{equation}\label{eq:curtailment}
	\text{Curtailed energy}_t \text{ (\%)} = \frac{\sum_{n,s} (\bar{g}_{n,s,t} G_{n,s} - g_{n,s,t})}{\sum_{n,s} \bar{g}_{n,s,t} G_{n,s}}
\end{equation}

where $\bar{g}_{n,s,t}$ is the capacity factor at time $t$ of a variable renewable generator $s$, $G_{n,s,t}$ is the installed power capacity, and $g_{n,s,t}$ is the actual electricity generation.\\

Open-cycle gas turbine (OCGT), combined cycle gas turbine (CCGT), nuclear, and coal power plants can be deployed if cost-optimal. Furthermore, in SC2 and SC3, the system can deploy combined heat and power (CHP) plants fueled with gas or biomass. Biomass and gas-fired CHP plants have the option of installing carbon capture at an additional investment and operational cost. Investment and O\&M cost predictions and lifetime estimates acquired from Refs. \cite{DEA2022_gen,DEA2022_carbon} of the power generation technologies subject to optimization are presented in Supplemental Table S3. Capacities and locations of hydropower generation and pumped-hydro storage units which are not optimized, but included exogenously, are acquired from Powerplantmatching v0.5.3 \cite{GOTZENS20191}. 

\subsection{Hydrogen}
H$_2$ can be produced with water electrolysis powered by electricity from the high-voltage grid. Furthermore, it can be produced from steam-methane reforming (SMR) with or without carbon capture (CC), corresponding to blue and grey H$_2$ respectively. H$_2$ network is installed if cost-optimal. Lastly, H$_2$ can be converted into methane through the Helmeth or Sabatier reactions. We disallow H$_2$ in the model to dispatch directly as electricity back into the high-voltage (HV) electricity (i.e., fuel cells are omitted) since we want to decouple the hydrogen production as part of a \textit{power-to-X} strategy and hydrogen storage used for grid balancing. A hydrogen storage disconnected from any other bus than the HV bus, charged with electrolyzers and discharged with fuel cells, is in this study represented by \mbox{\text{storage-X}}. In addition, a previous study with sector coupling suggests that only a minor fuel cell capacity for grid balancing is deployed (approximately 2~GW European-aggregate) in a sector-coupled European energy system \cite{PEDERSEN20221566}.W

\subsection{Storage}
Storage can be deployed in every node of the network. Existing PHS is included to the model before the optimization, whereas the capacity of stationary Li-ion batteries and \mbox{\text{storage-X}} is optimized. We do not include existing storage infrastructure that solely operates within other markets than the balancing, e.g., primary frequency response. Besides electricity storage, the model can select hot-water tanks and H$_2$ storage to be deployed as well, to shift the production from the heating and H$_2$ consumption in time. EV batteries with smart charging are included when coupling with the land transport sector and act as a flexible demand.

\subsubsection{Pumped-hydro storage}
Existing PHS in Europe constitutes approximately 55 GW power capacity \cite{iha2021} and 1.3~TWh \cite{GETH20151212}. Gimeno-Gutierrez and Lacal-Arantegui \cite{GIMENOGUTIERREZ2015856} identify a technical potential for new PHS projects in Europe, including social and environmental constraints, but here we assume the cost-optimal potential to be fully exploited. Thus, PHS is added exogenously to the model, keeping it fixed at today's capacity. Using plant-specific data from Powerplantmatching v0.5.3 \cite{GOTZENS20191}, this aggregates to 56 GW power capacity, and using discharge times from Geth et~al. \cite{GETH20151212}, this corresponds to 1.4~TWh energy capacity. Table \ref{tab:phs_storage_costs} summarizes the PHS specifications used in our calculation.


\begin{table}[h]
	\caption{\textbf{Technology cost and efficiency assumptions for PHS}. Assumptions are obtained from \cite{Schroder2013Current}.}
	\label{tab:phs_storage_costs}
	\begin{threeparttable}
		\begin{tabular}{@{}lccccccccc@{}}
			\toprule
			Parameters             & & & & PHS & & & & &\\ \midrule
		  Investment cost    & & & & 2208 €/kW & & & & &\\
		  Lifetime   & & & & 80 years & & & & &\\
		  FOM                & & & & 1.0 \% & & & & &\\
		  Round-trip efficiency & & & & 75 \% & & & & &\\
		  Total energy capacity & & & & 1.4 TWh & & & & &\\
		  Total power capacity & & & & 56 GW & & & & &\\
		\bottomrule
		\end{tabular}
	\end{threeparttable}
\end{table}

\subsubsection{Stationary battery storage}
Battery storage in the high-voltage grid is included in the objective of the optimization problem, where energy and power capacity are subject to independent optimization. The cost and performance assumptions reflect data from the Danish Energy Agency \cite{DEA2020} on utility-scale Li-ion batteries. 

Residential batteries are also subject to optimization and included in the low-voltage grid, with cost assumptions from Ram et~al. \cite{energywatchgroup2019}. Assumed investment and O\&M costs in 2030 are presented in Table \ref{tab:bat_storage_costs}.

\begin{table}[h]
	\caption{\textbf{Technology assumptions for battery storage}. Estimates are for the year 2030, acquired from Refs. \cite{DEA2020} and \cite{energywatchgroup2019}.}
	\label{tab:bat_storage_costs}
	\centering
	\begin{threeparttable}
		\begin{tabular}{@{}lcccc@{}}
			\toprule
			Storage                & Investment cost & Lifetime (Years) & FOM \\ \midrule
			Utility       & 142 €/kWh     & 25               & 0 \\
			-   Inverter\tnote{*}      & 160 €/kW      & 10               & 0.34 \% \\
			
			Residential  & 202.9 €/kWh   & 25               & 0 \\	
			-   Inverter \tnote{*} & 228.06 €/kW   & 10               & 0.34 \% \\
			\bottomrule
		\end{tabular}
		\begin{tablenotes}
			\item[*] Bidirectional. Battery round-trip efficiency $\eta = 96 \%$
		\end{tablenotes}
	\end{threeparttable}
\end{table}

\subsubsection{EV Battery}
For a total fleet of 217 million BEVs (corresponding to 85\% of the total number of cars according to JRC \cite{JRCIDEES2015}), assuming 50 kWh batteries with 11 kW chargers, an availability of 50\% for demand-side management, this represents a cumulative EV battery storage of 5.44~TWh energy capacity and 2.39 TW charge power capacity. Here, only uni-directional smart charging is considered a flexible option, i.e., BEV can charge when it is best for the system. The charging pattern is then subject to the optimization. The BEV battery's SOC is constrained to a minimum of 75\% at 7 am so that BEVs are ready for typical commuting. Dispatching electricity back to the HV electricity bus, i.e. vehicle-to-grid (V2G), is not allowed in this study, resembling a lack of user willingness to participate due to a reduced battery lifetime \cite{GONZALEZGARRIDO2019381}. 


\subsubsection{Storage-X}\label{subsec:storage_x}
Similar to the study by Sepulveda et~al. \cite{Sepulveda2021}, we enclose all feasible combinations into a design space of successful storage configurations. Here, we classify storage as successful if it is able to play a substantial role in the decarbonized energy system. To do so, we represent each of the six \mbox{\text{storage-X}} parameters with a discrete range (shown in Table \ref{tab:storage_reference}) and confine all combinations in a sample space $Q$, here defined as the Cartesian product of all parameter sets:

\begin{equation}
	Q = \{\hat{c}\} \times \{c_c\} \times \{c_d\} \times \{\eta_c\} \times \{\eta_d\} \times \{\tau_{SD}\}
\end{equation}

Every configuration within the sample space represented with sample $q \in Q$ is used as an input to calculate the optimum capacity and dispatch of the European energy system for the three system compositions (Electricity, Electricity + Heating + Land Transport, Fully sector-coupled). For every $q$, we assume a fixed operation and maintenance (FOM) cost of 1\% of investment cost for the storage and 2\% for charge and discharge components, equivalent to assumptions for a H$_2$ electricity storage in Ref. \cite{DEA2020}. Moreover, every configuration $q$ has a lifetime of 30 years. Similar to other technologies, a financial discount rate of 7\% is used in the annualization of the capital cost. 
\begin{table}[h]
	\caption{\textbf{Storage-X parameter sets}. Cost, efficiency, and self-discharge for a fixed storage configuration (first column) and the sample space (second column).}
	\label{tab:storage_reference}
	\centering
	\begin{threeparttable}
		\begin{tabular*}{0.5\textwidth}{@{}lcccc@{}}
			\toprule
			Parameter                           & \textcolor{white}{text} & Fixed & \textcolor{white}{text}	& Sets \\ \midrule
			$\hat{c}$      &                  	& 3 €/kWh 		& &  \{1,2\tnote{*},5, 10\tnote{**}, 20, 30, 40\}\\
			$c_c$    &                         	& 350 €/kW$_e$	& &  \{35, 350, 490, 700\} \\
			$\eta_c$ 		    &               & 50\%			& &  \{25, 50, 95\}\\
			$c_d$ &                         	& 350 €/kW$_e$ 	& &  \{35, 350, 490, 700\}\\ 
			$\eta_d$	    &                   & 50\% 			& &  \{25, 50, 95\}\\
			$\tau_{SD}$ 	&                   & 30 days 		& &  \{10, 30\}\\
			\bottomrule
		\end{tabular*}
		\begin{tablenotes}
			\item[*] 2030 cost of underground cavern H$_2$ storage \cite{DEA2020}
			\item[**] 2030 cost of H$_2$ storage tank \cite{DEA2020}
		\end{tablenotes}	
	\end{threeparttable}
\end{table}

As proxies for the potential of every $q$, we calculate the Europe-aggregate energy capacity $E$, 

\begin{equation}\label{eq:europeagg_E}
	E = \sum_{n=1}^{37} \eta_d E_{n} \,,
\end{equation}

and the load coverage $LC$, 

\begin{equation}\label{eq:load_coverage}
	LC = \frac{\sum_{n,t} |\Delta e^-_{n,t}|}{\sum_{n,t} l_{n,t}} \,,
\end{equation}

where, $E_n$ is the nodal energy capacity of \mbox{\text{storage-X}}, $l_{n,t}$ is the electricity load, and $|\Delta e^-_{n,t}|$ is electricity dispatched from \mbox{\text{storage-X}} in node $n$ at time $t$. The energy capacity $E_n$ is the maximum storable content in a considered storage in node $n$ in units of energy before discharging. The configuration space contains different discharge efficiencies, and to account for this, the optimal aggregated energy capacity $E$ is converted into units of dispatchable electricity by multiplying with the discharge efficiency $\eta_d$. We use the energy capacity as a metric to represent the storage market potential, similar to what was done by Ref. \cite{Parzen}. Here, we also use the load coverage, equal to the amount of electricity dispatched by the storage normalized with the total load, to indicate the potential contribution to covering power imbalances. Here, the total load accounts for the base load (national electricity load, including demand from the industry) and the additional load from sector coupling (e.g., the electrified heating supply, BEV in the land transport, the large H$_2$ production needed in a decarbonized sector-coupled energy system, etc.). 

Additional balancing from backup capacities has shown to be reduced substantially when an ideal storage (i.e., 100\% round-trip energy efficiency) with an energy capacity equivalent to 6 times the average hourly load (av.h.l.$_{el}$) is deployed in the system \cite{RASMUSSEN2012642}. This corresponds to approximately 2~TWh energy capacity for the European system prior to the inclusion of electricity demand for heating and land transport. Furthermore, 2~TWh energy capacity compares well with the average European storage requirement in the two ENTSO-E 2040-scenarios \textit{Distributed Generation} (1.348~TWh) and \textit{Sustainable Transition} (2.518~TWh) \cite{electronics8070729}. To encapsulate all successful storage configurations, we define, subsequent to the optimization, a lower threshold of $E\geq$2~TWh that the storage needs to fulfill. This is evaluated for all configurations within the sample space. The configurations from the sample space fulfilling this condition are collected and constitute the \mbox{\text{storage-X}} design space. The design space is the subset of the parameter values listed in Table \ref{tab:storage_reference}, where the system finds a substantial deployment of \mbox{\text{storage-X}}, more precisely an energy capacity of at least 2~TWh.\\

To reduce computation time, when repeating the calculations for the same sample space $Q$ in the three system compositions, we introduce a filter $M$ which omits the samples of which the ratio of capacity cost and efficiency leads to an insignificant capacity deployment in the first round of computations with the "Electricity" system, Table \ref{tab:model_assumptions}. With this approach, the uniform sample space $Q$ containing 2,016 configurations is reduced to 724 configurations (see \ref{appendix:filter}).

\subsection{Relative parameter importance}

To compare the relative importance of the parameters determining the optimal \mbox{\text{storage-X}} deployment, we perform a multivariable (Ordinary Least Squares) regression subsequent to the optimization. Here, a log-linear model $\log \hat{E}$ is used to relate the optimal energy capacity $E$ (the response variable) to the storage parameters $x$ (the features):
\begin{equation}\label{eq:GLM}
  \log \hat{E} = \beta_0 + \sum_{k=1}^{6} \beta_k \bar{x}_k
\end{equation}
where the vector $\beta$ contains the regression coefficients for the six parameters. The energy capacity is log-transformed with the natural logarithm to linearize the response. Since the parameters $x$ differ in range and units, we normalize them using a min-max scaling: 
\begin{equation}
	\bar{x}=\frac{x-\min{(x)}}{\max{(x)}-\min{(x)}}
\end{equation}

Likewise, a min-max scaling is applied to the log-transformed response variable. Following the regression, the coefficients are normalized:
\begin{equation}
	|\bar{\beta}| = \frac{\beta}{\max{(\beta)}} \,, 
\end{equation}
thus, $|\bar{\beta}|=1$ indicates the highest relative importance. 

The regression is validated based on the adjusted $R^2$. An additional measure of parameter importance $\Delta R^2_k$ of parameter $k$ is calculated based on a decomposition of the adjusted $R^2$: 

\begin{equation}
	\Delta R^2_k = R^2_{Z} - R^2_{k \notin Z} \,,
\end{equation}
where $Z$ indicates the full set of parameters, and $k \notin Z$ indicates the set of parameters omitting parameter $x_k$ in the regression. A large value of $\Delta R^2_k$ indicates that parameter $x_k$ explains a high share of the variance of the output (optimal energy capacity). 

\section{RESULTS - STORAGE-X}\label{sec:storage_x}

In the following subsections, the \mbox{\text{storage-X}} capacity deployment in the cost-optimal system layout is analyzed as a function of the parameters listed in Table \ref{tab:storage_reference}. Furthermore, we evaluate requirements on the cost and efficiency of \mbox{\text{storage-X}} to contribute significantly to the system layout, while competing with other more mature storage, backup, or flexibility options. The results are obtained for three system compositions, SC1: "Electricity", SC2: "Electricity + Heating + Land Transport", and SC3: "Fully sector-coupled". 

\subsection{Implication of sector coupling}

\begin{figure*} 
	\centering
	\includegraphics[width=0.8\textwidth]{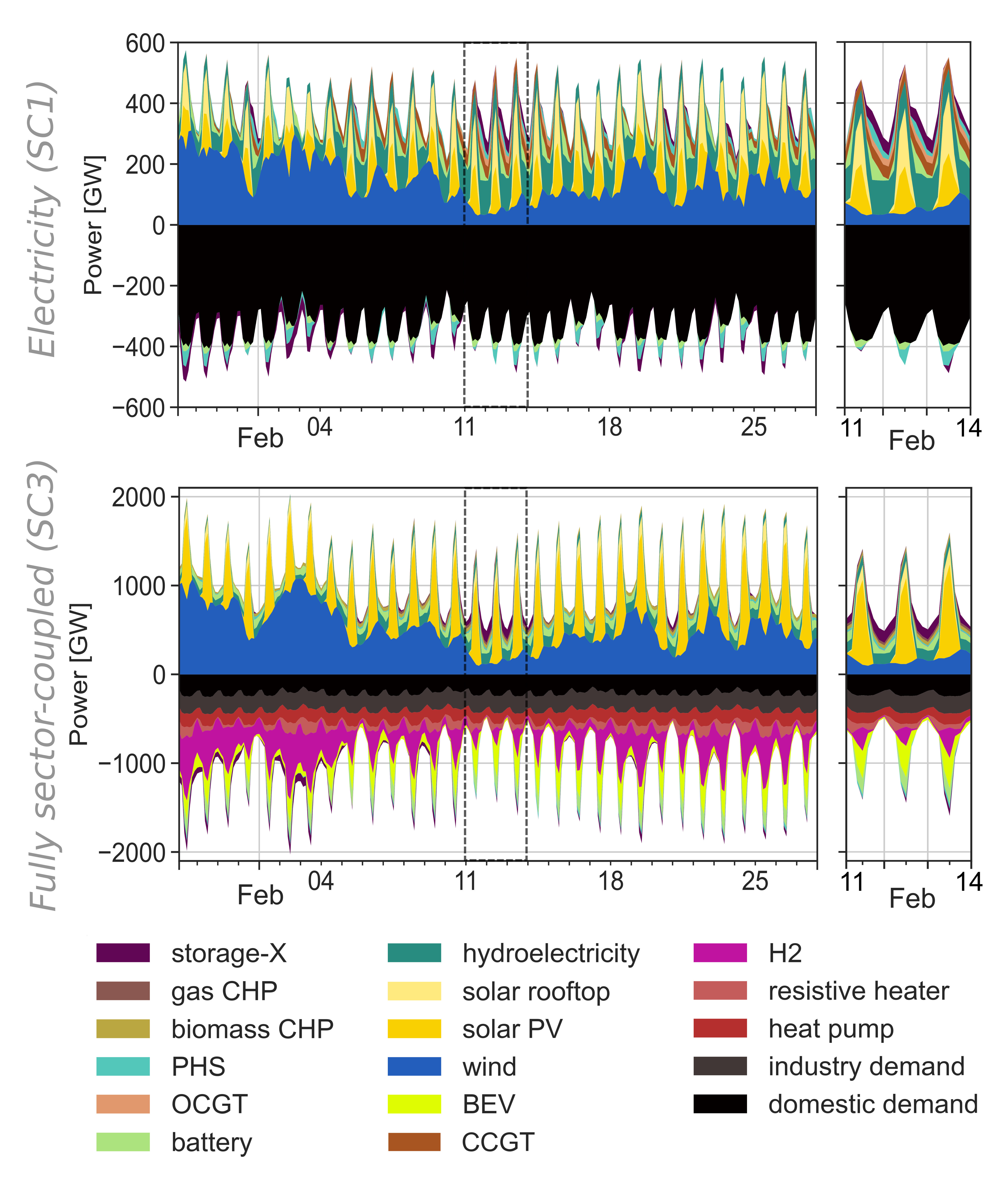}
	\caption{\textbf{Different mechanisms to balance renewable droughts}. Europe-aggregate energy balance in a period with low wind production for the Electricity system (top) and the Fully sector-coupled system (bottom). The results are obtained with the inclusion of the "Fixed" \mbox{\text{storage-X}} configuration (Table \ref{tab:storage_reference}) but with the discharge efficiency altered to 95\%. The figure indicates a renewable drought in which \mbox{\text{storage-X}} is discharged from the 11$^\text{th}$ to the 14$^\text{th}$ of February) caused by low wind production. Note that the y-axis range is different for the subfigures. See Supplemental Fig. S1 for the full year.}
	\label{fig:balancing_drought}
\end{figure*}

To gain insight into the differences in system compositions and their implications for \mbox{\text{storage-X}} deployment, we analyze how the same period with low renewable production, signified by a drop in wind resources, is balanced in two different systems. Specifically, Fig. \ref{fig:balancing_drought} compares the purely electricity-supply system (SC1) with the fully sector-coupled system (SC3). The results are obtained using the 'Fixed' \mbox{\text{storage-X}} configuration (Table \ref{tab:storage_reference}), with the discharge efficiency adjusted to 95\%. In SC1, the system response to the wind scarcity is to discharge \mbox{\text{storage-X}} together with PHS and to activate the fossil-fueled OCGT. In addition, the CCGT is used to cover imbalances over a longer period. In SC3, the main mechanisms to compensate for the drop in wind production are to discharge \mbox{\text{storage-X}} and to reduce the production of H$_2$. The difference in the balancing mix between the two systems can be explained by how the CO$_2$ emissions are distributed. The model does not impose sector-specific CO$_2$ emissions constraints, but instead, a global CO$_2$ emissions constraint, and the model then determines how to distribute the emissions across the sectors. The two systems, albeit the same CO$_2$ emissions constraint, distribute the emissions differently. In SC1, all of the allowed CO$_2$ emissions (5\% of 1990-levels) are allocated in the electricity supply to run fossil-fueled power plants in times with renewable droughts. When connecting with other sectors (S2/S3), a large share of the emissions cap is moved to other CO$_2$ sources that are more difficult to fully decarbonize (e.g., cement and steel production in the industrial sector) which consequently disallows a large share of the emissions in the electricity supply. Because of this, the model chooses fossil-fueled backup capacities in SC1, but not necessarily in SC2. For this reason, the sector-coupled systems in this study may require a higher volume of \mbox{\text{storage-X}}, despite the new flexibility options enabled by sector coupling.\\

To further investigate the impact of sector coupling, we examine how the system responds to different levels of CO$_2$ emissions constraints. We use the 'Fixed' \mbox{\text{storage-X}} configuration and conduct simulations by gradually reducing the CO$_2$ emissions from 10\% of the 1990-levels to 0\%. When more stringent CO$_2$ emissions constraints are imposed, the \mbox{\text{storage-X}} capacity deployment in SC1 exceeds that of the sector-coupled systems (Supplemental Fig. S2b). Upon achieving net-zero emissions, the storage energy capacity in SC1 reaches 12~TWh for the Fixed storage configuration. In contrast, the impact on storage energy capacity in the sector-coupled systems is less noticeable since their electricity supply is already almost fully decarbonized. For the Fixed storage configuration, storage energy capacity in SC2 and SC3 reaches 300 GWh at net-zero emissions.


\subsection{Single parametric sweep}


\begin{figure*}
	\centering
	\includegraphics[width=0.9\textwidth]{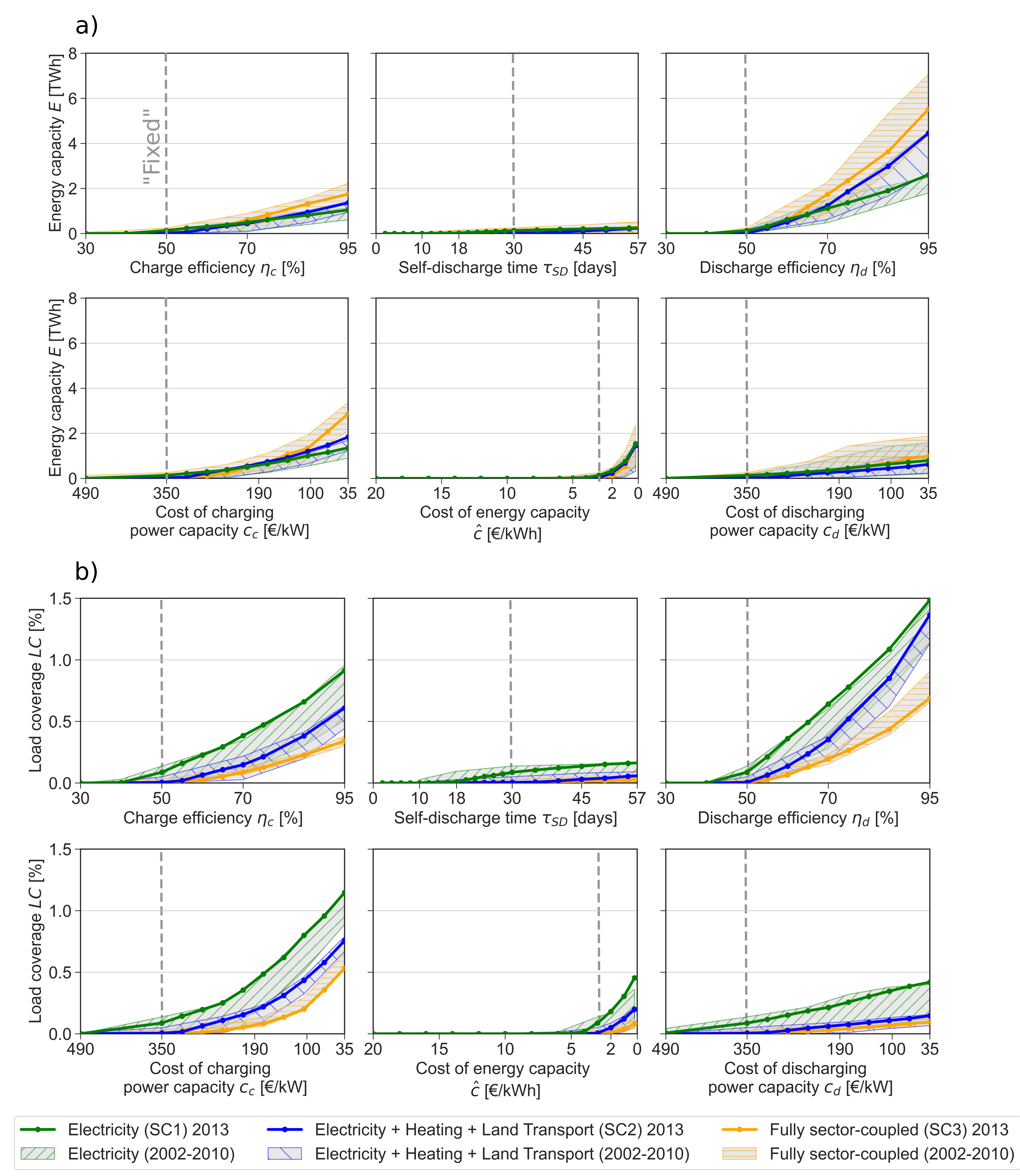}
	\caption{\textbf{Europe-aggregate (a) energy capacity, Eq. \ref{eq:europeagg_E}, and (b) Load coverage, Eq. \ref{eq:load_coverage} of \mbox{\text{storage-X}} in the cost-optimal system design}. Results are obtained by varying one \mbox{\text{storage-X}} parameter at a time for the 2013-weather year (solid lines) while keeping the remaining parameters fixed according to the "Fixed" configuration (Table \ref{tab:storage_reference}). This was repeated for an interval of different weather years from 2002 to 2010 (hatched area). The energy capacity is in units of dispatchable electricity. A comparison with existing PHS and optimized battery capacities are shown in Supplemental Figures S3-S5.}
	\label{fig:single_parametric_sweep}
\end{figure*}

Here, we consider the case in which one parameter is altered at a time while keeping the remainder fixed according to Table \ref{tab:storage_reference}. Fig. \ref{fig:single_parametric_sweep} depicts the resulting energy capacity deployment (top panel) and load coverage (bottom panel), used as proxies for storage potential, as function of the six \mbox{\text{storage-X}} parameters. The single-parametric sweep is performed for all three system compositions. As discussed, for all parametric ranges, the sector-coupled systems entail larger energy capacity deployment of \mbox{\text{storage-X}}. Concurrently, due to the increased electricity demand, the relative contribution to meeting the load, i.e., the load coverage, is reduced. The output shows different responses between the system compositions due to differences in the scaling of the electricity supply. This is consistent for all parameter levels. 

For the parameter alterations, we see a consistent response for all three system compositions: An improved discharge efficiency, as well as a reduction in the charge capacity cost, reveals the highest increase in both energy capacity deployment and load coverage, over the full parameter range. Conversely, storage self-discharge due to standing losses shows a minor impact. It is, however, noticeable for SC1 that to obtain a non-zero load coverage, the self-discharge time needs to be above 18 days (10 days when examining other weather years). The negligible gain of having low standing losses can be explained by \mbox{\text{storage-X}} not being used by the model as seasonal storage. Intermediate parameters, i.e., the parameters which cause higher response than self-discharge time but lower than discharge efficiency and charge capacity cost, are the charge efficiency, energy capacity cost, and discharge power capacity cost. Noteworthy is the response to reducing energy capacity cost which is negligible until reaching $\leq$5 €/kWh. Below this limit, the parameter improvement leads to a noticeable increase in energy capacity within a small domain, i.e. the response curve is subject to a steep slope, which explains why previous studies find this, together with discharge efficiency, to be the most determining parameter. 

The choice of weather year has shown to impact the optimal storage capacity deployment in the literature \cite{Ruhnau2021Storage}. To address this, we first perform a sweep across the years with the fixed \mbox{\text{storage-X}} configuration, defined in Table \ref{tab:storage_reference}) from 1997 to 2012 and compare this with the results obtained for 2013. Here, in accordance with literature, we observe a high sensitivity on the optimal storage energy capacity to the considered weather year (Supplemental Fig. S2a). The impact of weather input is however not consistent across the three system compositions. For SC1, the maximum \mbox{\text{storage-X}} deployment occurs using 2010 weather data, while for SC2, the same year leads to a minimum deployment. To account for this sensitivity, we include additional weather years in the single-parametric sweep to compare with the results obtained to the reference year (2013). The years (2002, 2003, 2006, 2008, 2010) are picked based on their distinct impact on the optimal energy storage capacity. These results are added to Fig. \ref{fig:single_parametric_sweep} (hatched areas). We see that, despite a considerable sensitivity to the weather year, the parametric sensitivity is stronger; thus, the observations made for one weather year still holds.\\

Enhancing the design of \mbox{\text{storage-X}} not only increases its capacity build-out, but also affects the deployment of other storage options due to competition. This is illustrated in Supplemental Figures S3 - S5, which compares the capacities of the available electricity storage in the three systems. Here, stationary battery capacity is the only electricity storage competitor to \mbox{\text{storage-X}} since PHS capacity is fixed. The reduction in battery power capacity is particularly noticeable when improving the discharge efficiency of \mbox{\text{storage-X}}. At the highest level of \mbox{\text{storage-X}} discharge efficiency improvement (from 30\% to 95\%), the battery power capacity is reduced by 64\% in SC1 (27\% and 23\% in SC2 and SC3). We conclude that a storage with a high discharge efficiency has the highest competitiveness with Li-ion batteries. 

\subsection{Design space of storage-X}\label{subsection:designspace}

\begin{figure}[!t]
	\centering
	\includegraphics[width=0.45\textwidth]{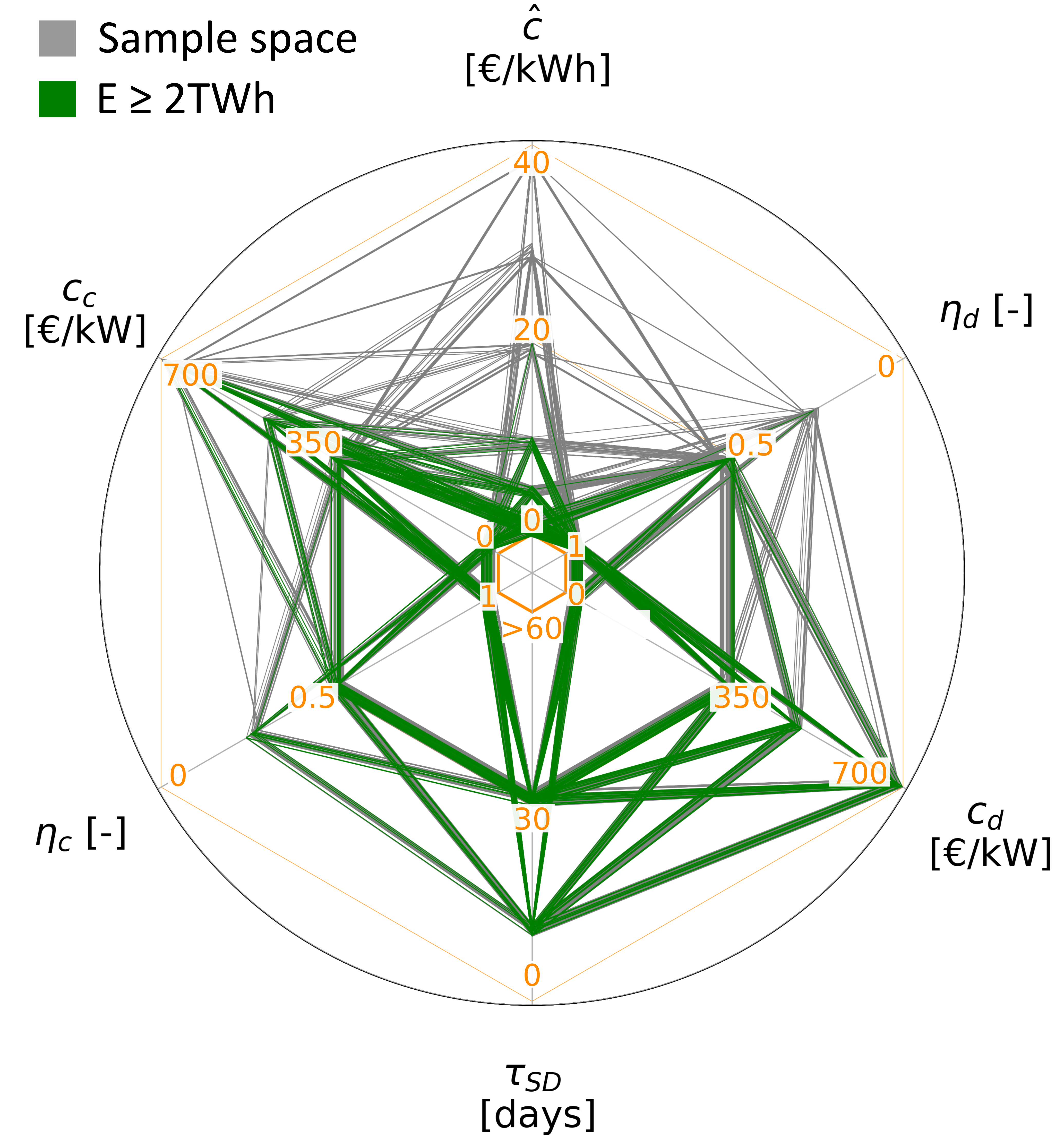}
	\caption{\textbf{Configurations fulfilling $\geq$2~TWh for the Electricity system (SC1)}. Sample space (grey) and the derived design space (green) containing the configurations entailing $E \geq 2$~TWh. See Supplemental Fig. S6 for a similar depiction of the sector-coupled systems, and Supplemental Fig. S7 for the frequency of each parameter level within the design space.}
	\label{fig:design_space_E}
\end{figure}

The single-parametric sweep indicates a sensitivity of the resulting storage energy capacity and load coverage to individual storage parameter adjustments. As the results are based on specific values of five fixed parameters from Table \ref{tab:storage_reference}, changing these parameters could lead to different indications. In order to generalize our findings, we run a multi-parametric sweep based on the predefined sample space of 724 storage configurations as described in section \ref{subsec:storage_x}. With the requirement of $E\geq 2$~TWh, we show how the sample space is reduced to the resulting design space in Fig. \ref{fig:design_space_E}. The figure depicts the initial configurations of the sample space and the 205 configurations that qualify for the design space in SC1 (for depictions of SC2 and SC3, see Supplemental Fig. S6). A key selection criterion is a low energy capacity cost as the number of combinations in the design space dramatically drops when increasing the level of this parameter. Only a few combinations are present at an energy capacity cost of 20~€/kWh. A similar observation can be made for the discharge efficiency which shows only a few combinations at 25\%. The remaining axes do not, in this depiction, indicate similar limitations. The same finding is made for the sector-coupled systems, but in that case, a higher energy capacity cost is allowed (configurations are observed with 30~€/kWh for SC2 and 40~€/kWh for SC3), inviting a higher number of configurations to enter the design space.


From the linear regression (Eq. \ref{eq:GLM}), the variation of the optimal capacity deployment can also to a large extent be explained by the energy capacity cost and the discharge efficiency. This is indicated in Table \ref{tab:coefficients} by the normalized regression coefficients and the increase in $R^2$ (shown in percentage points) by adding a parameter as a descriptor to the regression. The largest coefficients are observed for the energy capacity cost $\hat{c}$ and discharge efficiency $\eta_d$ across all system compositions. They furthermore lead to the largest increase in the adjusted $R^2$. Thus, the regression suggests that the highest relative importance is within these two parameters. Following $\hat{c}$ and $\eta_d$, the third largest coefficients are observed for the charge capacity cost $c_c$.

\begin{table}[!h]
	\caption{\textbf{Relative importance of \mbox{\text{storage-X}} parameters estimated by normalized regression coefficients}. The table shows the normalized coefficients from the linear regression model (Eq. \ref{eq:GLM}) of the optimal storage energy capacity of 724 storage samples. Prior to the regression, the parameters are scaled to range from 0 to 1 (min-max scaling) and the response variable is transformed with a natural logarithmic. The regression shows an adjusted $R^2$ of 74.6\%, 74.5\%, and 76.7\% for SC1, SC2, and SC3, and all coefficients are significant ($p<0.05$). The grey annotations show the increase in percentage points of the adjusted $R^2$ by adding the parameter to the fit, $\Delta R^2_k$. The color bands highlight the parameters with the highest coefficients and $\Delta R^2_k$.}
	\label{tab:coefficients}
	\includegraphics[width=0.5\textwidth]{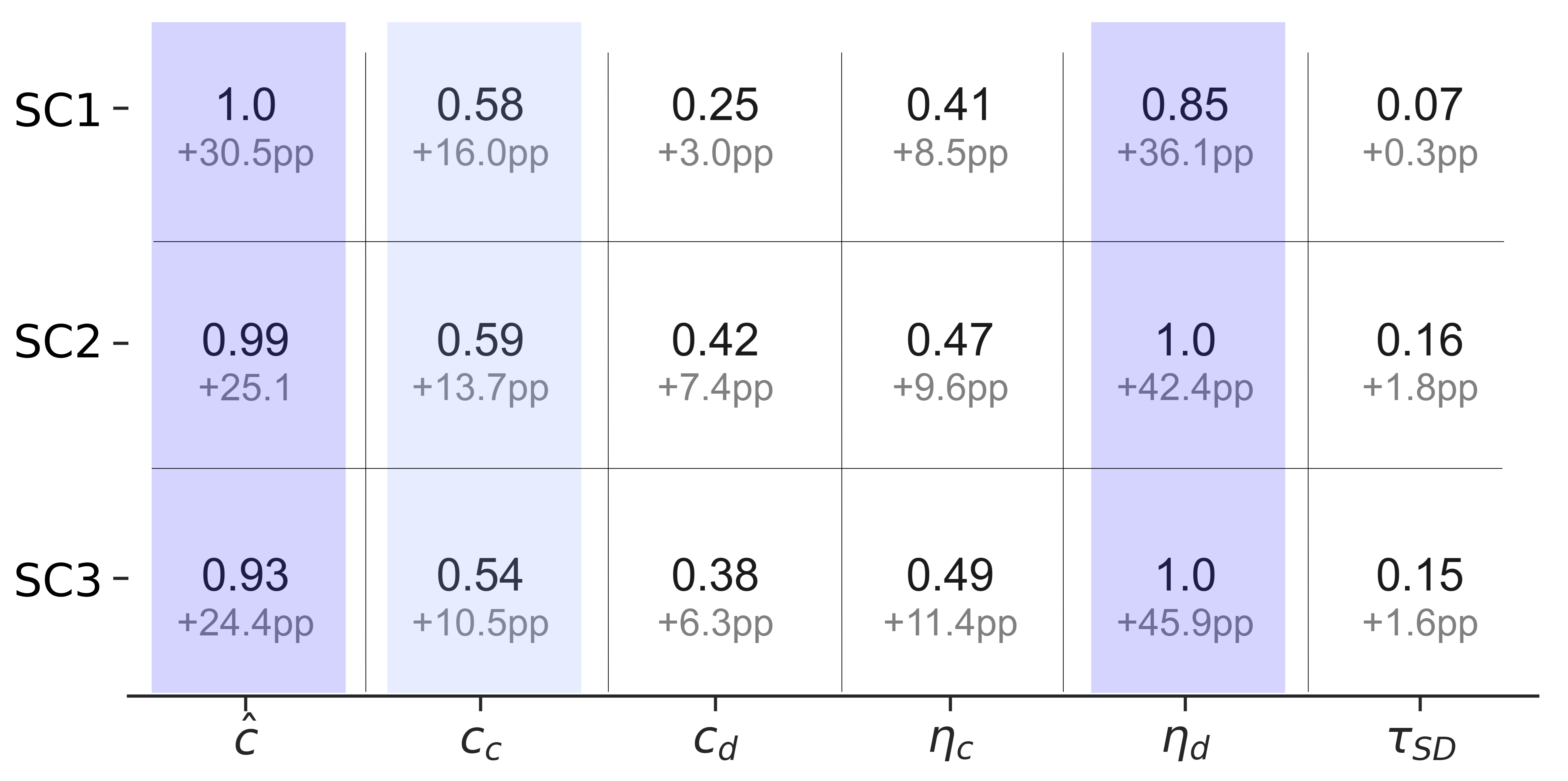}
\end{table}

\begin{figure*}[t]
	\centering
	\includegraphics[width=0.75\textwidth]{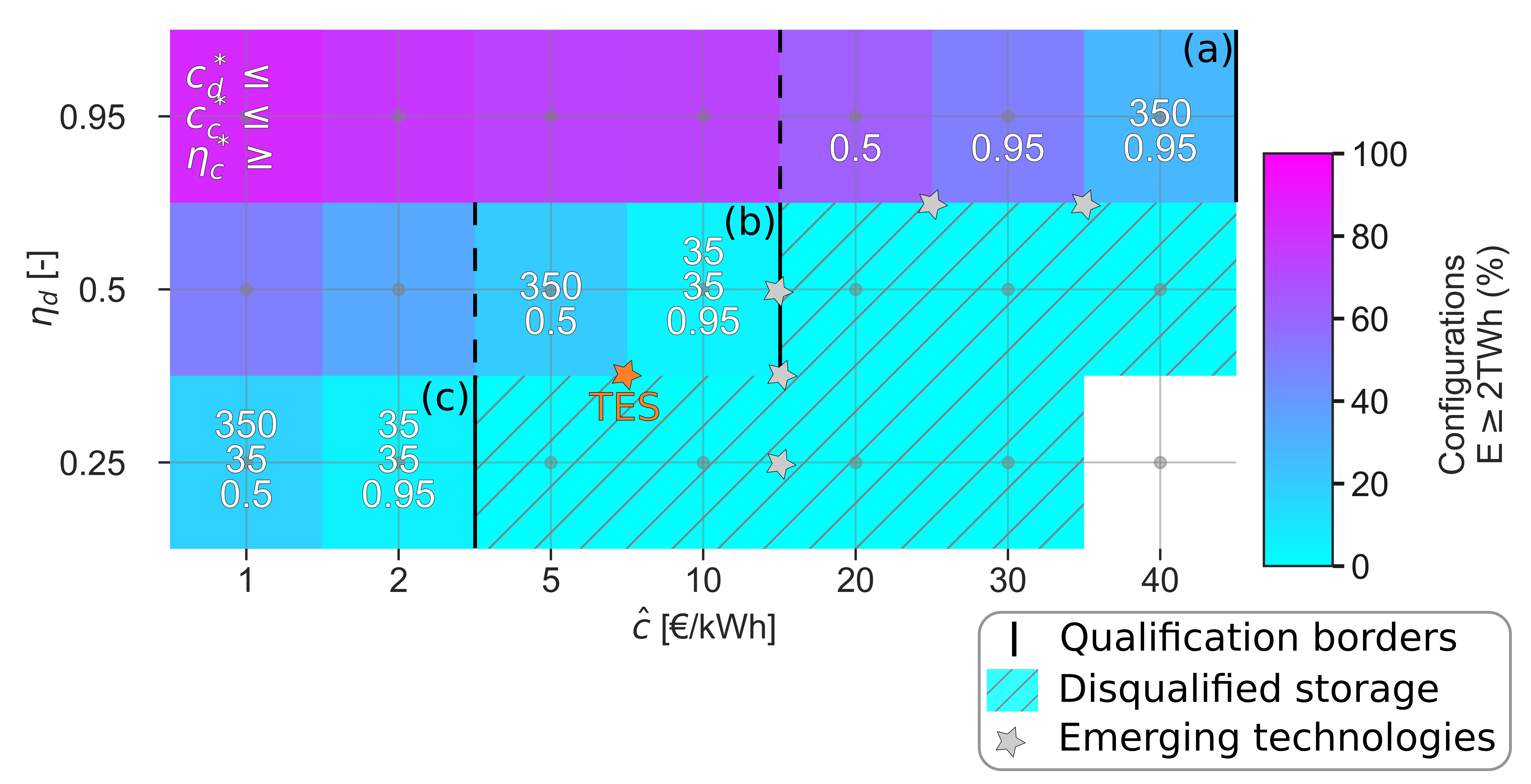}
	\caption{\textbf{Cost and efficiency requirements for the Fully sector-coupled system}. 
	Share of configurations from the sample space (indicated by color) fulfilling $E\geq2$~TWh at combinations of energy capacity costs (primary axis) and discharge efficiencies (secondary axis). The hatched area indicates combinations at which no configuration entails energy capacity $E\geq2$~TWh. The white annotations are additional conditions that need to be met for the considered combination to qualify. If no annotation is present, the combination is not conditioned by additional parameter requirements. The border at which the storage only just qualifies is indicated for each discharge efficiency level. This is done for conditioned (solid lines) and unconditioned (dashed lines) qualifications. The lines also delineate three design strategies (a), (b), and (c). The attributes of the emerging technologies in Table \ref{tab:existing_storage_assumptions_short} are indicated with stars. Due to the discreteness and the multidimensionality of the parameter space, the emerging technology indicators are located in between the investigated cells, unless they exactly match. See Supplemental Fig. S8 for combinations of the power capacity costs ($c_c$ and $c_d$) and the efficiencies ($\eta_c$ and $\eta_d$).}
	\label{fig:result_matrix}
\end{figure*}

By evaluating the output of the optimization for all combinations of $\hat{c}$ and $\eta_d$, we can identify critical parameter values that a generic storage technology is required to fulfill. These values determine the border of the optimal \mbox{\text{storage-X}} design space. We can divide the parameter requirements in two: 1) Unconditioned requirements that are true in all cases of the considered parameter space, and 2) Conditioned requirements that depend on further conditions besides the $\hat{c}$ and $\eta_d$ (e.g., power capacity costs $<$700~€/kW and efficiencies $>$25\%). Fig. \ref{fig:result_matrix} presents the combinations of $\hat{c}$ and $\eta_d$, and evaluates them based on whether $E\geq2$~TWh is satisfied. The color represents the share of configurations within the sample space at the particular combination that satisfies this condition. The white annotations indicate conditioned requirements for the remaining parameters ($c_d$, $c_c$, and $\eta_c$) at a given combination. Here, $\tau_{SD}$ is not shown due to its low impact. One obvious option to maintain a high share of qualified configurations, according to the figure, is to select a high discharge efficiency and a low energy capacity cost; at such combination, no noticeable reduction (i.e., the exclusion of configurations not satisfying the qualification condition) is observed. In that case, storage is deployed in the cost-optimal system independent of the other cost and efficiency parameters. This is in alignment with the indications in Fig. \ref{tab:coefficients}. When deviating from the best combination in either direction of the chart, the share of qualifying configurations drops. If the deviation is too large, the share drops to 0 (indicated with the hatches); thus, the combination disqualifies from entering the optimal design space. We use the point at which the storage parameters just meet the minimum requirements for deployment (qualification border) to describe the necessary storage parameter values. Fig. \ref{fig:result_matrix} depicts three qualification borders, each corresponding to a certain discharge efficiency level. If the storage is attributed with a high discharge efficiency ($\eta_d=95\%$), the qualification border (a) is located at the far right on the primary axis. This means that the full range of the energy capacity cost $\hat{c}$ can satisfy $E\geq2$~TWh when discharge efficiency is high, although the share of qualified configurations drops noticeably when $\hat{c}\geq20$~€/kWh. However, when $\hat{c}\geq20$~€/kWh, the qualification is conditioned by a mid-range charge capacity cost and mid/high discharge efficiency. In the case of having a mid-range discharge efficiency ($\eta_d=50\%$), the energy capacity cost qualification criterion is shifted leftwards towards border (b) which omits configurations with $\hat{c}\geq20$~€/kWh. However, this limit is conditioned by having perfect remaining parameters, meaning that both the charge efficiency ($\eta_c$) is high (95\%) and the charge capacity cost ($c_c$) and discharge capacity cost ($c_d$) are very low (35~€/kW). This can be eluded by reducing energy capacity cost to $\hat{c}<5$~€/kWh to reach the unconditioned region. At a low discharge efficiency ($\eta_d=25\%$), the energy capacity cost requirement (c) is even more stringent, resulting in the exclusion of all configurations with an energy capacity cost ($\hat{c}$) of 5~€/kWh or greater. A very low energy capacity cost may compensate for the low discharge efficiency, but this is also conditioned by a low charge capacity cost ($c_c$) and a charge efficiency of at least 50\%. These conditions suggest that low-efficiency storage technologies may be viable options but provided that they come at an exceedingly low cost.

\begin{figure}[!h]
	\centering
	\hspace{-1.5cm}
	\includegraphics[width=0.45\textwidth]{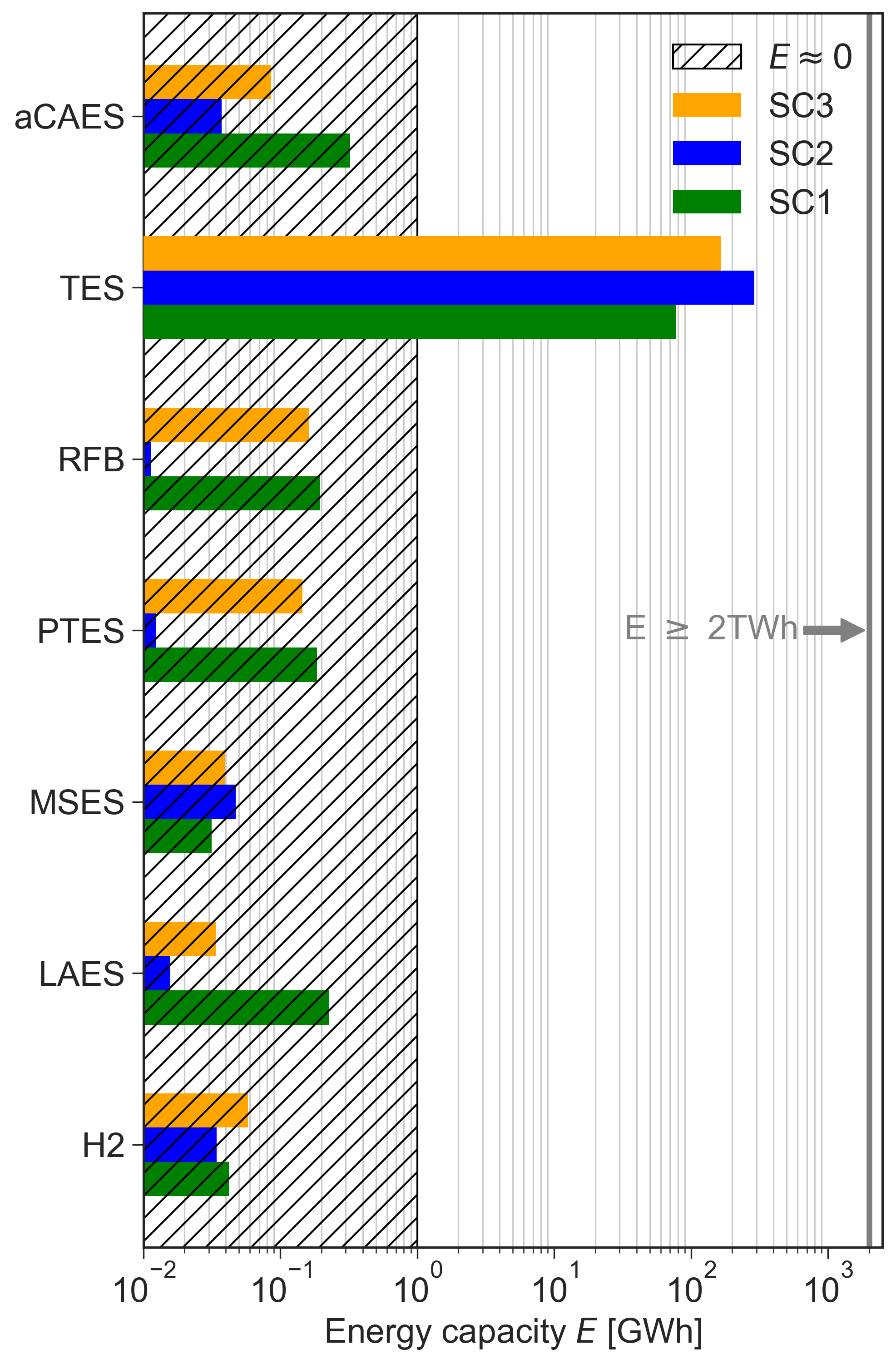}
	\caption{\textbf{Optimal energy capacity of \mbox{\text{storage-X}} candidates}. Comparison of the energy capacity deployment if the technology listed in Table \ref{tab:existing_storage_assumptions_short} was made available in PyPSA-Eur-Sec. Due to its low energy capacity cost and charge power capacity cost, TES is the only storage technology that entails a noticeable storage deployment on a European scale (100-300 GWh) but does not qualify for the imposed threshold of $\geq$2~TWh. The rest of the candidates entails a negligible energy capacity ($<$1 GWh).}
	\label{fig:candidates}
\end{figure}

The depicted borders of the design space, derived from cost optimization, can serve as optimal design approaches for novel storage technologies. Specifically, the three borders can signify three optimal design strategies that storage developers should pursue. The energy capacity cost and discharge efficiency for the emerging technologies listed in Table \ref{tab:existing_storage_assumptions_short} are also shown in Fig. \ref{fig:result_matrix}, alongside the optimization results. Among the emerging storage technologies, aCAES and LAES are found to be situated near design strategy (a) based on their combination of energy capacity cost and discharge efficiency. The remainder is located near design strategies (b) or (c). The RFB technology is absent from the graph as its corresponding energy capacity cost ($\hat{c}=115$~€/kWh) falls beyond the limit of the sample space considered in this study. We find that none of the candidates meet the unconditioned qualification criteria solely based on their energy capacity cost and discharge efficiency. To qualify for the conditioned region, the storage needs, on top of having either high discharge efficiency or low energy capacity cost, additional well-performing attributes. The TES technology benefits from not only having the lowest energy capacity cost but also the lowest charge capacity cost ($c_c=38$~€/kW) concurrent with a high charge efficiency ($\eta_c=98$~\%), which makes it a promising configuration. Based on its location in the chart, it still needs either a higher discharge efficiency or lower energy capacity cost to clearly qualify. Due to its low discharge efficiency and concurrent mid-range energy capacity cost, PTES follows a mixed design strategy between (b) and (c) which positions it in the disqualified region. The two design strategy-(a) technologies, aCAES and LAES, have the highest discharge efficiencies ($\eta_d=65$\%). This is still 30 percentage points lower than the considered discharge efficiency in (a). Since a large variation in optimal energy capacity deployment is explained by the discharge efficiency (Table \ref{tab:coefficients}), this reduces its chance to qualify. To test these hypotheses, we evaluate the energy capacity deployment with the exact parameters for the candidates in Table \ref{tab:existing_storage_assumptions_short}, which leads to the same conclusion (Fig. \ref{fig:candidates}) that only TES shows a noticeable deployment, mainly due to its low energy capacity cost and low charge power capacity cost. It does, however, not entail the optimal energy capacity sufficient to qualify for the design space.

\subsection{Storage-X impact on the energy system design}

\begin{figure}[!ht]
	\centering
	\includegraphics[width=0.45\textwidth]{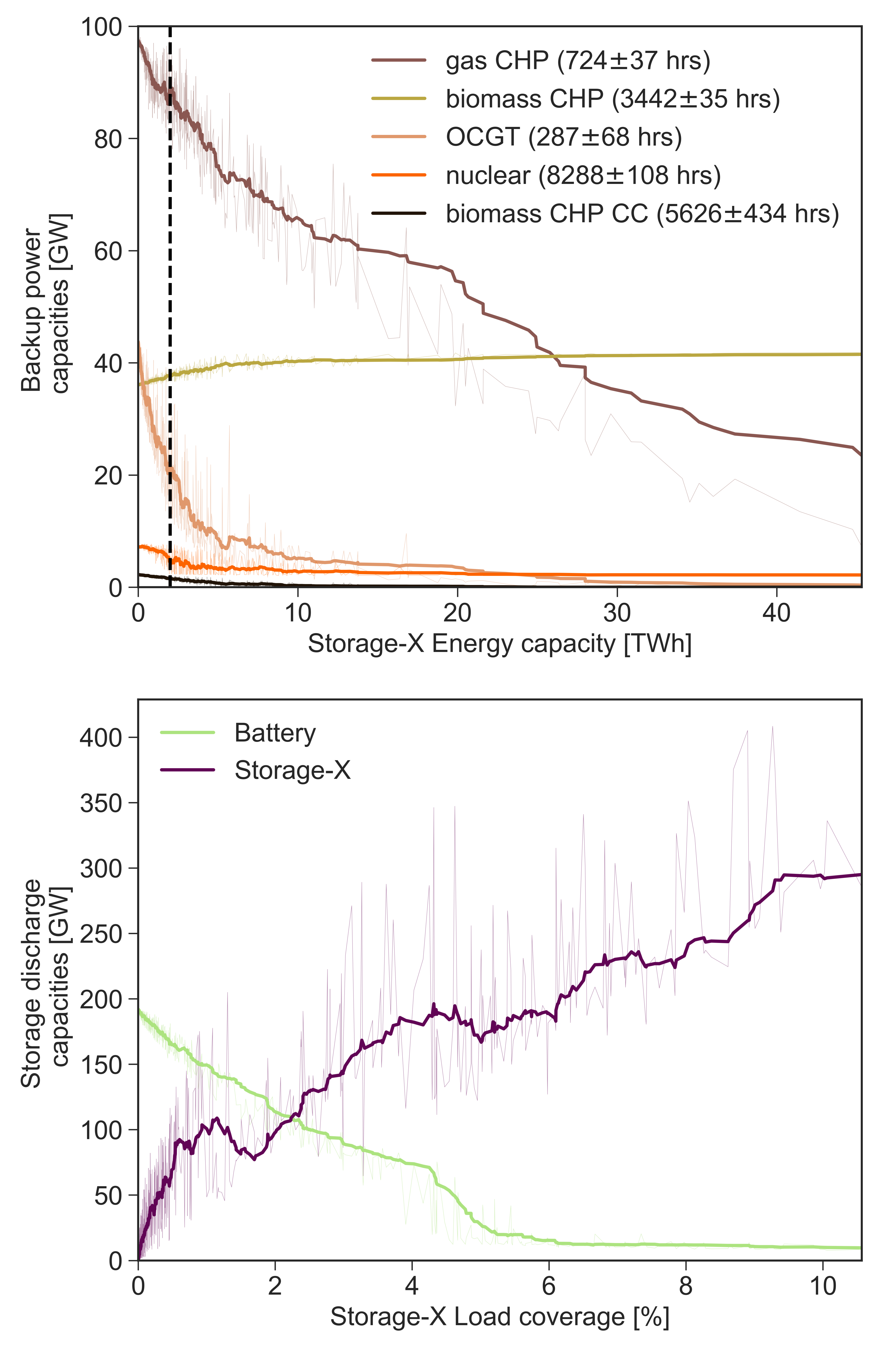}
	\caption{\textbf{Backup power and storage discharge capacity}. (top) Power capacities of dispatchable thermal power plants for the 724 storage samples sorted by the \mbox{\text{storage-X}} energy capacity, and (bottom) storage discharge power capacities sorted by the \mbox{\text{storage-X}} load coverage. Numbers in parentheses indicate the average full load hours and the standard deviation across all storage scenarios. Thick lines show the moving average of the 15 previous configurations. The dashed line indicates the $\geq2$~TWh threshold imposed to derive the design space. Gas power plants (CHP and OCGT) and battery capacity drop continuously as \mbox{\text{storage-X}} is deployed. For the results of SC1 and SC2, see Fig. S10.}
	\label{fig:genmix}
\end{figure}

Every system composition and all storage integration scenarios have been obtained for a CO$_2$ emissions constraint of 5\%. This involves a large roll-out of non-carbon-emitting generation technologies. Across all storage scenarios, the electricity generation is on average supplied by $56\%$ wind, $34\%$ solar, and $6\%$ hydropower. The remaining 4\% is covered with fossil-fueled power plants, biomass CHP, or nuclear (Supplemental Fig. S9). The mix of backup reserves and storage (excluding PHS) is a result of the optimization and is impacted by the deployment of \mbox{\text{storage-X}} (Fig. \ref{fig:genmix} top). At low levels of \mbox{\text{storage-X}} penetration, based on the deployed energy capacity, a high share of flexible generation is needed, mainly provided by gas-powered OCGT and CHP (supplying both heat and power), both showing low load hours equivalent to 8.3\% and 3.3\% utilization rates throughout a full year. In comparison, in SC1, fossil-fueled generation can be deployed with a much higher utilization rate: 43.4\% for CCGT (Supplemental Fig. S10). Increasing the \mbox{\text{storage-X}} deployment reduces these capacities noticeably. CHP fueled with biomass is weakly affected since the model finds it optimal to utilize the full potential of biomass across all scenarios.


Battery capacity is replaced as \mbox{\text{storage-X}} increases its load coverage, indicated by the linear decay of battery discharge capacity (Fig. \ref{fig:genmix} bottom). For comparison, the discharge capacity of \mbox{\text{storage-X}} is also depicted to indicate that the battery capacity deployment is linearly dependent on the \mbox{\text{storage-X}} load coverage but does not show the same proportionality with \mbox{\text{storage-X}} discharge capacity. This is due to the fact that load coverage is also determined by other \mbox{\text{storage-X}} attributes. While \mbox{\text{Storage-X}} is capable of replacing battery capacities with increasing load coverage, it does not adapt the temporal short-term characteristics since most of the \mbox{\text{storage-X}} configurations result in a duration (i.e., energy-to-power ratio) of approximately 50 hours (see Supplemental Fig. S11. The maximum duration within the configurations equals 190.3 hours equivalent to 7.6 days). Concurrently, most batteries show durations of 3 or 6 hours (Supplemental Fig. S12).\\

\begin{figure}[!ht]
	\centering
	\includegraphics[width=0.5\textwidth]{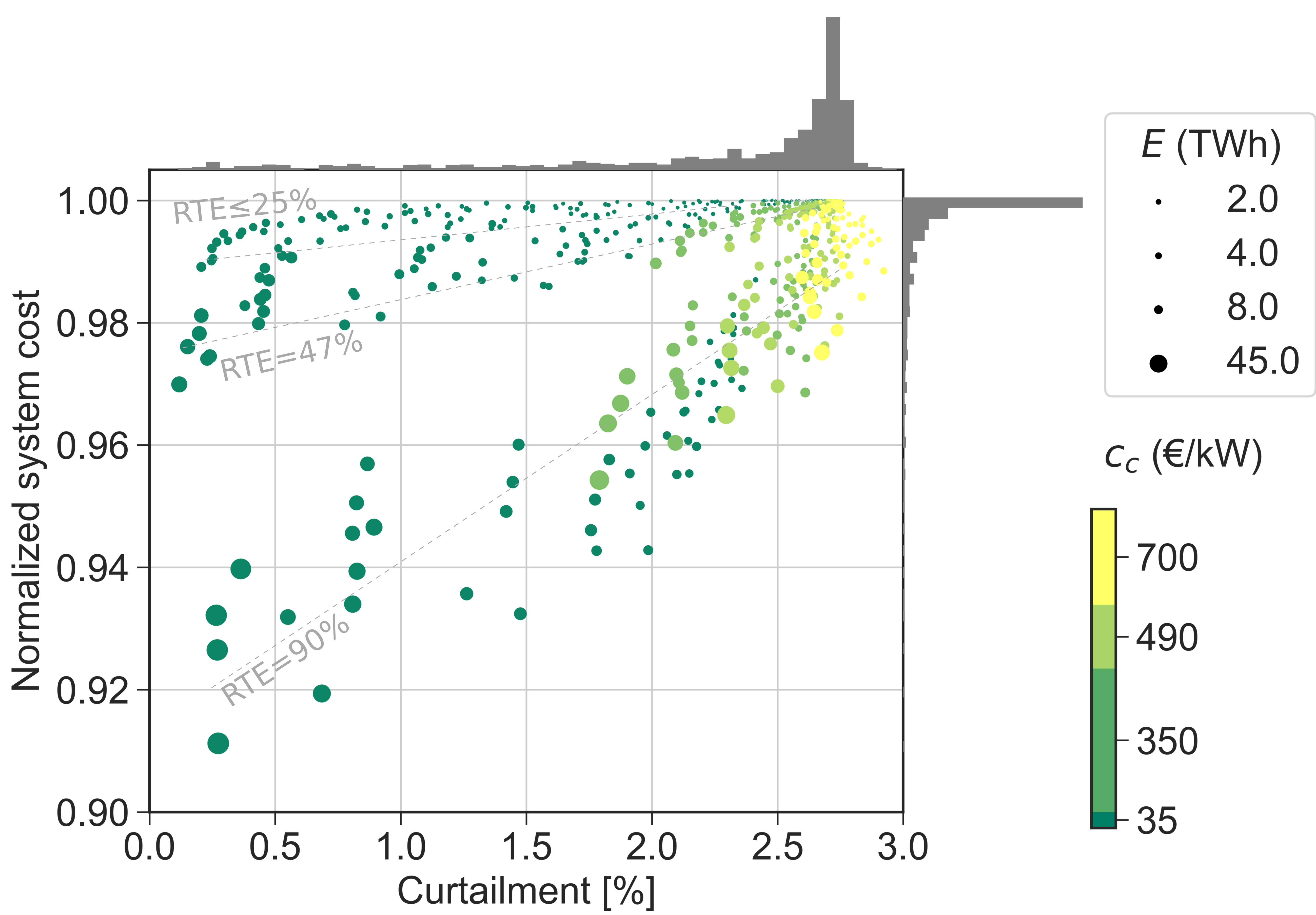}
	\caption{\textbf{System cost reduction and renewable curtailment}. Normalized system cost (secondary axis) ordered by the level of renewable curtailment (primary axis) for the 724 storage samples. The scatters are colored according to the corresponding charge capacity cost ($c_c$) and are furthermore sized according to the optimal energy capacity. The results are obtained with the fully sector-coupled system (SC3). See Supplemental Fig. S13 for SC1 and SC2. 
	}
	\label{fig:system_cost}
\end{figure}

Another alternative to new storage facilities is deploying excess capacity of renewable generators, which entails a less energy-efficient system as more energy is curtailed instead of being utilized. Whether this entails a more expensive system is evaluated in Fig.\ref{fig:system_cost}. Here, the normalized system cost (i.e., 1 corresponds to the most expensive system in which \mbox{\text{storage-X}} is not deployed) is acquired for each storage configuration and compared with the level of renewable curtailment (Eq. \ref{eq:curtailment}). In general, a low curtailment is observed ($<$3\% for all configurations). Some observations can, however, be made. 

All configurations with a high charge capacity cost are concentrated within the same upper level of renewable curtailment. Thus, expensive charge components can be associated with more curtailment. As the charge capacity cost is reduced, the system encounters lower levels of renewable curtailment. Concurrently, the system cost reduction potential is improved. The best \mbox{\text{storage-X}} configuration reduces the total system cost by 9\%. 

Additionally, the level of system cost reduction is shown to depend on the round-trip efficiency ($RTE=\eta_c\eta_d$). Two distinct groups of scenarios form (an artifact of the parameter discreteness): one group containing mid-low RTE ($\leq$47\%) configurations and another one with the high RTE (90\%). A high RTE is a prerequisite to obtaining the steepest reduction proportional to the curtailment reduction. A high RTE is, however, not a guarantee for obtaining the lowest system cost. E.g., low charge capacity cost configurations with mid-range RTE are capable of achieving a lower system cost than expensive charge capacity with high RTE. Equivalently, low curtailment is not a guarantee for low system cost either since it depends on the portfolio of storage technologies. Thus, obtaining a highly energy-efficient system, in terms of low energy curtailment, is not necessarily the cost-optimal strategy. 

For the other two system compositions, curtailment of renewable energy is higher, with less than 7\% in SC1 and less than 4\% in SC2 (Supplemental Fig. S13). The variation in curtailment between the three system compositions can be explained by the flexible demand that arises from sector coupling. In particular, the large-scale hydrogen production in SC3 results in a higher utilization rate of renewable energy, as it can be produced at any time, independently of the concurrent demand, provided that hydrogen storage or transport is available. For SC1 and SC2, the largest reduction in system cost is 12\% and 9\%, respectively, at the best \mbox{\text{storage-X}} configuration. When considering 95\% of the configurations (i.e., the 95$^\text{th}$ percentile), the largest system cost reductions decrease to 6\%, 4\%, and 4\% for SC1, SC2, and SC3, respectively.

\section{DISCUSSION}\label{sec:discussion}

This study aims to find the requirements for an additional electricity storage technology, \mbox{\text{storage-X}}, to be deployed in the cost-optimal sector-coupled energy system. In the following, we discuss our findings and the corresponding implications.\\

First, our results support the findings of Refs. \cite{Sepulveda2021, ZIEGLER20192134, Dowling2020, TONG2020101484} that the discharge efficiency and storage energy capacity cost are the parameters that potentially, if improved, entail the highest rise in storage deployment. The significance of the discharge efficiency can to some extent be explained by the serial connection of the \mbox{\text{storage-X}} components of which the discharge stage is the last element. As an example, we imagine the case in which the discharge efficiency was worsened, while the other input parameters were fixed. In that case, to ensure the same storage electricity output, the storage energy capacity and power capacity for charging had to be enlarged. This requires higher investments in both of the components. As a comparison, the capacity cost in general (charge, storage, discharge) does not have such a serial-accumulating impact but only alters the investments in the considered component. The energy capacity cost is linked to the cost-optimal size of the storage, which is why such high sensitivity is observed in the optimal energy capacity deployment to this parameter. As large energy capacity allows balancing low-frequency droughts where batteries are a very expensive option, this makes it a parameter of high importance. 

Here, we also highlight the importance of a third parameter; the charge capacity cost. Renewable curtailment is related to the cost of charging (Fig. \ref{fig:system_cost}). The system can either deploy, if cost-optimal, sufficient charge capacity to utilize the full renewable potential, or, in lack of storage charge capacity, curtail some of it. Thus, it is a trade-off between investments in larger charge capacity or more renewable curtailment, which can explain the significance of this input parameter. This was not shown in prior literature \cite{Sepulveda2021} inspecting the parameter importance of electricity storage. A share of the discrepancy can be related to the distinct assumptions concerning electricity transmission. Using a copperplate model, i.e., ignoring electricity transmission constraints over large regions, can lead to a misestimation of the renewable electricity generation \cite{BRINKERINK2022105336}. This also implies misestimating the renewable curtailment and, consequently, underestimating the importance of the storage charge capacity cost.\\

Second, our study identifies the trade-offs between cost and efficiency. A storage with high discharge efficiency and very low energy capacity cost would have the best chances of being competitive. If this is not attainable, a trade-off between cost and efficiency arises, i.e., a high cost can be compensated by a high efficiency. Our results show that energy capacity cost requirement is driven by the attainable discharge efficiency, and vice versa. Here, we have identified three design strategies for storage to play a prominent role in future renewable energy systems. The optimal design strategies are to select either:
\begin{enumerate}[label=\alph*)]
	\item high discharge efficiency ($\eta_d\geq$95\%)
	\item mid-range discharge efficiency ($\eta_d\geq$50\%) if energy capacity cost $\hat{c}\leq$10~€/kWh.
	\item low discharge efficiency ($\eta_d\leq$25\%) if energy capacity cost $\hat{c}\leq$2~€/kWh. 
\end{enumerate}

None of the emerging technologies fulfills these requirements. Flow-redox batteries are discarded due to their high energy capacity cost. The remainders are not good enough in any of the a, b, or c design strategies. The only exception is TES which is attributed with the lowest energy capacity cost of all candidates. For this reason, as the only candidate, it shows a noticeable optimal storage deployment. However, to fully qualify for design strategy b (and increase its optimal deployment by one order of magnitude), it would need to alter its discharge efficiency from 38\% to 50\%. This might not be attainable with the applied discharge technology (Brayton cycle), depending on the available temperature ratio. In that case, it needs to reduce its energy capacity cost from 8~€/kWh to $<$5~€/kWh. Another technology, PTES, is discarded due to not complying with the optimal trade-off. It has the combination of low discharge efficiency (design strategy c) and mid-range energy capacity cost (b). Thus, our results also show that it is optimal to pursue one of these design strategies, while hybrid strategies have both low potentials and can be difficult to realize.\\

In this paragraph, the limitations of this study are discussed. 

Applying the chosen temporal resolution with 3-hourly time steps implies that the variation within these time steps is perfectly balanced without any cost. For this reason, this framework does not cover nor calculate the potential for storage used for such short-term purposes. In addition to this, with the applied 37-node network resolution, we do not include bottlenecks in the transmission connecting different regions within the countries but only account for the cross-border transmission. A higher storage capacity would be expected to account for such additional transmission constraints in a more detailed network representation. 

We do not allow grid services from BEV batteries. Since this assumption mainly determines the volume of stationary battery storage, it is a question of whether storage should be located in the low-voltage grid provided by BEV batteries or provided at the high-voltage level by utility-scale battery storage. In this model framework,  batteries are primarily used to smoothen the diurnal fluctuation in solar generation. At a certain level of \mbox{\text{storage-X}} penetration, we do see that the stationary batteries are replaced by \mbox{\text{storage-X}}. For this reason, and due to the competition between V2G and batteries, the inclusion of V2G would also impose an upper limit on the \mbox{\text{storage-X}} deployment. Another assumption related to the inclusion of hydrogen (H$_2$). The hydrogen bus in our model does not have a direct link to the electricity bus (i.e., fuel cells or H$_2$-fired gas power plants are excluded). This assumption offsets some flexibility that would have been contained by the additional usage option of the H$_2$ carrier.

As we investigate flexibility options and suggest gas power plants as new backup capacities, this is established on cost assumptions that do not reflect the recent development of commodity prices of, e.g., gas. We perform an overnight optimization that does not correspond to a specific year but is to resemble a possible near-future system prior to full decarbonization. In the system today, a large number of central fossil-fueled power plants are still either active or kept on stand-by. Thus, some of these flexibility options might still be valuable assets in commission which the backup capacity in our study also could represent.

As this is an overnight optimization, we do not consider the transition toward the final system. The electricity generation capacity (including fossil-fueled and renewable generators) and the storage capacity are co-optimized concurrently. From the literature, we have seen large storage quantities only at the late stage of the renewable penetration. For this reason, a transitional optimization would entail the renewables to be rolled out, eventually reaching a certain level at which storage would be added to the system. In that sense, the storage ensemble does not influence the share of generation sources, which is the case in this study. In pathway optimization, there is also the option of learning. The considered portfolio of storage candidates has differences within modularity and technological simplicity and is expected to gain differently from scale-up. A pathway optimization with endogenous learning would shed light on this and could make the differences in the potential of the different storage candidates even more pronounced. 

In this study, we evaluate the integration of single configurations of \mbox{\text{storage-X}} in parallel scenarios. Multiple \mbox{\text{storage-X}} do not appear simultaneously in the same scenario. Therefore, the modeled competition is only related to existing storage technologies (stationary Li-ion battery storage) and other flexibility options (backup power plants and demand-side management) in the system, and it does not account for the potential competition between multiple \mbox{\text{storage-X}} technologies. Such competition would better represent reality and it could furthermore shed light on potential alliances of storage that in combination could be more beneficial for the system than one unique dominant option. Future work should address this.

\section{CONCLUSION}\label{sec:conclusion}


To explore the requirements of a successful additional storage technology, \mbox{\text{storage-X}}, we attained a design space established on 724 storage samples to identify the configurations that would lead to substantial storage deployment (here, equivalent to $\geq2$~TWh energy capacity) in a highly renewable sector-coupled energy system. This was performed when accounting for the competition with stationary battery storage, flexible generation technologies, and possible demand-side management.\\

We find that energy capacity cost and discharge efficiency are the parameters leading to the highest impact on the optimal \mbox{\text{storage-X}} deployment. Energy capacity cost is linked to the ability to provide low-frequency balancing of renewable droughts, and the importance of discharge efficiency is related to the serial dependence of the storage components. On top of this, the charge capacity cost shows high relative importance as well, here explained by its noticeable impact on the level of renewable curtailment.\\

For a cost-optimal deployment, an additional electricity storage technology is required to have (a) discharge efficiency of at least 95\%, (b) discharge efficiency of at least 50\% in combination with a low energy capacity cost (10€/kWh), or (c) a very low energy capacity cost (2€/kWh) if discharge efficiency is 25\%. Comparing our findings with seven emerging technologies reveals that none of them fulfill these requirements. The only noticeable deployment is observed for the thermal energy storage (TES) which is a candidate for design strategy b, but to fully qualify, it needs to increase its discharge efficiency from 38\% to 50\% or reduce its energy capacity cost from 8~€/kWh to $<$5~€/kWh.\\

Finally, we assessed the system impact of the integration of \mbox{\text{storage-X}}. Our analysis shows that the integration of an additional storage technology can lead to a system cost reduction of up to 9\%, but only if the technology has a high round-trip efficiency (RTE$\geq$90\%) and a low charge capacity cost ($c_c=$35~€/kW).

\section{ACKNOWLEDGEMENTS}
Ebbe K. G\o{}tske, Marta Victoria, and Gorm B. Andresen are partially or fully funded by the GridScale project supported by the Danish Energy Technology Development and Demonstration Program under the grant number 64020-2120. We are thankful for the helpful comments from Fredrik Hedenus, Lina Rechenberg, Martin Greiner, Stefan Lechner, and Sleiman Farah. 

\appendix

\section{Disaggregation of reported storage \\ technology cost assumptions}\label{App:Disagg_storage_assumptions}

This appendix describes the procedure of converting reported data on emerging storage technologies to fit the representation of \mbox{\text{storage-X}} in this paper, with the results presented in Table \ref{tab:existing_storage_assumptions_short}. 

In some of the considered technology costs reviews, reported power capacity costs cover expenses related to both charging and discharging. It is given by the aggregate power capacity cost $C_P$, accompanied by a power capacity cost ratio $r_C$. To disaggregate $C_P$ into a charge and discharge power capacity cost, $c_c$ and $c_d$, we define the following two equations:

\begin{equation}\label{seq:1}
	C_P = \frac{1}{\eta_{RT}}c_c + c_d
\end{equation}

\begin{equation}\label{seq:2}
	r_C = \frac{c_c}{c_d}
\end{equation}

where $\eta_{RT}$ is the round-trip efficiency acquired from the literature. From the two above equations, the charge and discharge capacity costs can be explicitly determined:
\begin{equation}
	c_c = r_Cc_d
\end{equation}

\begin{equation}
	c_d = C_P - \frac{1}{\eta_{RT}}c_c
\end{equation}
	
\vspace{\baselineskip}
Along with the ratio $r_\eta$ between the charge and discharge efficiency, $\eta_c$ and $\eta_d$, we define the following two equations: 

\begin{equation}\label{seq:3}
	\eta_{RT} = \eta_c \eta_d
\end{equation}

\begin{equation}\label{seq:4}
	r_\eta = \frac{\eta_c}{\eta_d}
\end{equation}

The round-trip efficiency is then split into the charge and discharge efficiencies:
\begin{equation}
	\eta_c = r_\eta \eta_d
\end{equation}

\begin{equation}
	\eta_d = \frac{\eta_{RT}}{\eta_c}
\end{equation}
	
Standing loss is reported as a percentage, given as an energy loss per day relative to the energy content of the previous day. The state of charge is assumed to follow an exponential decay. The percentage can be converted into a time constant, which represents the time passed before the storage reaches a substantially low state of charge (here, 36.8\%). We call this time constant the self-discharge time $\tau_{SD}$, and it is determined by following equation (Eq. \ref{eq:standing_loss}):
\begin{equation}\label{seq:5}
	\tau_{SD} = -\ln (1-\Delta \bar{e}_{\text{loss}})^{-1}
\end{equation}

where $\Delta \bar{e}_{\text{loss}} = \bar{e}_{t} - \bar{e}_{t-1}$ is the relative daily energy loss which in the reporting is assumed to be constant.

\section{PYPSA-EUR-SEC}\label{sec:Appendix}

The system is modeled using the open energy system model PyPSA-Eur-Sec \cite{BROWN2018720}. The dispatch and capacity of generators, storage units, and conversion links in each node are optimized such that the total system cost is minimized which mathematically is written as:
\begin{equation}
	\begin{array}{rrclcl}
		\displaystyle \min_{G_{n,s},E_{n,s},F_{l},g_{n,s,t}} 
			\bigg[ \sum_{n,s} c_{n,s}G_{n,s} + \sum_{n,s} \hat{c}_{n,s}E_{n,s}\\
			+ \sum_l c_l F_l + \sum_{n,s,t} o_{n,s,t}g_{n,s,t} \bigg]\\
	\end{array}
\end{equation}
\noindent
where $c_{n,s}$ and $\hat{c}_{n,s}$ are the annualized costs for generator and storage power capacity $G_{n,s}$ and storage energy capacity $E_{n,s}$ for technology $s$ in node $n$. $c_l$ are the fixed annualized costs for the capacities $F_l$ of the links $l$. $o_{n,s,t}$ are the marginal costs of generation and storage dispatch $g_{n,s,t}$ at time $t$. 

The optimization is subject to a list of linear equality and inequality constraints, of which two of them are listed here:

\begin{equation}\label{eq:energy_balance}
	\begin{split}
		\sum_s g_{n,s,t} + \sum_l \alpha_{n,l,t} f_{l,t} = d_{n,t} &\leftrightarrow \lambda_{n,t} \hspace{0.5cm} \forall n,t
	\end{split}
\end{equation}

\begin{equation}\label{eq:co2_constraint}
	\begin{split}
		\sum_{n,s,t} \varepsilon_s \frac{g_{n,s,t}}{\eta_{n,s}} \leq \text{CAP}_{CO_2} &\leftrightarrow \mu_{n,t} \hspace{0.5cm} \forall n,t
	\end{split}
\end{equation}

Eq. \ref{eq:energy_balance} is an equality constraint that ensures that the energy generation, dispatch/charging of storage, and import/export $g_{n,s,t}$ and $f_{l,t}$ is exactly balanced with the demand $d_{n,t}$ in all nodes $n$ at all times $t$. Here, $f_{l,t}$ is the power flow through link $l$ at time $t$, and $\alpha_{n,l,t}$ includes the direction and efficiency of the flow in the links. The equality constraint entails a Lagrange multiplier $\lambda_{n,t}$ which is the shadow price of the energy carrier. 

Eq. \ref{eq:co2_constraint} is an inequality constraint that enforces an upper bound on the total annual CO$_2$ emissions in the system (i.e., a global CO$_2$ emissions constraint) which is given as a percentage of the 1990-emissions level. Here, $\varepsilon_s$ is the CO$_2$ intensity in tonne CO$_2$ per MWh$_\text{th}$, and $\eta_{n,s}$ is the efficiency. The PyPSA-Eur-Sec model is capable of tracing direct emissions from power generation, energy conversion, and industrial processes. It can furthermore deploy negative-emissions technologies (direct air capture or carbon capture on point sources such as thermal power plants) if cost-optimal. The calculations do not include life-cycle emissions on each of the technologies. The inequality constraint entails a Lagrange multiplier $\mu_{n,t}$ which is the CO$_2$ shadow price.\\

The history of charging and discharging in combination with the level of standing loss determines the state of charge of the storage at any time $t$. The state of charge is constrained to not exceeding the storage energy capacity. Here, the energy content $e_{n,s,t}$ of an energy storage technology $s$ in node $n$ at time $t$ is given by the following energy balance, when accounting for standing loss ($\eta_0$) and efficiency losses ($\eta_c$ and $\eta_d$) and given the energy content at the previous time $t$:

\begin{equation}
	\begin{split}
		e_{n,s,t} &= \eta_0 e_{n,s,t-1} + \eta_c |\Delta e^+_{n,t}| - \eta_d |\Delta e^-_{n,s,t}|\\
		0 &\geq e_{n,s,t} \geq E_{n,s}
	\end{split}
\end{equation}

\noindent
where $|\Delta e^+_{n,s,t}|$ is the absolute electricity stored and $|\Delta e^-_{n,s,t}|$ is the dispatched electricity in node $n$ at time $t$. The standing loss is in this equation represented by an efficiency $\eta_0$ which equals 1 in case $\tau_SD=\inf$. The energy content can not exceed the storage energy capacity, i.e., $e_{n,s,t} \leq E_{n,s}$. 

All technology investment capacity costs are annualized assuming a lifetime of each asset and a discount rate $r$ of 7\%:

\begin{equation}
	c_{n,s} = c_{n,s}^\text{inv.} \frac{r(1+r)^n}{(1+r)^n-1}
\end{equation}
\noindent
where $c_{n,s}$ are the annualized costs and $c^{\text{inv.}}_{n,s}$ are the capacity costs. 

\section{Reduction of sample space based on filter $M$}\label{appendix:filter}

To reduce computation time, when repeating the calculations for the same sample space $Q$ in the three system compositions, we introduce a filtering metric $M$ (Eq. \ref{eq:filter_metric}). The $M$ metric represents the ratio between the capacity costs and the efficiencies, If the ratio exceeds a threshold, the storage becomes non-competitive and will, for this reason, most probably not be part of the optimal design space. The threshold is derived from the optimization results obtained with SC1. We compute the ratio of capacity cost and efficiency as:

\begin{equation}\label{eq:filter_metric}
	M = \frac{\frac{c_c}{\lambda_1}\frac{c_d}{\lambda_2}+\hat{c}}{\eta_c \eta_d} \leq \max\{M(E=2\text{ TWh})\}
\end{equation}

\begin{figure}[!h]
	\centering
	\includegraphics[width=0.45\textwidth]{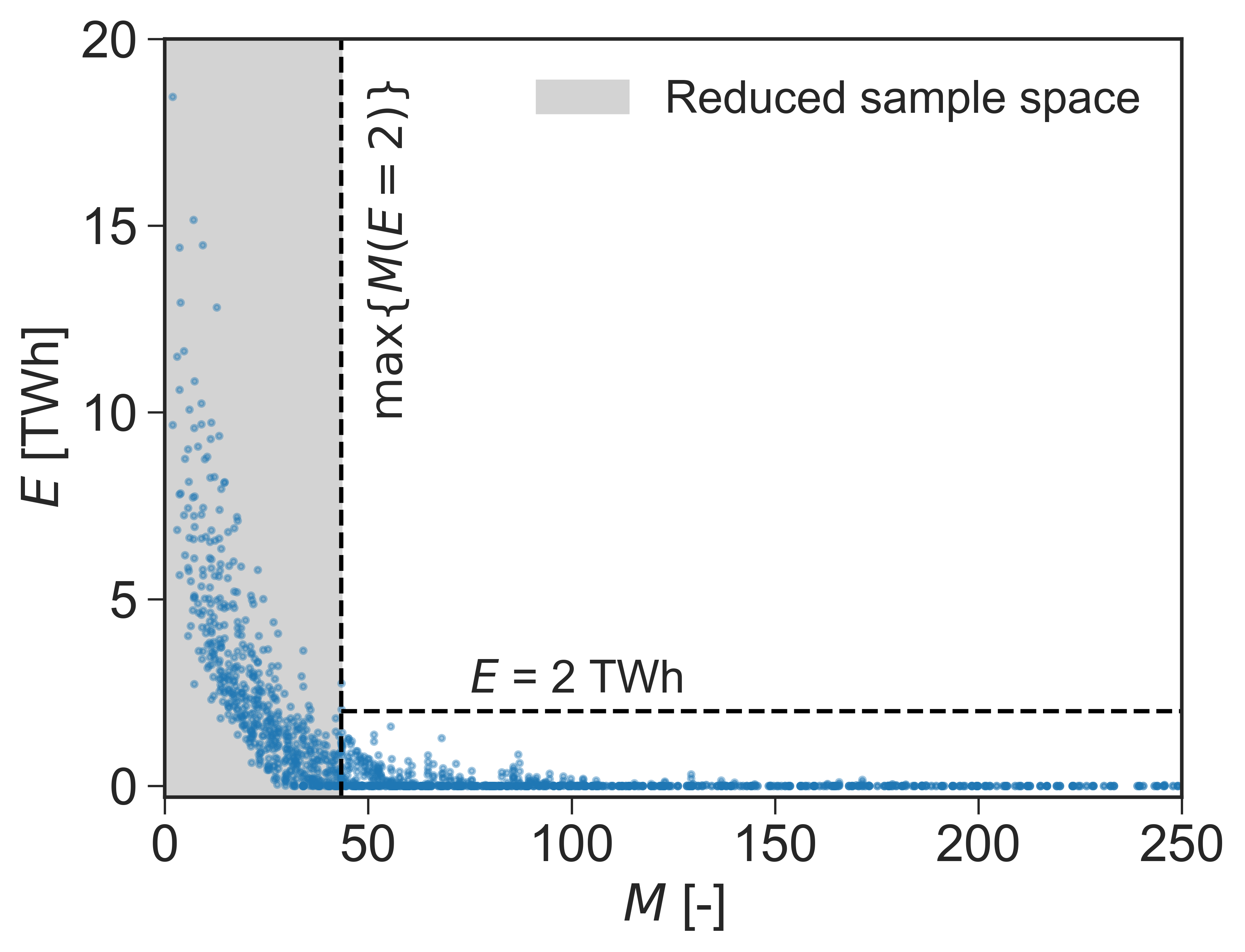}
	\caption{\textbf{Reduction of sample space}. Reduction based on $M$ defined in Eq. \ref{eq:filter_metric} obtained with results for the Electricity system.}
	\label{fig:reduction_sample_space}
\end{figure}

Here, $\lambda_1$ and $\lambda_2$ are scaling factors obtained by fitting Eq. \ref{eq:filter_metric} to the output ($E$) in SC1. By matching the size of the design space ($Q(E\geq2\text{ TWh})$) and the number of configurations that meet the condition in Eq. \ref{eq:filter_metric}, the scaling factors $\lambda_1=57$ and $\lambda_2=175$. The obtained reduction of the sample space is illustrated in Fig. \ref{fig:reduction_sample_space}. The sample space $Q$ containing 2,016 configurations is reduced to 593 configurations. Subsequently, to retain the exterior of the considered sample space (combinations in which the upper parameter limits are included), 131 additional configurations are included, entailing a final sample space of 724 configurations. 




\clearpage 
\onecolumn 
\beginsupplement 
\textbf{\textcolor{gray}{Supplemental Material}}\\

\section{Transmission lines}

\begin{table}[!ht]
	\caption{\textbf{AC lines added exogenously to the model}.}
	\label{tab:ac_lines}
	\centering
	\begin{tabular}{@{}llcc:lllcc@{}}
		\toprule
		\textbf{Bus0} & \textbf{Bus1} & \textbf{Capacity {[}MVA{]}} & \textbf{Type} &  & \textbf{Bus0} & \textbf{Bus1} & \textbf{Capacity {[}MVA{]}} & \textbf{Type} \\ \midrule
		AL0 0         & GR0 0         & 1698     				   & AC            &  & DE0 0         & DK0 0         & 4535                     & AC            \\
		AL0 0         & ME0 0         & 2267                     & AC            &  & DE0 0         & FR0 0         & 9629                    & AC            \\
		AL0 0         & MK0 0         & 1698                     & AC            &  & DE0 0         & LU0 0         & 3984                     & AC            \\
		AL0 0         & RS0 0         & 2267                     & AC            &  & DE0 0         & NL0 0         & 10189                    & AC            \\
		AT0 0         & CH0 0         & 6792                     & AC            &  & DE0 0         & PL0 0         & 6792                     & AC            \\
		AT0 0         & CZ0 0         & 4535                     & AC            &  & DK3 0         & SE3 0         & 3396                     & AC            \\
		AT0 0         & DE0 0         & 16431                    & AC            &  & EE6 0         & LV6 0         & 3175                     & AC            \\
		AT0 0         & HU0 0         & 4535                     & AC            &  & ES0 0         & FR0 0         & 4535                     & AC            \\
		AT0 0         & IT0 0         & 569.                     & AC            &  & ES0 0         & PT0 0         & 15292                    & AC            \\
		AT0 0         & SI0 0         & 569                      & AC            &  & FI3 0         & SE3 0         & 3396                     & AC            \\
		BA0 0         & HR0 0         & 5673                     & AC            &  & FR0 0         & IT0 0         & 7362                     & AC            \\
		BA0 0         & ME0 0         & 2836                     & AC            &  & GB4 0         & IE4 0         & 1138                     & AC            \\
		BA0 0         & RS0 0         & 2267                     & AC            &  & GR0 0         & MK0 0         & 3396                     & AC            \\
		BE0 0         & FR0 0         & 11327                    & AC            &  & HR0 0         & HU0 0         & 6792                     & AC            \\
		BE0 0         & LU0 0         & 1138                     & AC            &  & HR0 0         & RS0 0         & 1698                     & AC            \\
		BE0 0         & NL0 0         & 6792                     & AC            &  & HR0 0         & SI0 0         & 6233                     & AC            \\
		BG0 0         & GR0 0         & 1698                     & AC            &  & HU0 0         & RO0 0         & 3396                     & AC            \\
		BG0 0         & MK0 0         & 1698                     & AC            &  & HU0 0         & RS0 0         & 1698                     & AC            \\
		BG0 0         & RO0 0         & 6792                     & AC            &  & HU0 0         & SK0 0         & 3396                     & AC            \\
		BG0 0         & RS0 0         & 1698                     & AC            &  & IT0 0         & SI0 0         & 2267                     & AC            \\
		CH0 0         & DE0 0         & 20377                    & AC            &  & LT6 0         & LV6 0         & 4233                    & AC            \\
		CH0 0         & FR0 0         & 13604                    & AC            &  & ME0 0         & RS0 0         & 2836                     & AC            \\
		CH0 0         & IT0 0         & 8500                     & AC            &  & MK0 0         & RS0 0         & 3396                     & AC            \\
		CZ0 0         & DE0 0         & 6792                     & AC            &  & NO3 0         & SE3 0         & 7362                     & AC            \\
		CZ0 0         & PL0 0         & 4535                     & AC            &  & PL0 0         & SK0 0         & 3396                     & AC            \\
		CZ0 0         & SK0 0         & 6233                     & AC            &  & RO0 0         & RS0 0         & 5094                    & AC            \\ \bottomrule
	\end{tabular}
\end{table}

\newpage
\begin{table}[!ht]
	\caption{\textbf{DC links added exogenously to the model}.}
	\label{tab:dc_links}
	\centering
	\begin{tabular}{@{}llcc:lllcc@{}}
		\toprule
		\textbf{Bus0} & \textbf{Bus1} & \textbf{Capacity {[}MW{]}} & \textbf{Type} &  & \textbf{Bus0} & \textbf{Bus1} & \textbf{Capacity  {[}MW{]}} & \textbf{Type} \\ \midrule
		IT0 0         & GR0 0         & 500             & DC            &  & FI3 0         & EE6 0         & 650             & DC            \\
		ES2 0         & ES0 0         & 400             & DC            &  & SE3 0         & FI3 0         & 800             & DC            \\
		ME0 0         & IT0 0         & 1000            & DC            &  & SE3 0         & FI3 0         & 500             & DC            \\
		IT0 0         & IT1 0         & 1000            & DC            &  & ES0 0         & FR0 0         & 2200            & DC            \\
		FR0 0         & ES0 0         & 2000           & DC            &  & IT0 0         & FR0 0         & 1000           & DC            \\
		DE0 0         & BE0 0         & 1000            & DC            &  & FR0 0         & GB5 0         & 1000            & DC            \\
		GB5 0         & NL0 0         & 1000            & DC            &  & GB5 0         & FR0 0         & 1400           & DC            \\
		GB5 0         & BE0 0         & 1000            & DC            &  & GB5 0         & DK0 0         & 1400            & DC            \\
		GB5 0         & FR0 0         & 2000           & DC            &  & IT0 0         & CH0 0         & 1000           & DC            \\
		DK3 0         & DK0 0         & 600             & DC            &  & SE3 0         & DE0 0         & 700             & DC            \\
		DK3 0         & DE0 0         & 600             & DC            &  & NO3 0         & GB5 0         & 1400            & DC            \\
		PL0 0         & SE3 0         & 600             & DC            &  & GB5 0         & FR0 0         & 2000            & DC            \\
		NL0 0         & DK0 0         & 700             & DC            &  & GB5 0         & FR0 0         & 1400            & DC            \\
		NL0 0         & NO3 0         & 700             & DC            &  & GB5 0         & DE0 0         & 1400            & DC            \\
		DE0 0         & NO3 0         & 1400           & DC            &  & LT6 0         & SE3 0         & 700             & DC            \\
		GB5 0         & GB4 0         & 250             & DC            &  & EE6 0         & FI3 0         & 350             & DC            \\
		GB5 0         & GB4 0         & 250             & DC            &  & PL0 0         & LT6 0         & 2000            & DC            \\
		GB5 0         & IE4 0         & 500             & DC            &  & IT1 0         & IT0 0         & 300             & DC            \\
		SE3 0         & DK0 0         & 250             & DC            &  & DK0 0         & NO3 0         & 940             & DC            \\
		SE3 0         & DK0 0         & 600             & DC            &  &               &               &                   &               \\ \bottomrule
	\end{tabular}
\end{table}

\newpage
\section{Technology cost assumptions}
\begin{table}[h]
	\caption{Cost assumptions for electricity generation technologies for 2030 from \cite{DEA2022_gen} and \cite{DEA2022_carbon}.}
	\label{tab:generator_costs}
	\centering
	\begin{threeparttable}
		\begin{tabular}{@{}lcccc@{}}
			\toprule
			Generators                        & Investment cost & Lifetime (Years) & FOM     & VOM         \\ \midrule
			OCGT\tnote{a}                     & 435 €/kW   & 25               & 1.78 \% & 4.5 €/MWh  \\
			CCGT\tnote{b}                    & 830 €/kW    & 25               & 3.35 \% & 4.2 €/MWh  \\
			Nuclear\tnote{c}                & 7940 €/kW  & 40               & 1.4 \% & 3.5 €/MWh  \\
			Coal\tnote{d}                  & 3846 €/kW  & 40               & 1.6 \%  & 3.5 €/MWh  \\
			Biomass CHP \tnote{e}         & 3210 €/kW   & 25               & 3.58 \% & 2.1 €/MWh  \\
			Gas CHP \tnote{f}            & 560 €/kW   	& 25               &  3.32 \% & 4.2 €/MWh  \\
			Carbon capture \tnote{g}	  & 2,700,000 €/(tCO$_2$/h) & 25   & 3.0 \% & 2.1 €/MWh \\
			Solar (utility)                   & 376 €/kW   & 35               & 1.93 \% & 0           \\
			Solar (rooftop)                   & 784 €/kW   & 30               & 1.24 \% & 0            \\
			Onshore wind                      & 1036 €/kW  & 30               & 1.22 \% & 1.35 €/MWh \\
			Offshore wind 				   	  & 1573 €/kW  & 30 			   & 2.29 \% & 2.67 €/MWh \\
			offwind-ac-connection-submarine   & 2685 €/MW/km  & 30               & 0       & 0           \\
			offwind-ac-connection-underground & 1342 €/MW/km  & 30               & 0       & 0            \\
			offwind-ac-station                & 250 €/kW      & 30               & 0       & 0            \\
			offwind-dc-connection-submarine   & 2000 €/MW/km  & 30               & 0       & 0            \\
			offwind-dc-connection-underground & 1000 €/MW/km  & 30               & 0       & 0            \\
			offwind-dc-station                & 400 €/kW      & 30               & 0       & 0            \\ \bottomrule
		\end{tabular}
		\begin{tablenotes}
			\item[a] OCGT efficiency $\eta = 41\%$, fuel cost =  20.1 €/MWh
			\item[b] CCGT efficiency $\eta = 58\%$, fuel cost = 20.1 €/MWh
			\item[c] Nuclear efficiency $\eta = 33\%$, fuel cost = 2.6 €/MWh
			\item[d] Coal efficiency $\eta = 33\%$, fuel cost = 8.15 €/MWh
			
			\item[e] Biomass CHP $\eta_{electricity} = 33\%$, $\eta_{heat} = 69\%$, fuel cost = 25.2 €/MWh
			\item[f] Gas CHP $\eta_{electricity} = 41\%$, $\eta_{heat} = 41\%$, fuel cost = 20.1 €/MWh
			\item[g] Carbon capture rate = $90\%$
		\end{tablenotes}
	\end{threeparttable}
\end{table}

\newpage
\section{Supplemental Figures}

\begin{figure}[!h]
	\centering
	\includegraphics[width=0.7\textwidth]{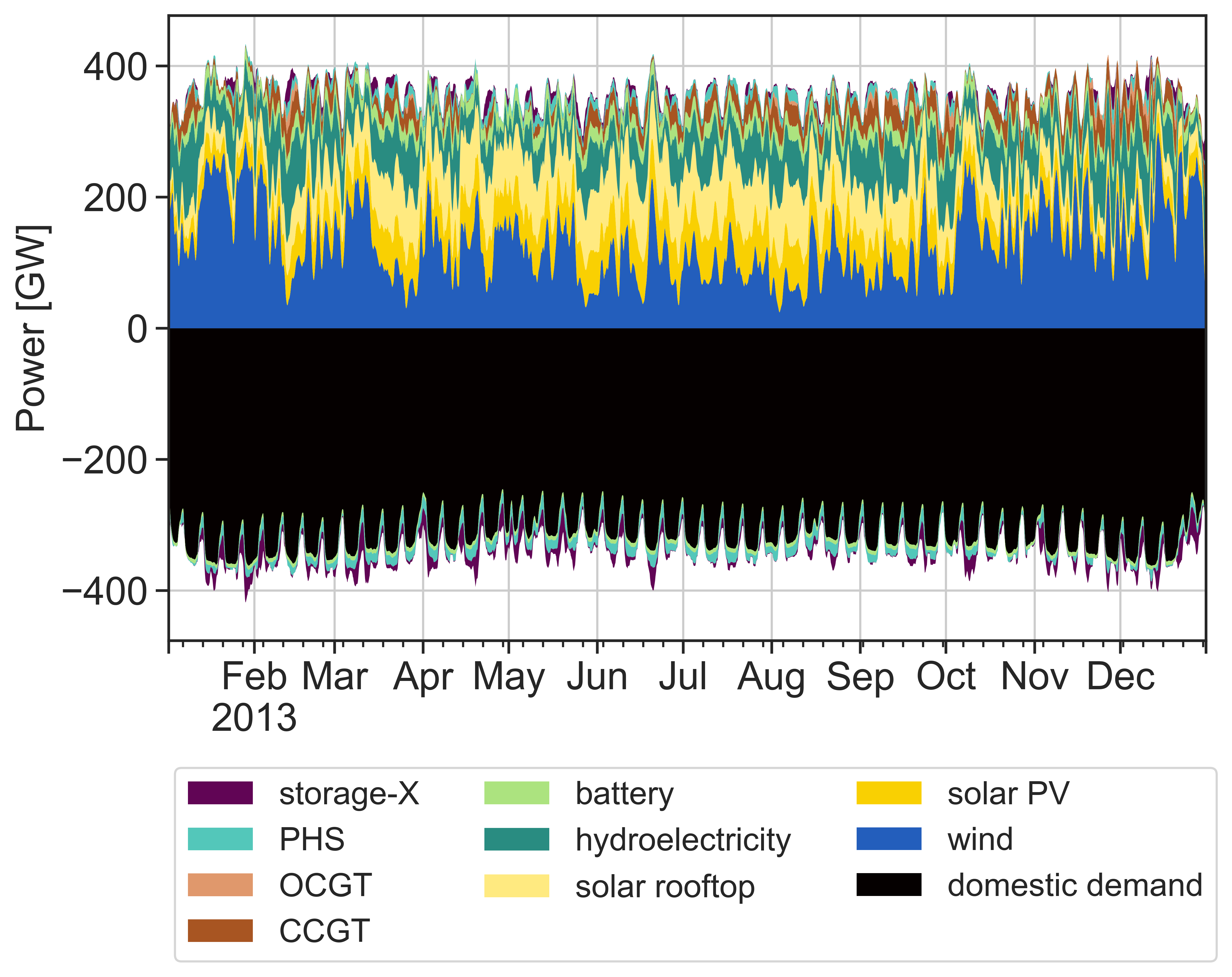}
	\includegraphics[width=0.7\textwidth]{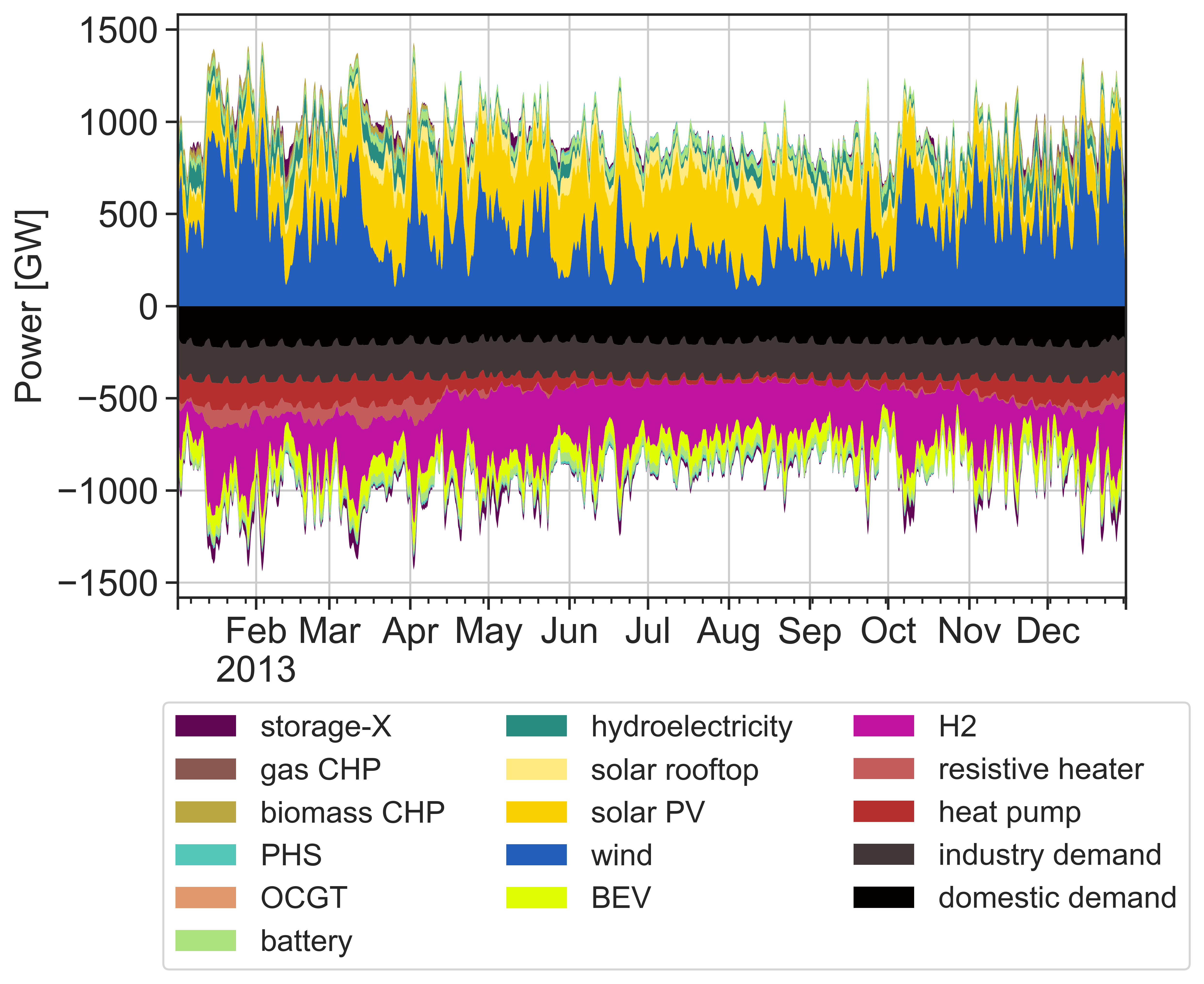}
	\caption{\textbf{Balancing of supply and demand}. Results are 24-hourly moving averages of the balance of the power supply and demand for the (top) Electricity and (bottom) fully sector-coupled system, obtained with the best storage-X in the single parametric sweep, i.e., with a discharge efficiency of 95\% and the remaining parameters according to the "Fixed" configuration.}
	\label{sfig:balancing}
\end{figure}

\newpage
\begin{figure}[!ht]
	\centering
	\begin{subfigure}[b]{0.7\textwidth}
		\includegraphics[width=1\linewidth]{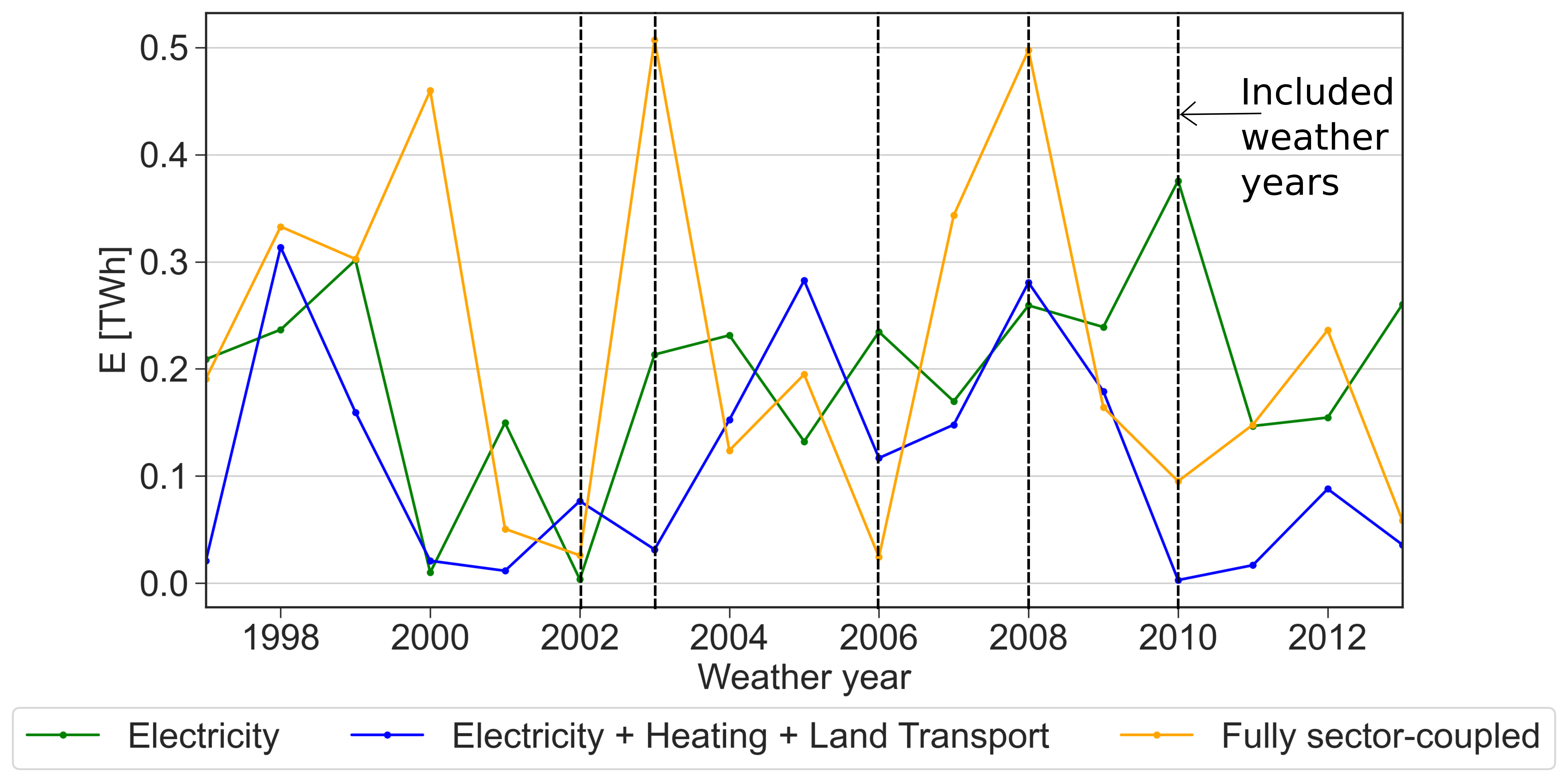}
		\caption{}
	\end{subfigure}%

	\begin{subfigure}[b]{0.7\textwidth}
		\includegraphics[width=1\linewidth]{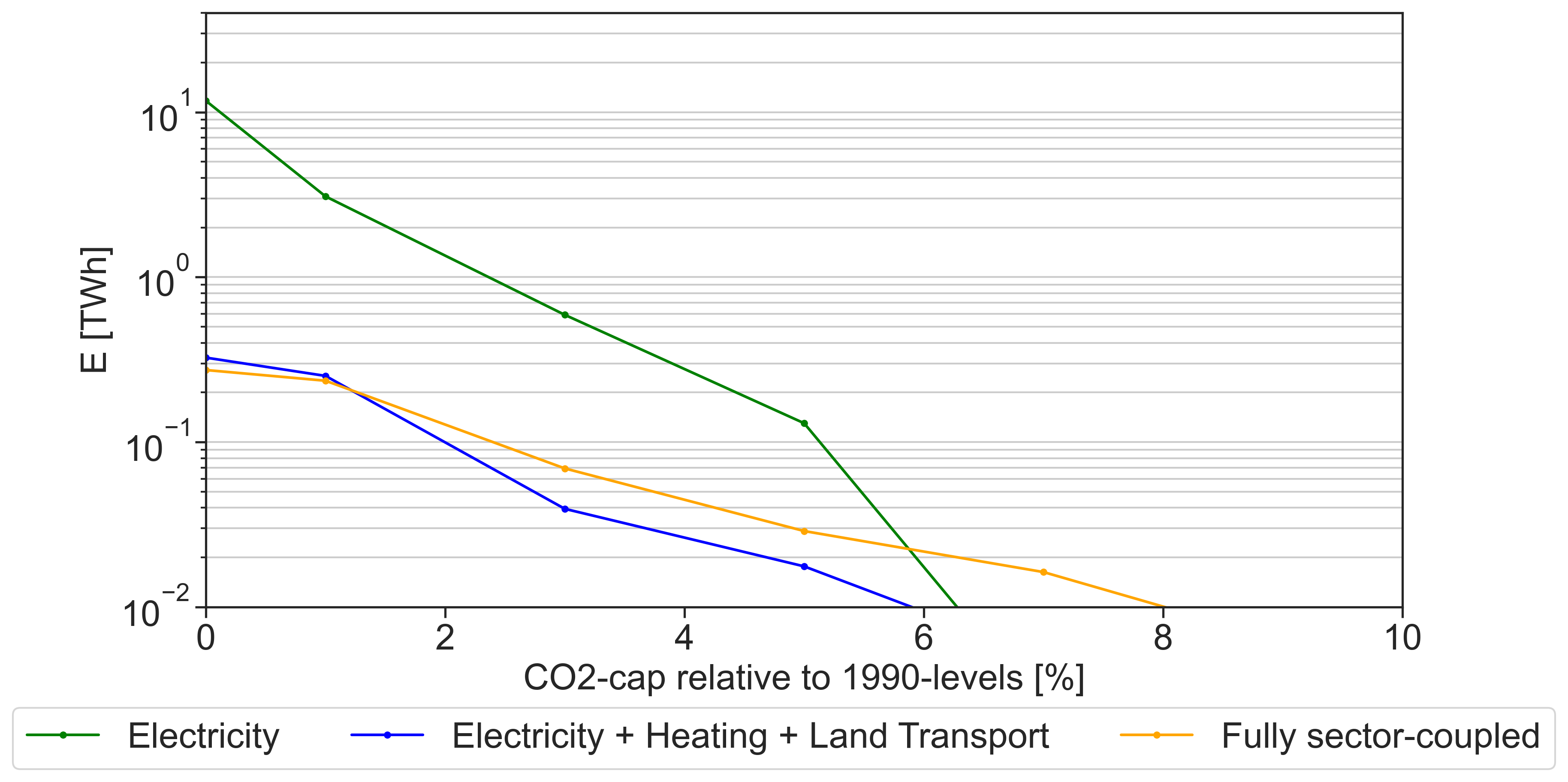}
		\caption{}
	\end{subfigure}
	\caption{\textbf{\mbox{\text{Storage-X}} energy capacity at different (a) weather years and (b) CO$_2$-emissions constraints}. Shown is the energy capacity obtained with \mbox{\text{storage-X}} with fixed parameters from Tab. \ref{tab:storage_reference}, for the three system compositions. 
	\newline Weather years are varied from 1997 to 2012 and compared with the base year of 2013. The weather year determines the available renewable resources and the heat demand. Choosing different weather years impacts differently the storage-X deployment depending on the system composition.
	\newline The CO$_2$ constraint is varied from 10\% to 0\%. Without changes in the \mbox{\text{storage-X}} parameters, sector-coupling reduces the optimal energy capacity by more than an order of magnitude, when considering a fully decarbonized system.
	}
	\label{sfig:structural_senstivity}
\end{figure}

\newpage
\begin{figure}[!htbp]
	\centering
	\includegraphics[width=0.9\textwidth]{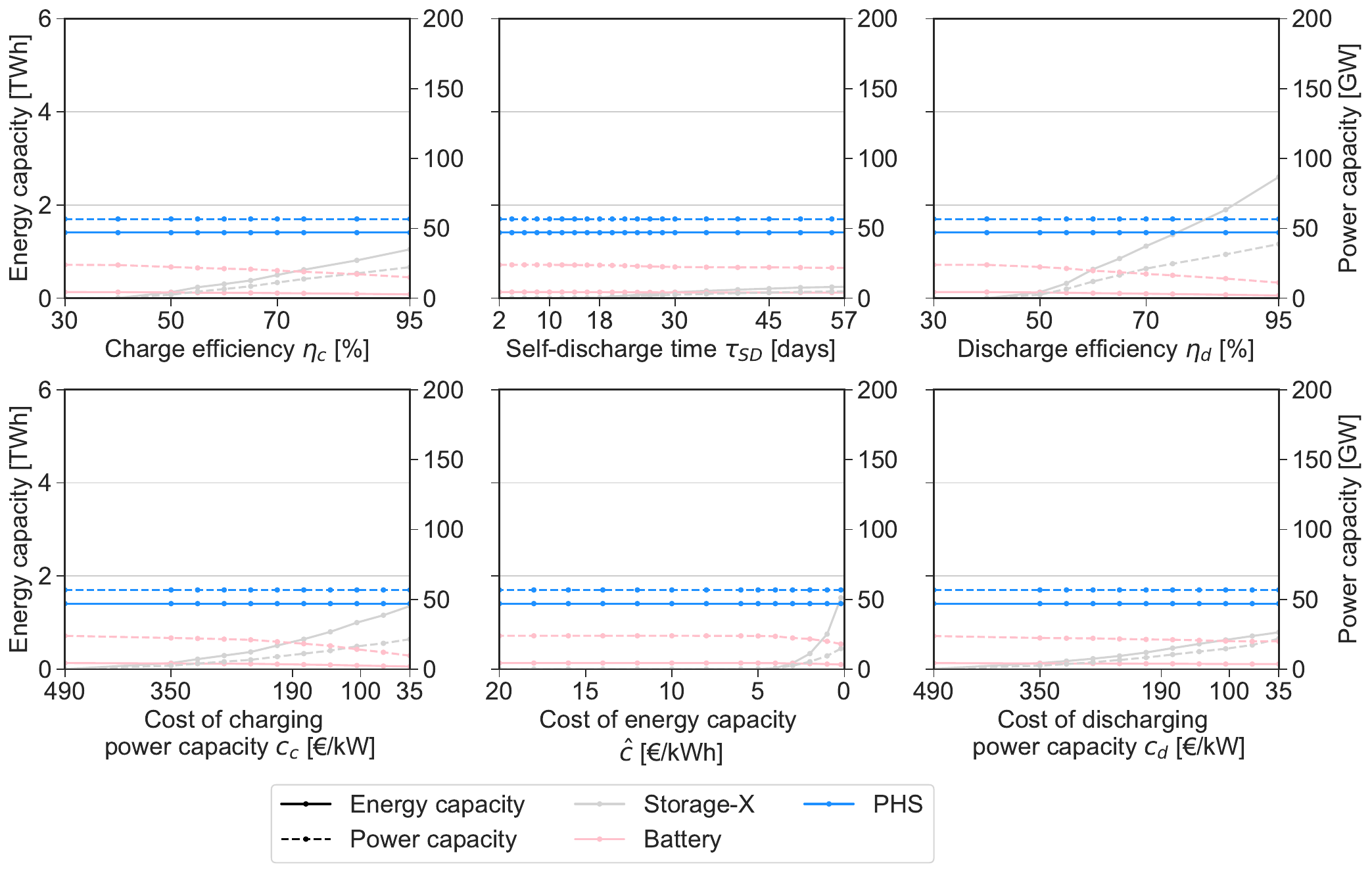}
	\caption{\textbf{Stationary battery, PHS, and storage-X capacities for the Electricity system}. Energy capacities (left axis) and power capacities (right axis) at variable \mbox{\text{storage-X}} parameters. }
	\label{sfig:PHS_battery_caps}
\end{figure}

\newpage
\begin{figure}[!htbp]
	\centering
	\includegraphics[width=0.9\textwidth]{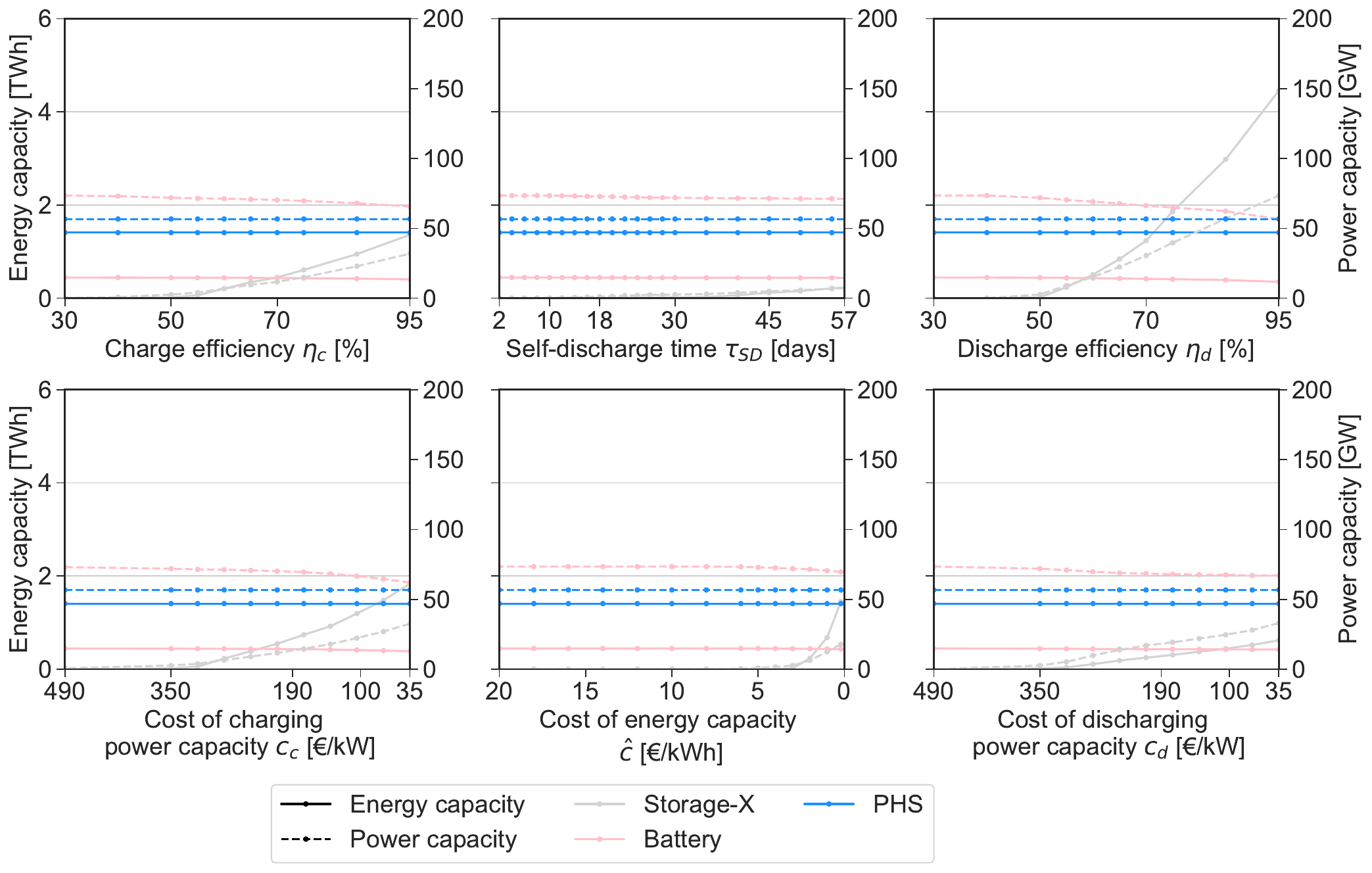}
	\caption{\textbf{Stationary battery, PHS and storage-X capacities for the Electricity + Heating + Land Transport system}. Energy capacities (left axis) and power capacities (right axis) at variable \mbox{\text{storage-X}} parameters. }
	\label{sfig:PHS_battery_caps_T_H}
\end{figure}

\newpage
\begin{figure}[!htbp]
	\centering
	\includegraphics[width=0.9\textwidth]{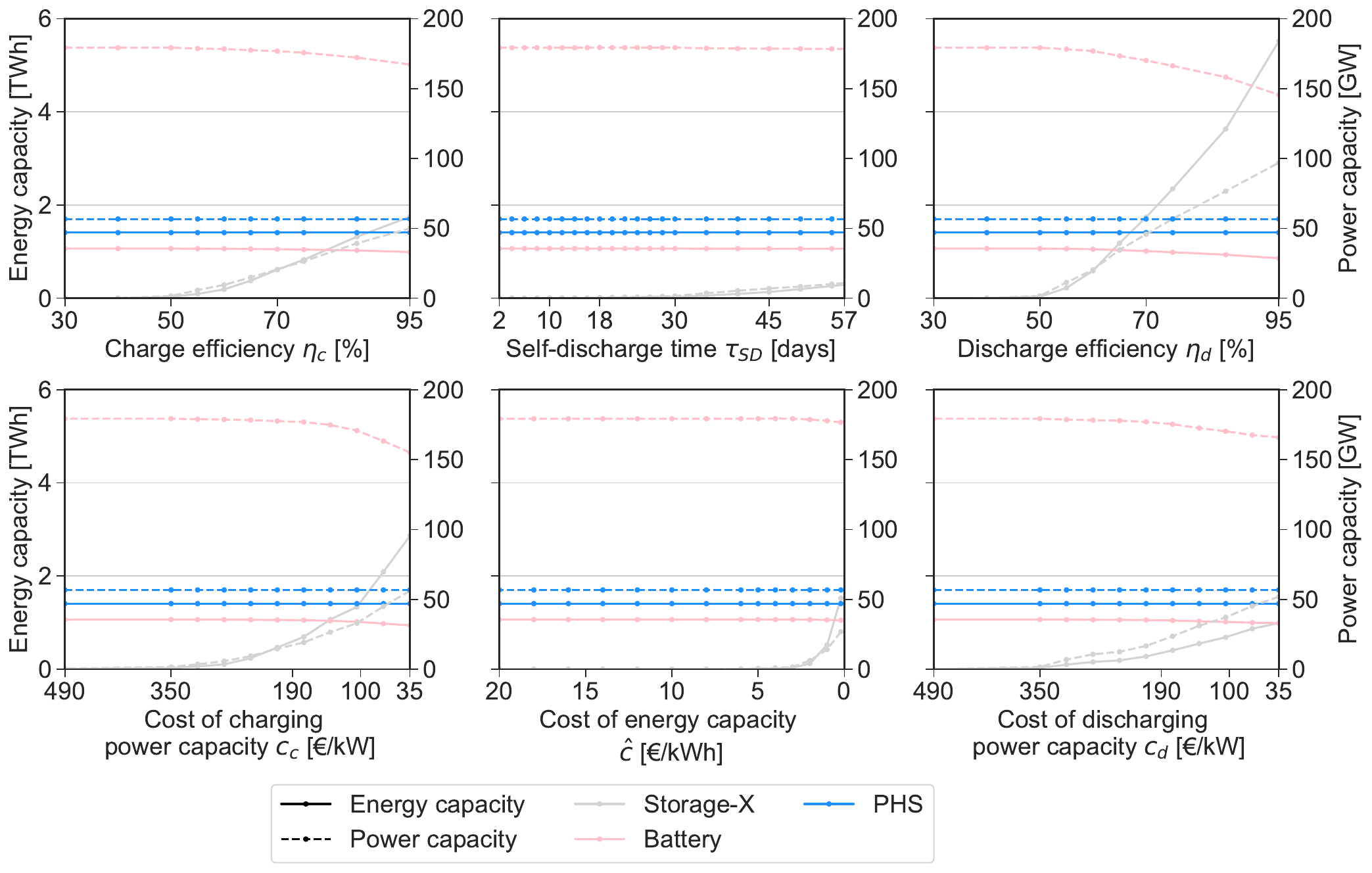}
	\caption{\textbf{Stationary battery, PHS and storage-X capacities for the Fully sector-coupled system}. Energy capacities (left axis) and power capacities (right axis) at variable \mbox{\text{storage-X}} parameters.}
	\label{sfig:PHS_battery_caps_T_H_I_B}
\end{figure}

\newpage
\begin{figure}[!ht]
	\centering
	\includegraphics[width=0.45\textwidth]{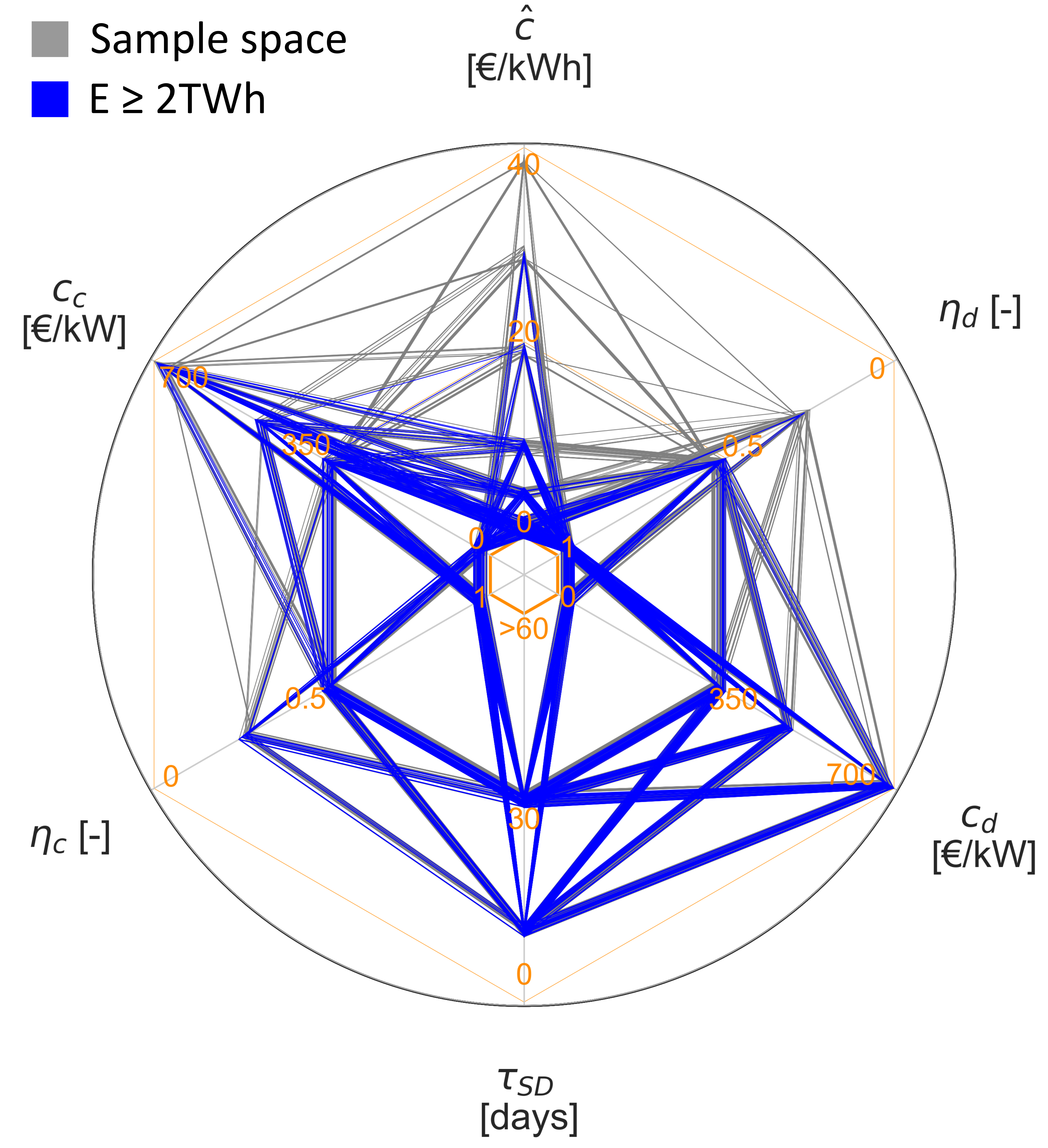}
	\includegraphics[width=0.45\textwidth]{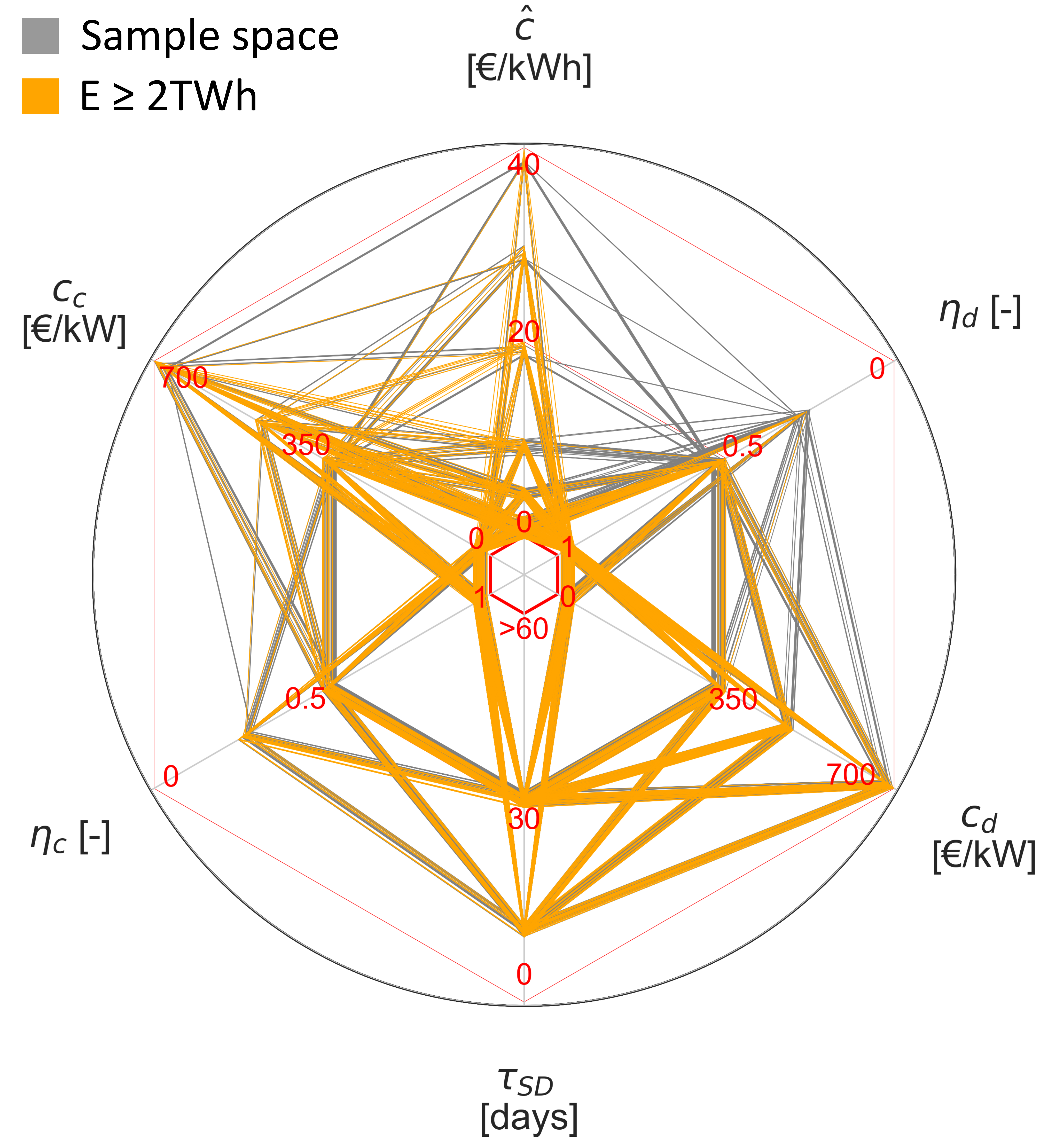}
	\caption{\textbf{Configurations fulfilling $\geq$2~TWh}. (Left) SC1 Electricity + Heating + Land Transport and (right) SC2 Fully sector-coupled system. See Fig. \ref{fig:design_space_E} for the Electricity system. Out of 758 samples, 275 and 344 configurations qualify for the design space in SC2 and SC3.) }
	\label{sfig:design_space_E_sectors}
\end{figure}

\newpage
\begin{figure}[!ht]
	\centering
	\includegraphics[width=0.8\textwidth]{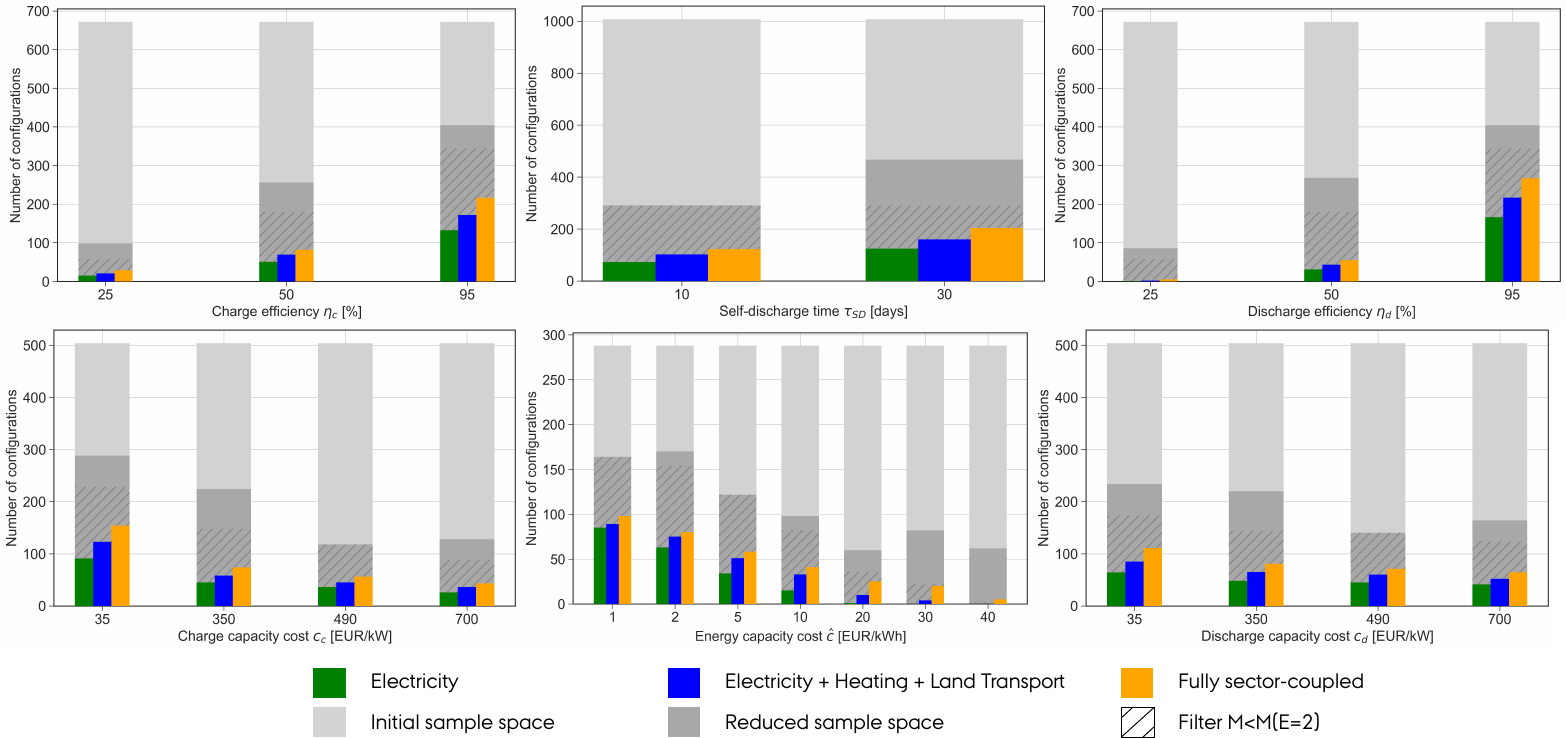}
	\caption{\textbf{Frequency of parameters}. Shown is the number of configurations within the design space ($E\geq2$~TWh) across all parameter ranges and sectors (colored bins). The "Initial sample space" is a uniform sampling, i.e., the values within each parameter set occur equally frequent. For computation purposes, the sample space is reduced subsequent to the calculation of the Electricity system ("Filter $M<M(E=2)$") marked with the hatches. The "Reduced sample space" is marginally larger than the "Filter $M<M(E=2)$" since it preserves the exterior of the sample space. See \ref{appendix:filter}. Only the energy capacity cost has 0 configurations (for the "Electricity" and "Electricity + Heating + Land Transport" systems) at its worst values (30 and 40 €/kWh).}
	\label{sfig:frequency_of_parameters}
\end{figure}

\newpage
\begin{figure}[!ht]
	\centering
	\includegraphics[width=0.7\textwidth]{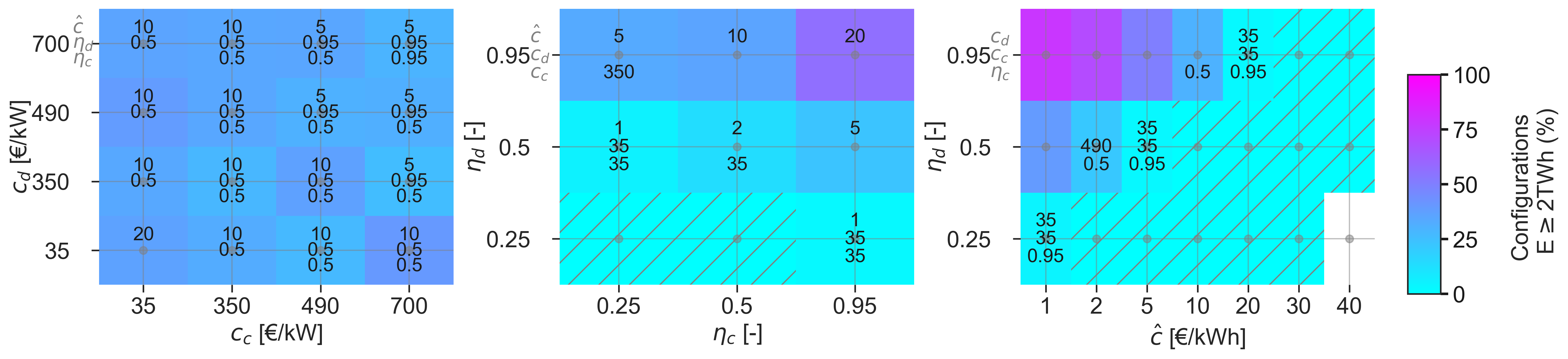}
	\includegraphics[width=0.7\textwidth]{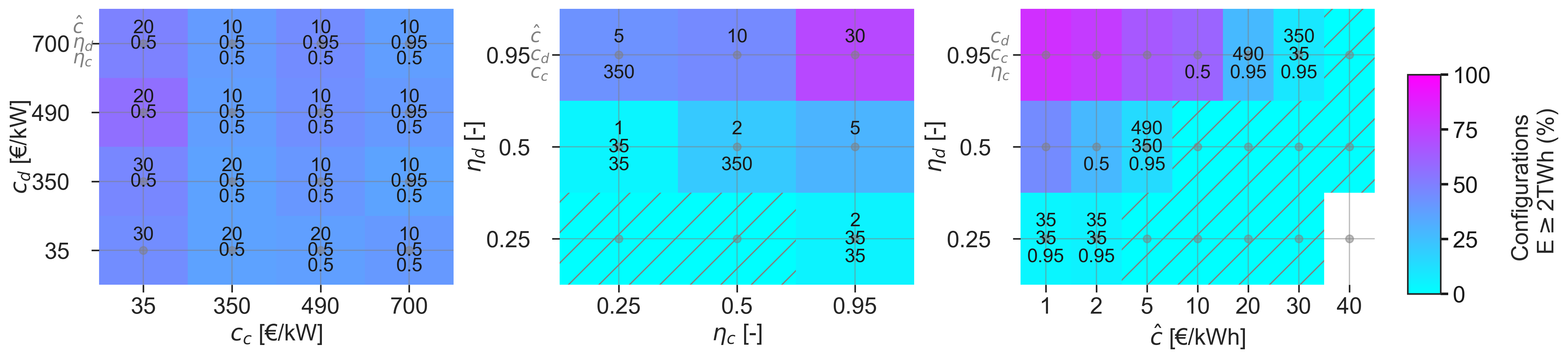}
	\includegraphics[width=0.7\textwidth]{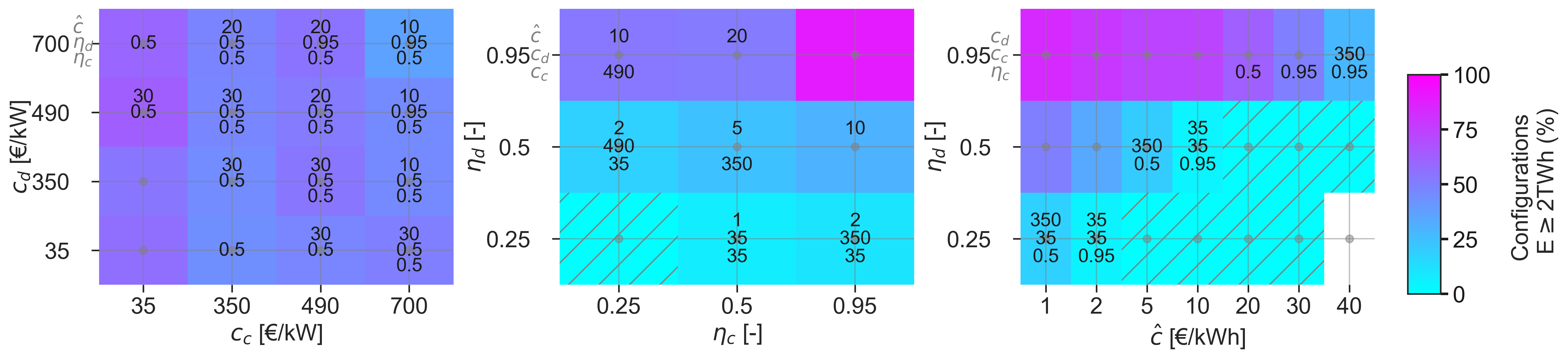}
	\caption{\textbf{Cost and efficiency requirements}. Requirements of storage-X to appear in (top) SC1, (middle) SC2, and (bottom) SC3. The requirements (on both axes) refer to the cost and efficiency combination that entails a cumulative deployment of $E\geq2$~TWh. The crossed areas indicate that no storage at the considered combinations entails sufficiently large deployment. The white area does not contain data due to combinations being omitted in the reduction of the sample space, explained in \ref{appendix:filter}.}
	\label{sfig:matrix_E_T_H_I_B}
\end{figure}

\newpage
\begin{figure}[!ht]
	\centering
	\includegraphics[width=0.9\textwidth]{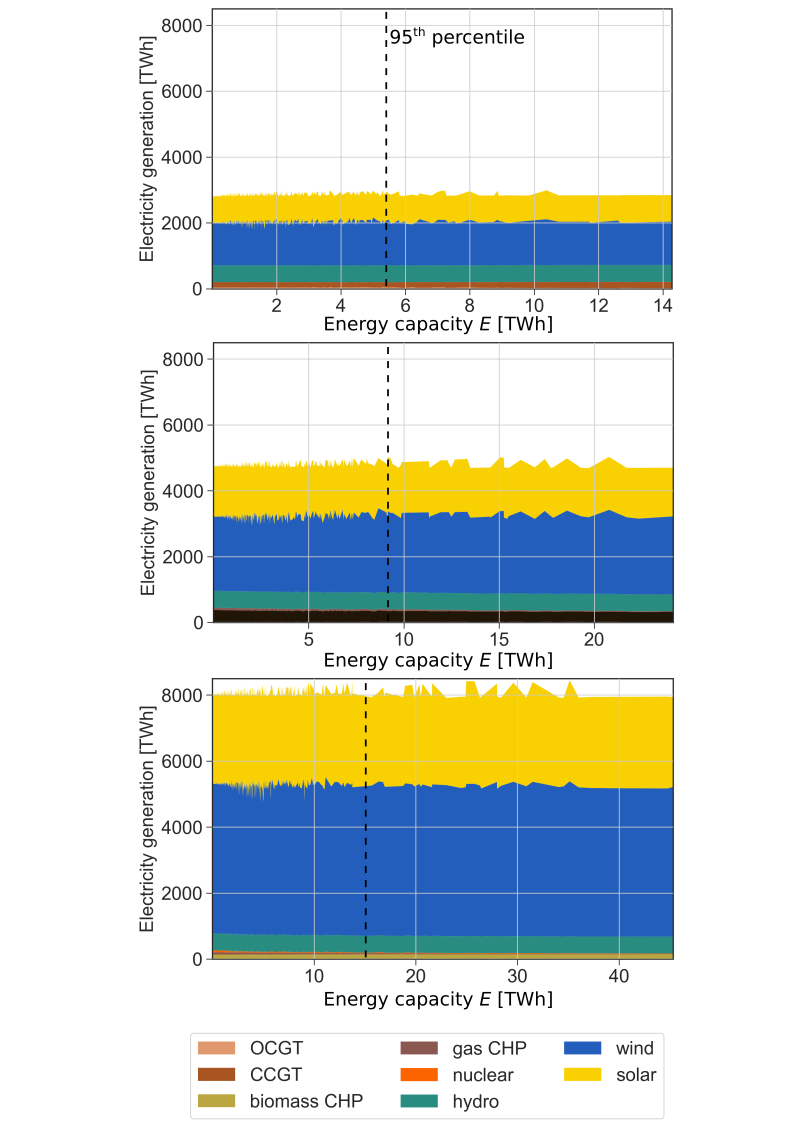}
	\caption{\textbf{Annual electricity generation and capacity for the Electricity system}. Here, electricity is supplied from wind ($45.7\pm2.9\%$, $47.7\pm2.2\%$, $56.3\pm1.9\%$), solar ($29.2\pm2.7\%$, $32.5\pm2.1\%$, $34.2\pm2.3\%$), and hydropower ($18.1\pm0.6\%$, $10.8\pm0.6\%$, $6.4\pm0.1\%$) for SC1, SC2, and SC3.}
	\label{sfig:gen_mix_E}
\end{figure}

\newpage
\begin{figure}[!ht]
	\centering
	\includegraphics[width=0.45\textwidth]{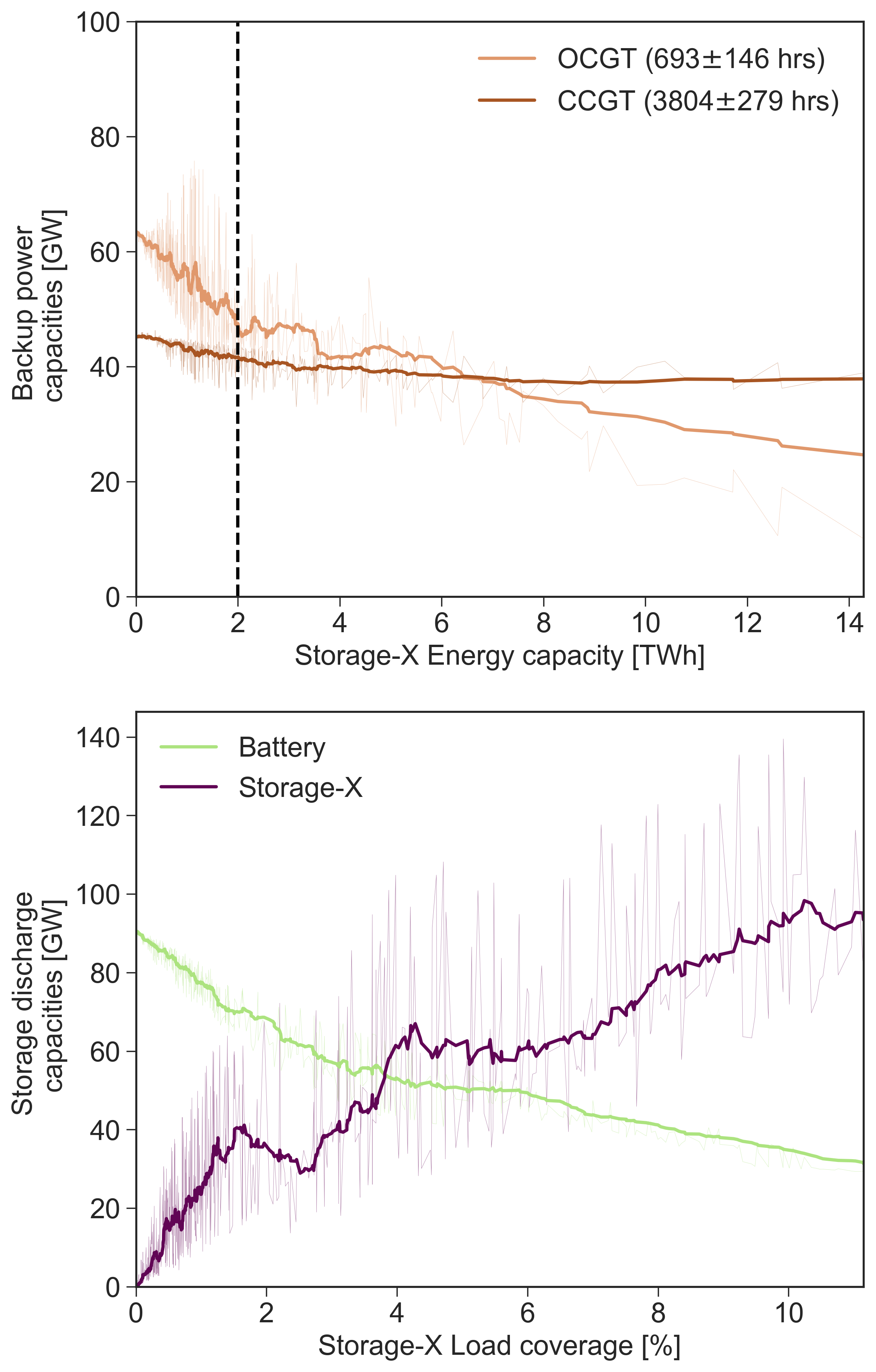}
	\includegraphics[width=0.45\textwidth]{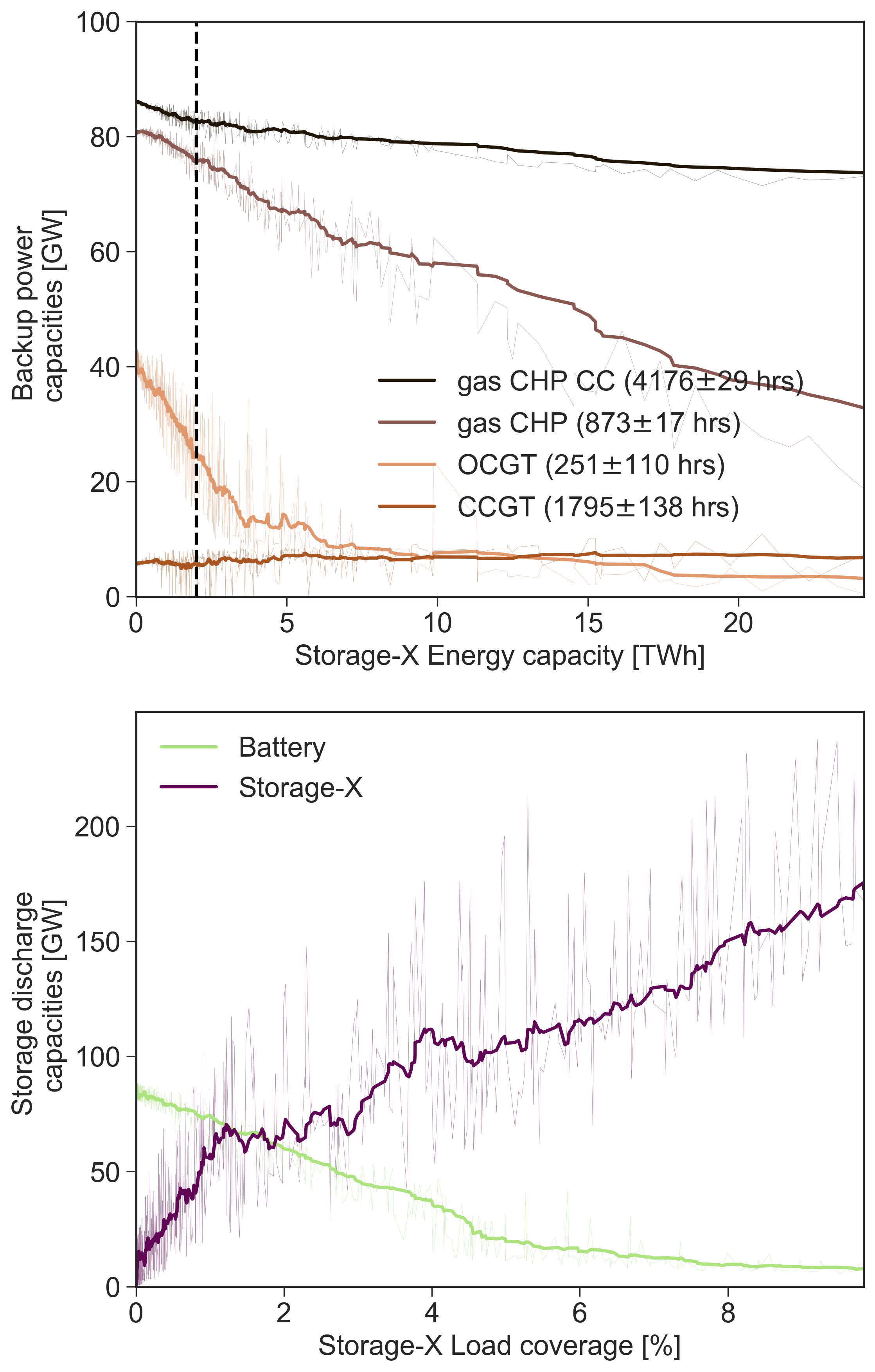}
	\caption{\textbf{Backup power and storage discharge capacity}. Results for (left) SC1 and (right) SC2.}
	\label{sfig:backup_mix}
\end{figure}

\newpage
\begin{figure}[!ht]
	\centering
	\includegraphics[width=0.55\textwidth]{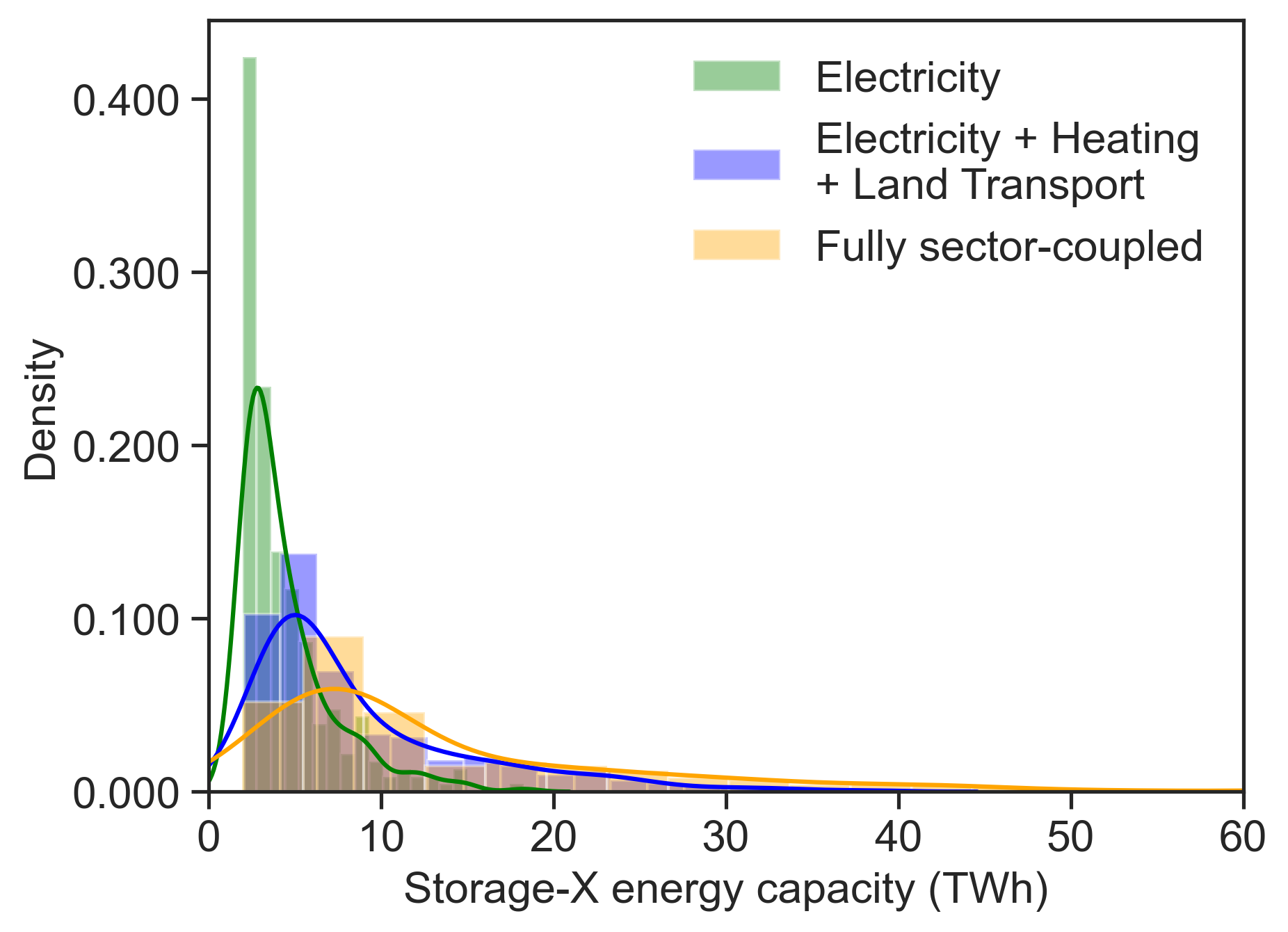}
	\includegraphics[width=0.55\textwidth]{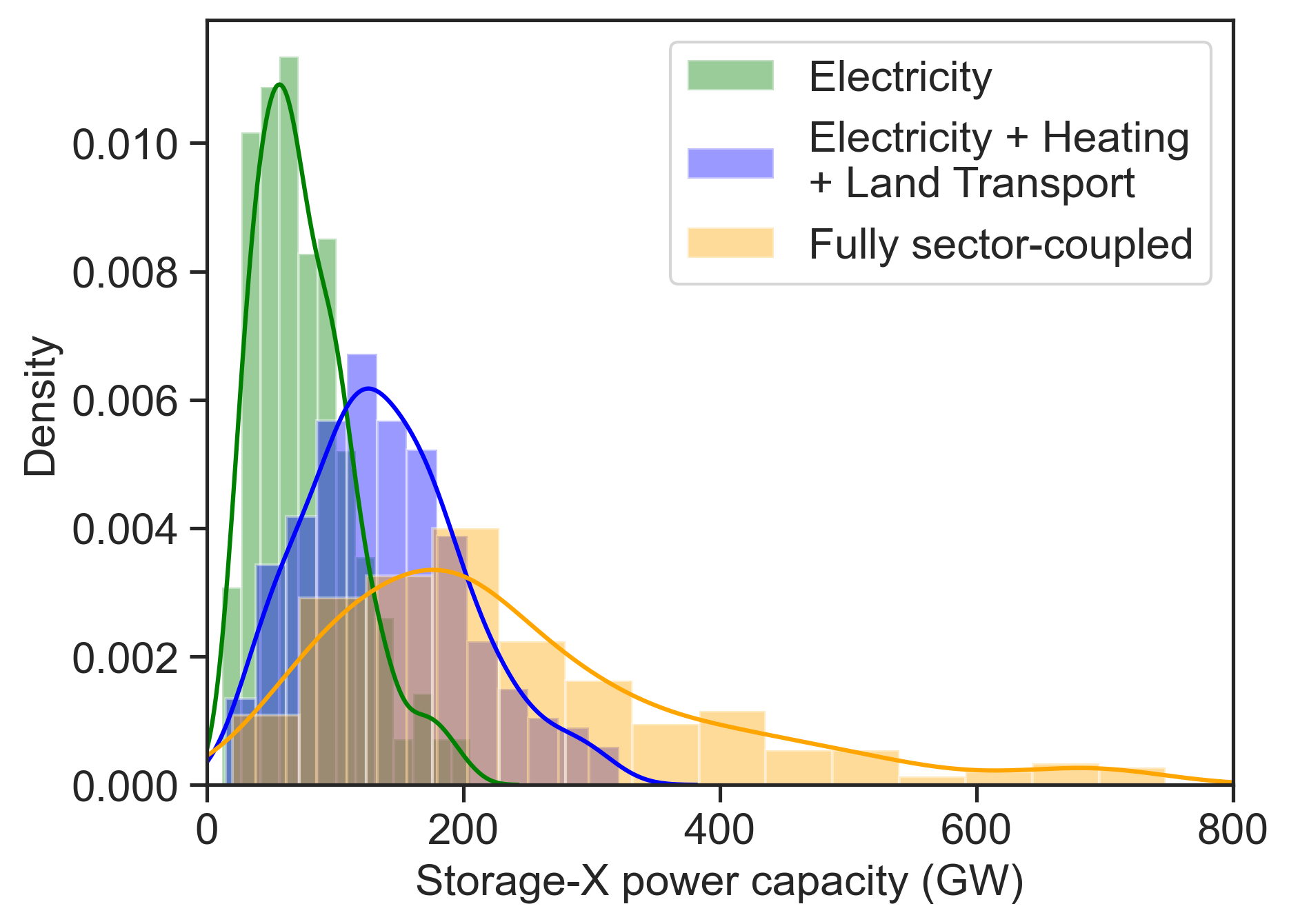}
	\includegraphics[width=0.55\textwidth]{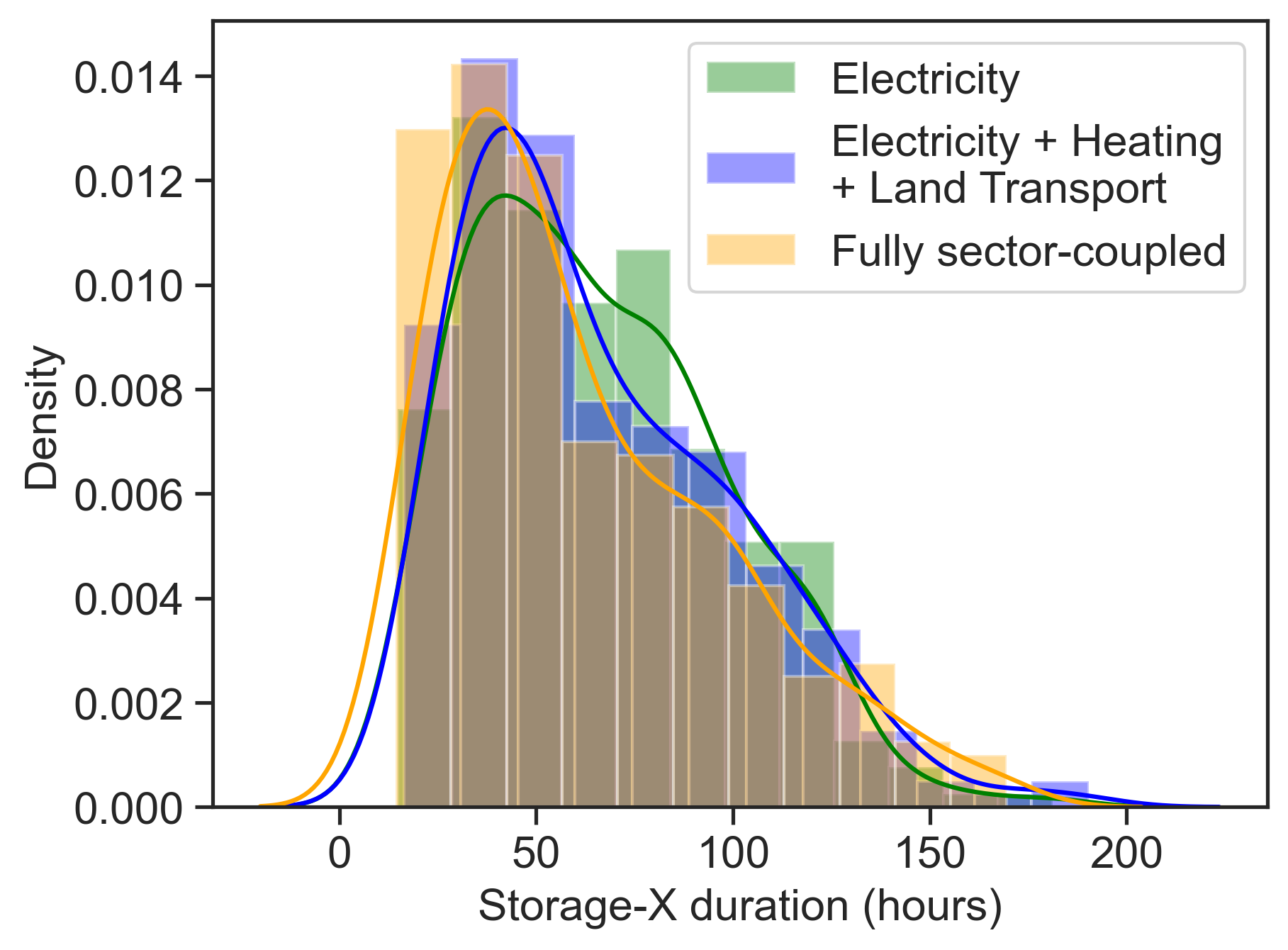}
	\caption{\textbf{Storage-X energy capacity, discharge power capacity, and duration}. Histogram of Europe-aggregate Storage-X (top) energy capacity in units of dispatchable electricity, (middle) discharge power capacity in units of dispatchable electricity, and (bottom) the duration (i.e., the ratio of the energy capacity and the discharge power capacity).}
	\label{sfig:storagex_volume}
\end{figure}

\newpage
\begin{figure}[!ht]
	\centering
	\includegraphics[width=0.55\textwidth]{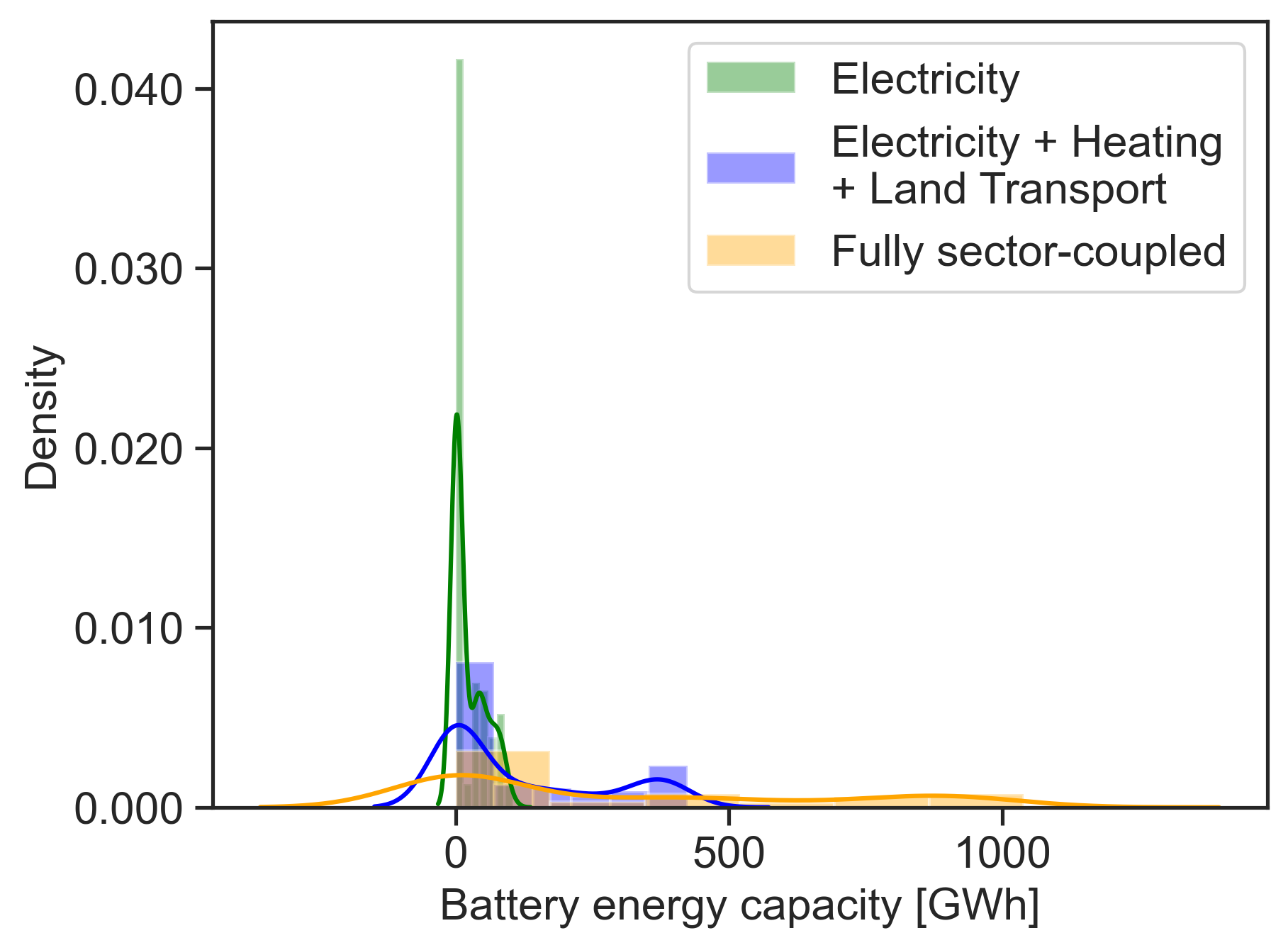}
	\includegraphics[width=0.55\textwidth]{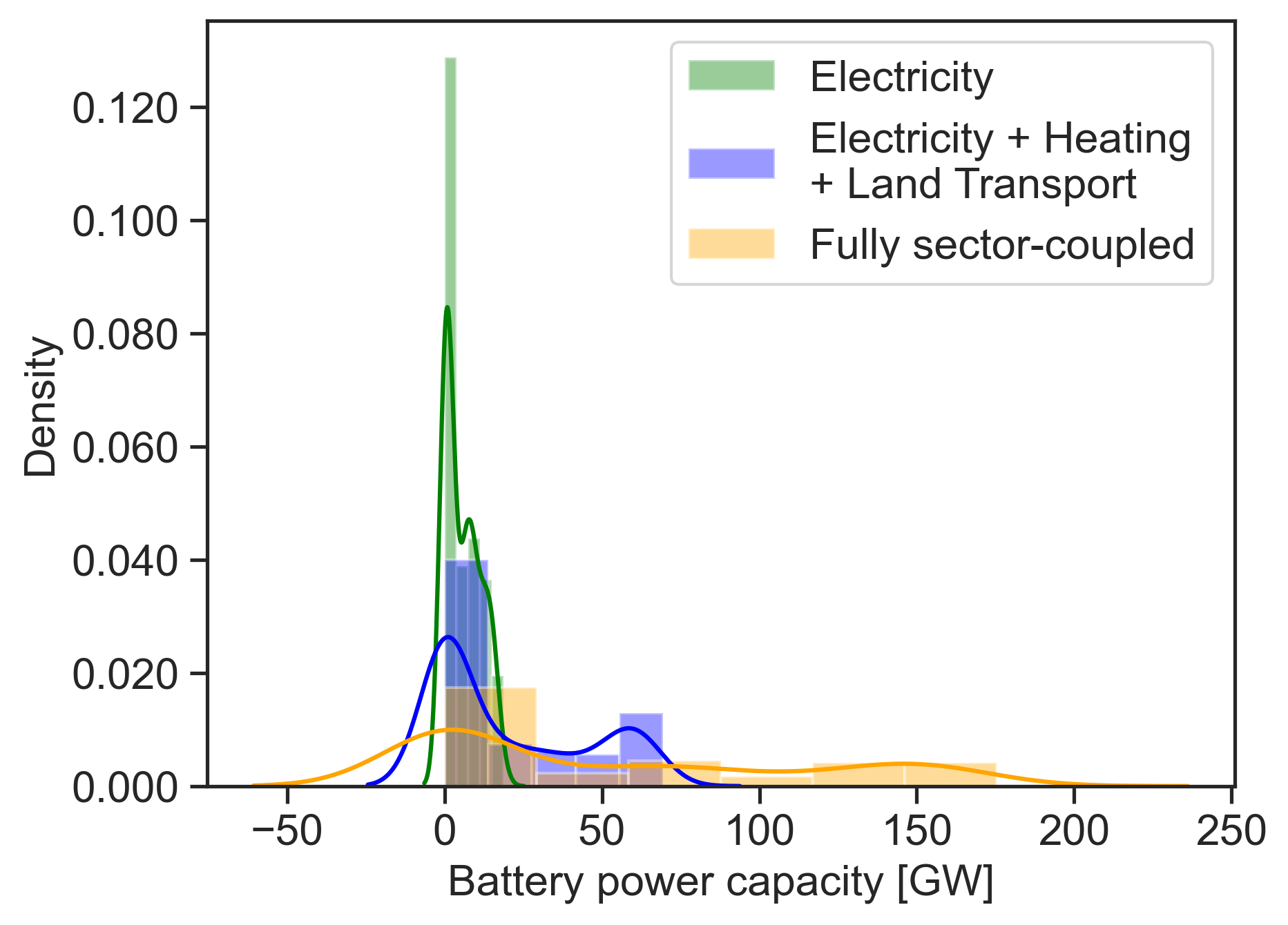}
	\includegraphics[width=0.55\textwidth]{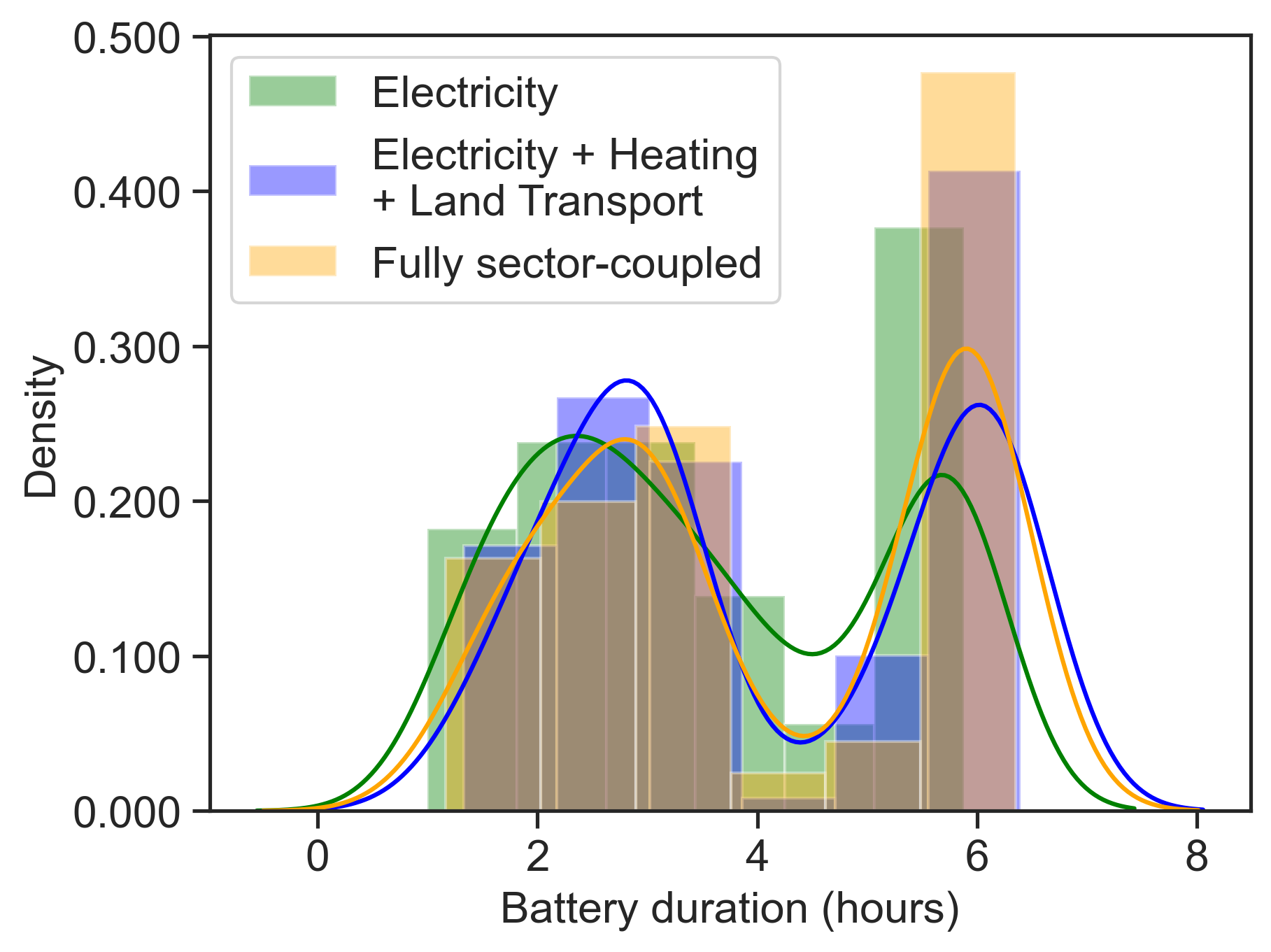}
	\caption{\textbf{Battery energy capacity, discharge power capacity, and duration}. Histogram of Europe-aggregate battery (top) energy capacity in units of dispatchable electricity, (middle) discharge power capacity in units of dispatchable electricity, and (bottom) the duration (i.e., the ratio of the energy capacity and the discharge power capacity).}
	\label{sfig:battery_volume}
\end{figure}

\newpage
\begin{figure}[!ht]
	\centering
	\includegraphics[width=0.45\textwidth]{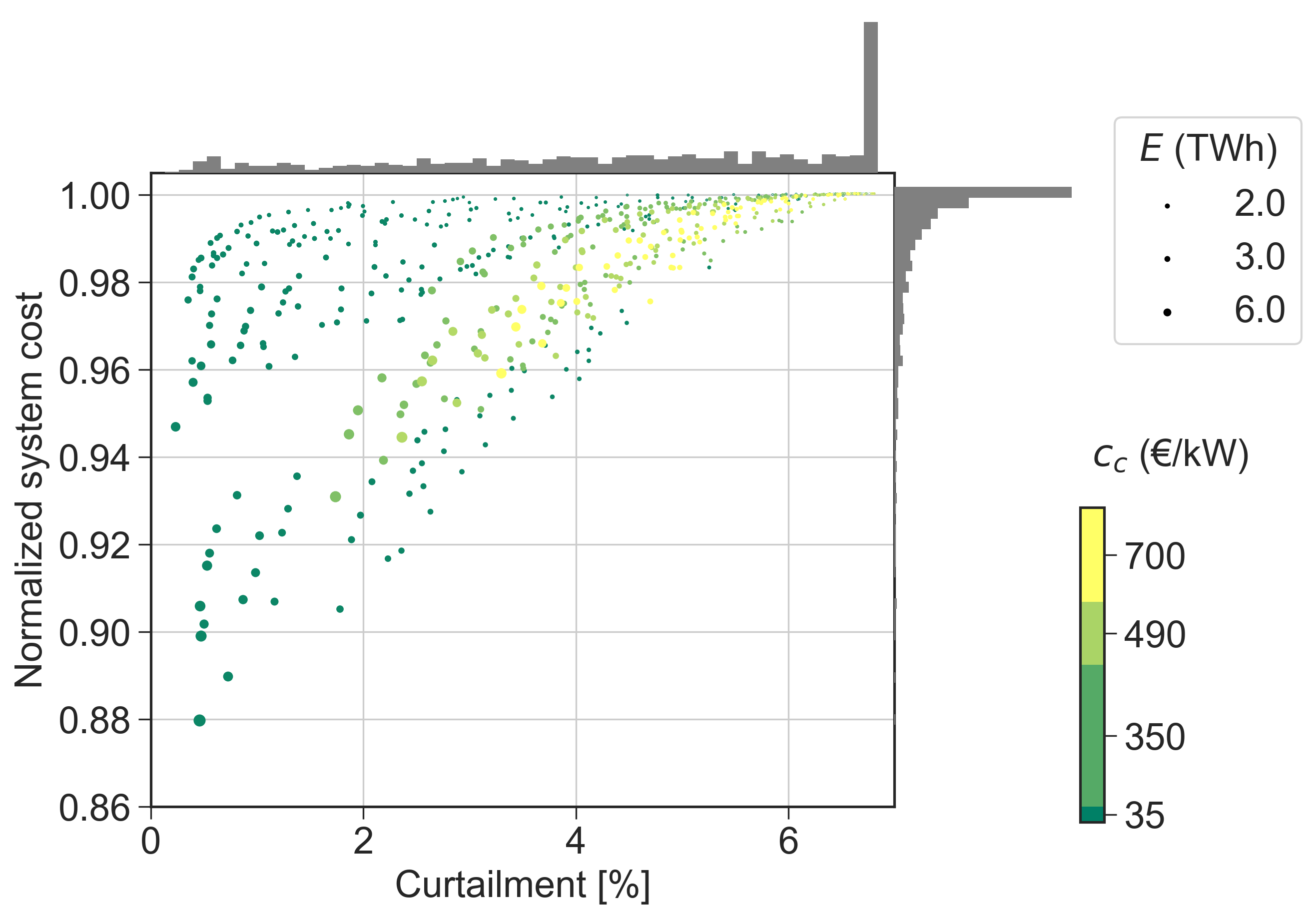}
	\includegraphics[width=0.45\textwidth]{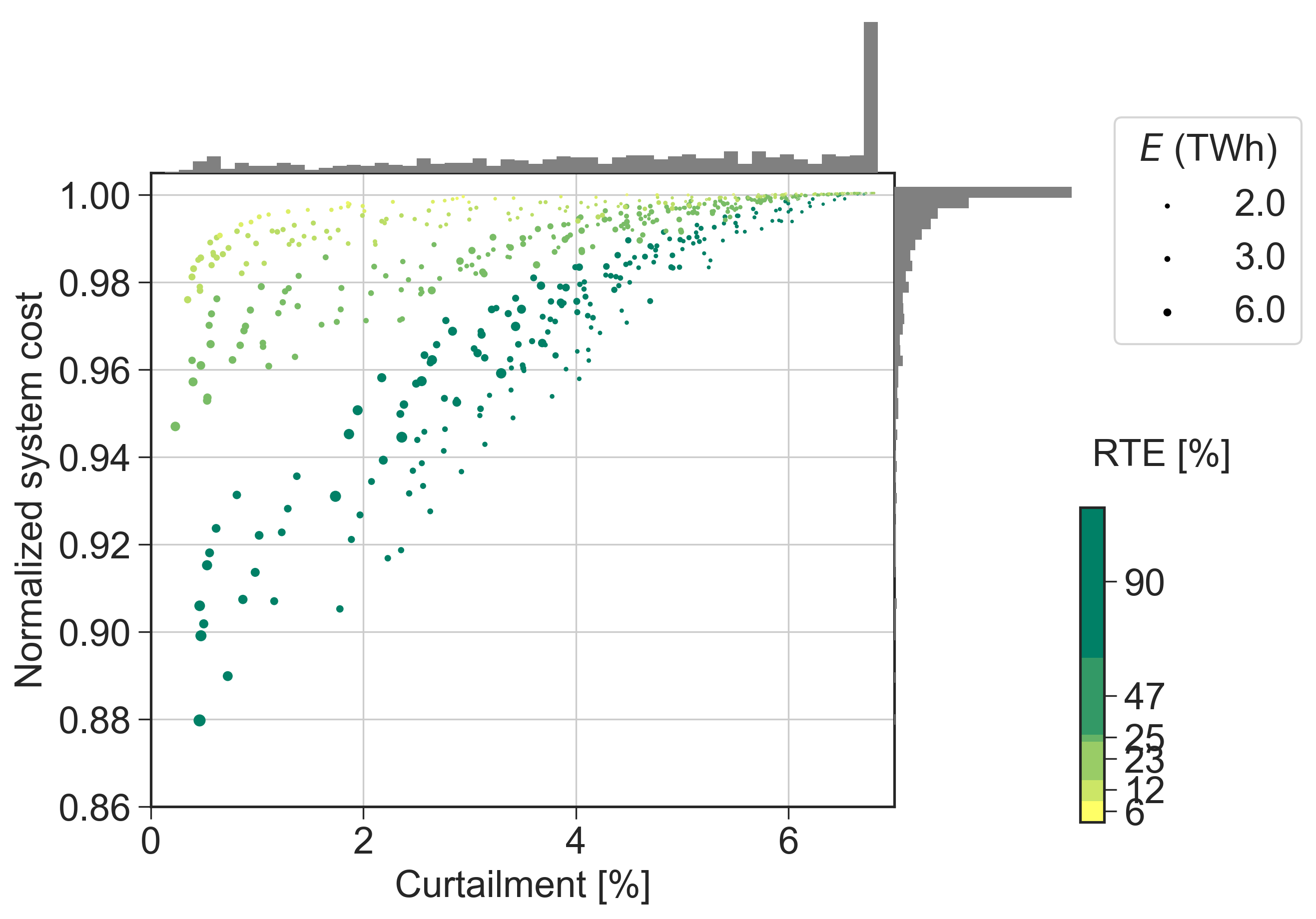}
	\includegraphics[width=0.45\textwidth]{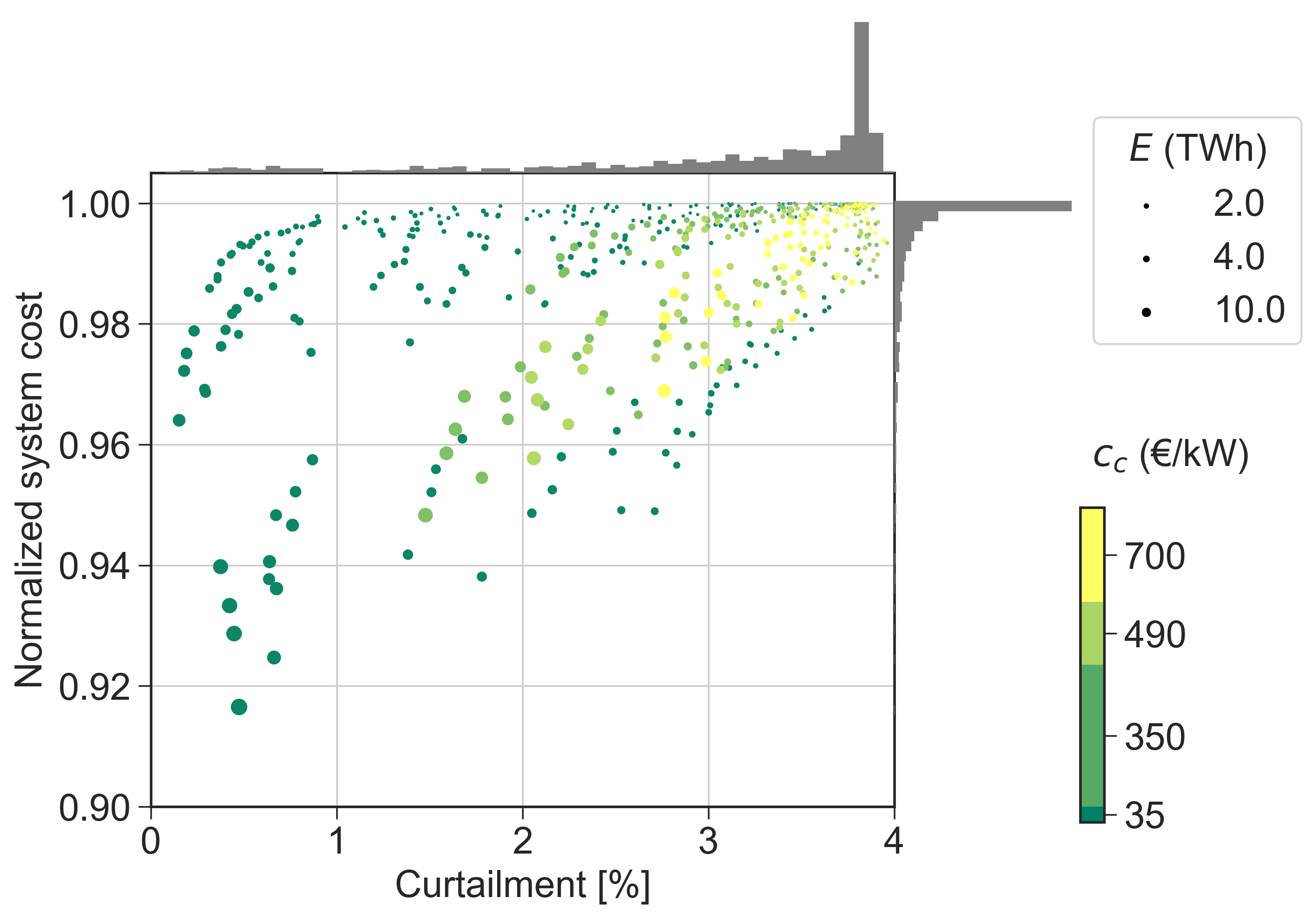}
	\includegraphics[width=0.45\textwidth]{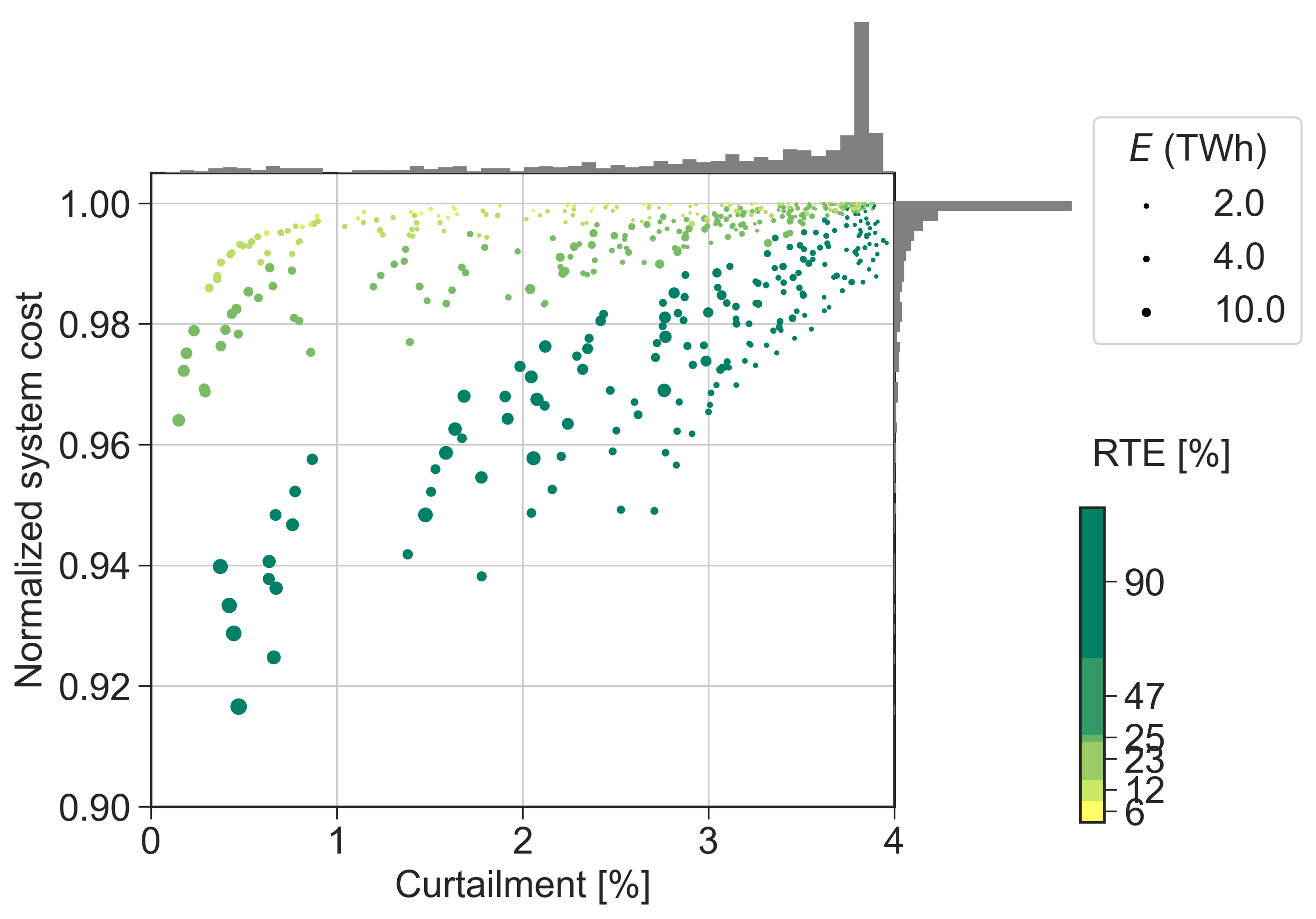}
	\includegraphics[width=0.45\textwidth]{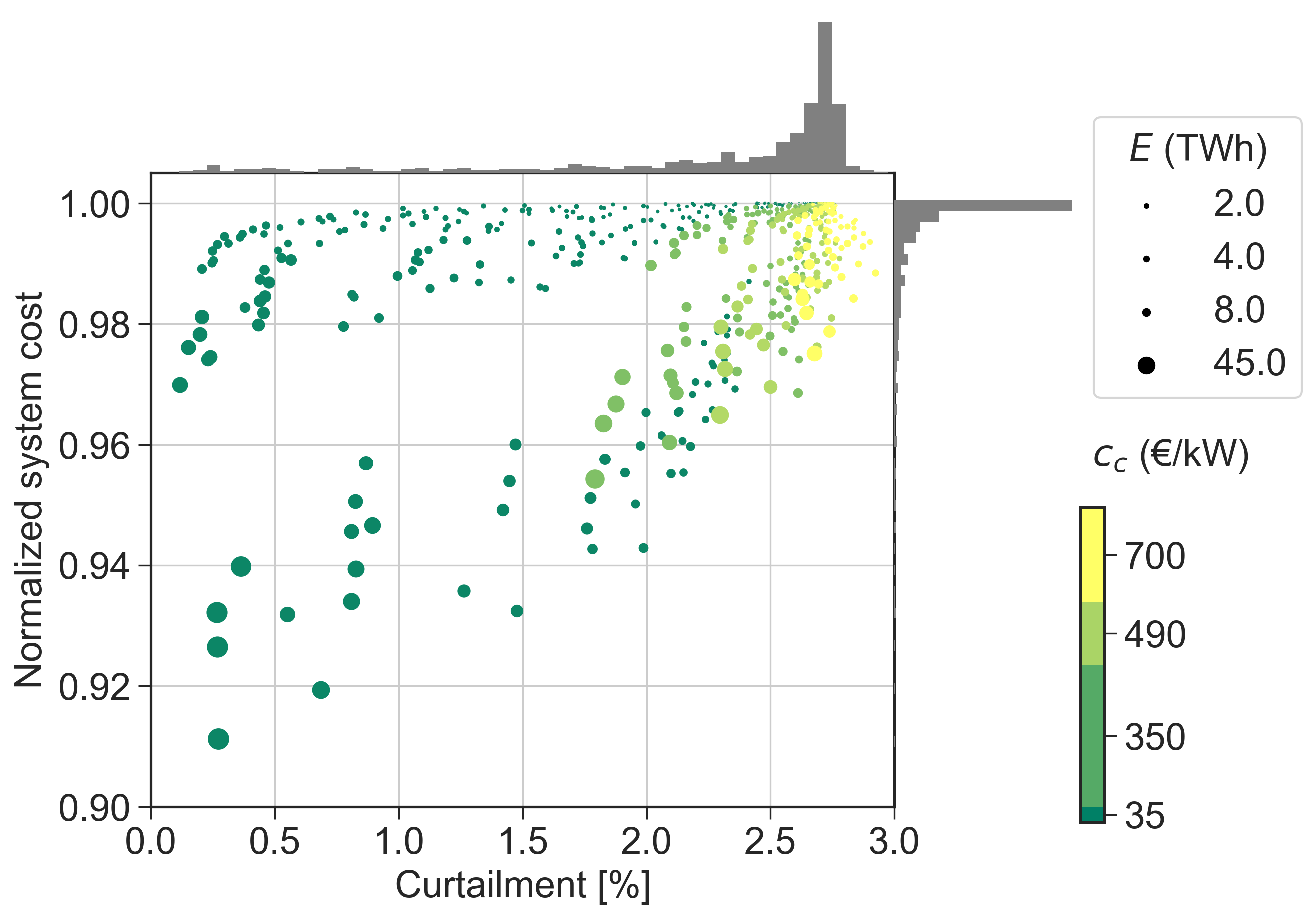}
	\includegraphics[width=0.45\textwidth]{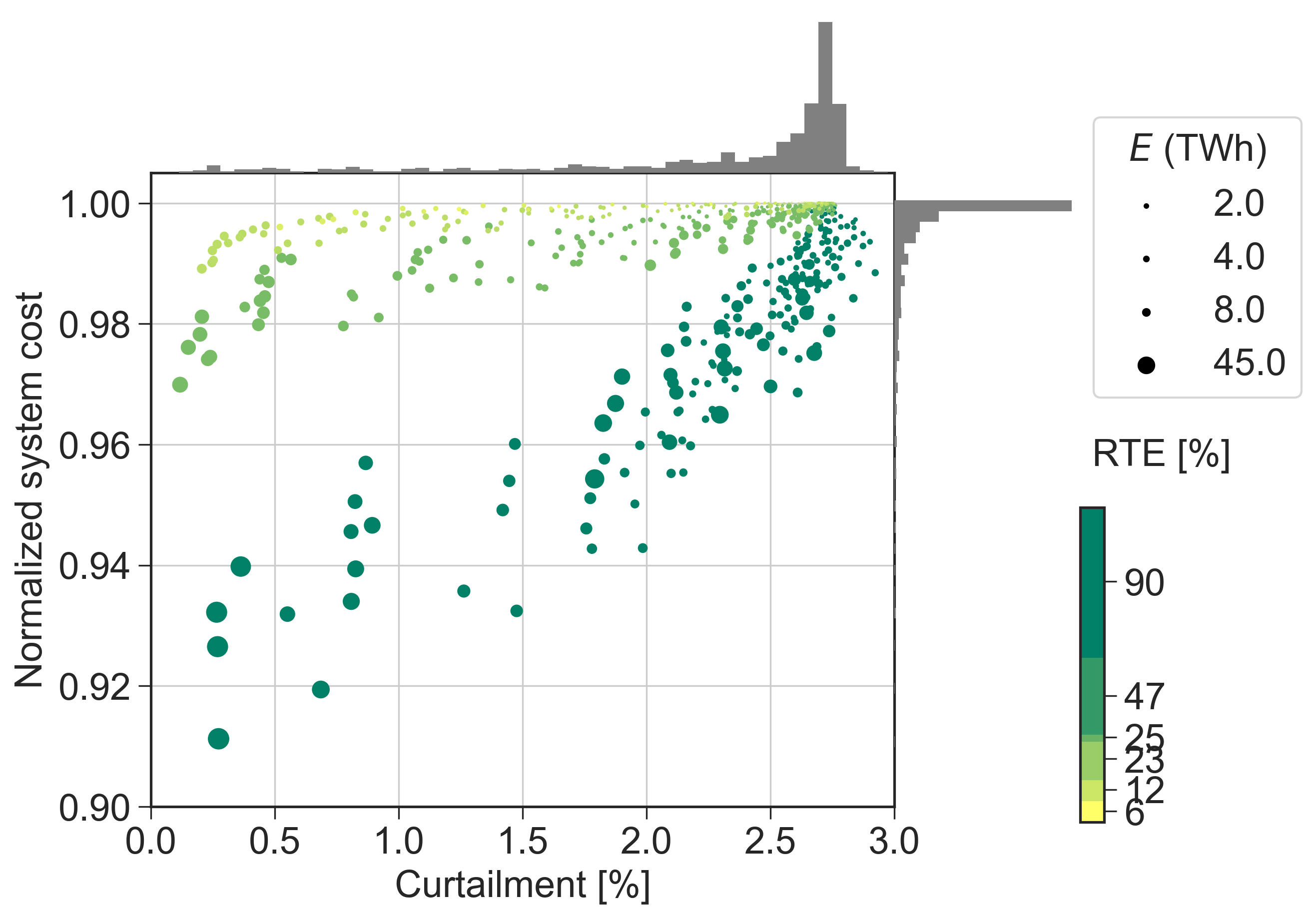}
	\caption{\textbf{System cost and renewable curtailment}. Shown results are Europe-aggregate system cost and renewable curtailment for the (top) Electricity, (middle) Electricity + Heating + Land Transport, and (bottom) the Fully sector-coupled system.}
	\label{sfig:scatter_E}
\end{figure}

\newpage
\begin{figure}[!h]
	\centering
	\includegraphics[width=0.7\textwidth]{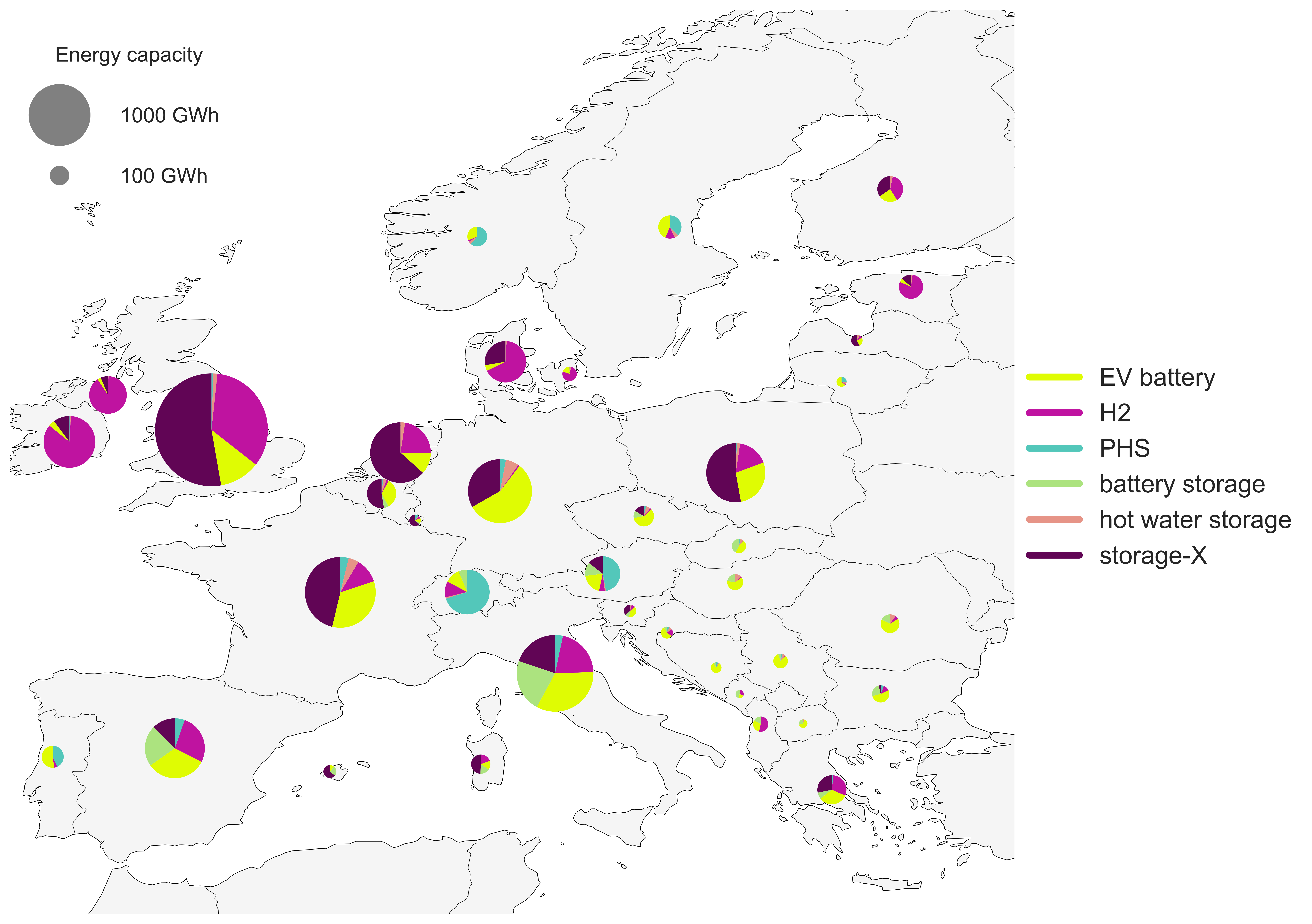}
	\includegraphics[width=0.7\textwidth]{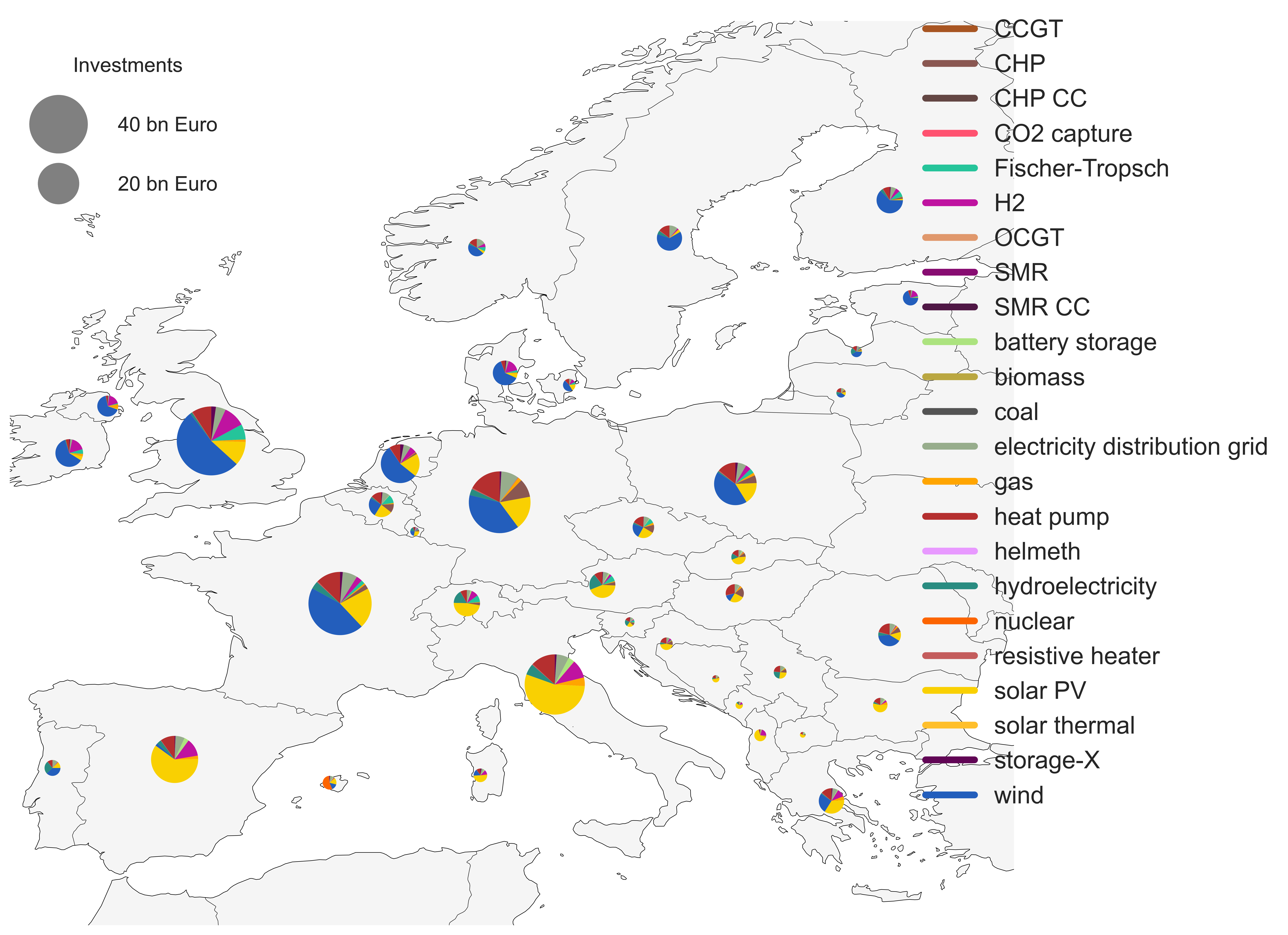}
	\caption{\textbf{Storage capacities and technology investments in the Fully sector-coupled system}. (top) Storage capacities and (bottom) investments in the scenario obtained with the best storage-X in the single parametric sweep, i.e., with a discharge efficiency of 95\% and the remaining parameters according to the "Fixed" configuration.}
	\label{sfig:investments}
\end{figure}

\newpage
\begin{figure}[!ht]
	\centering
	\includegraphics[width=0.5\textwidth]{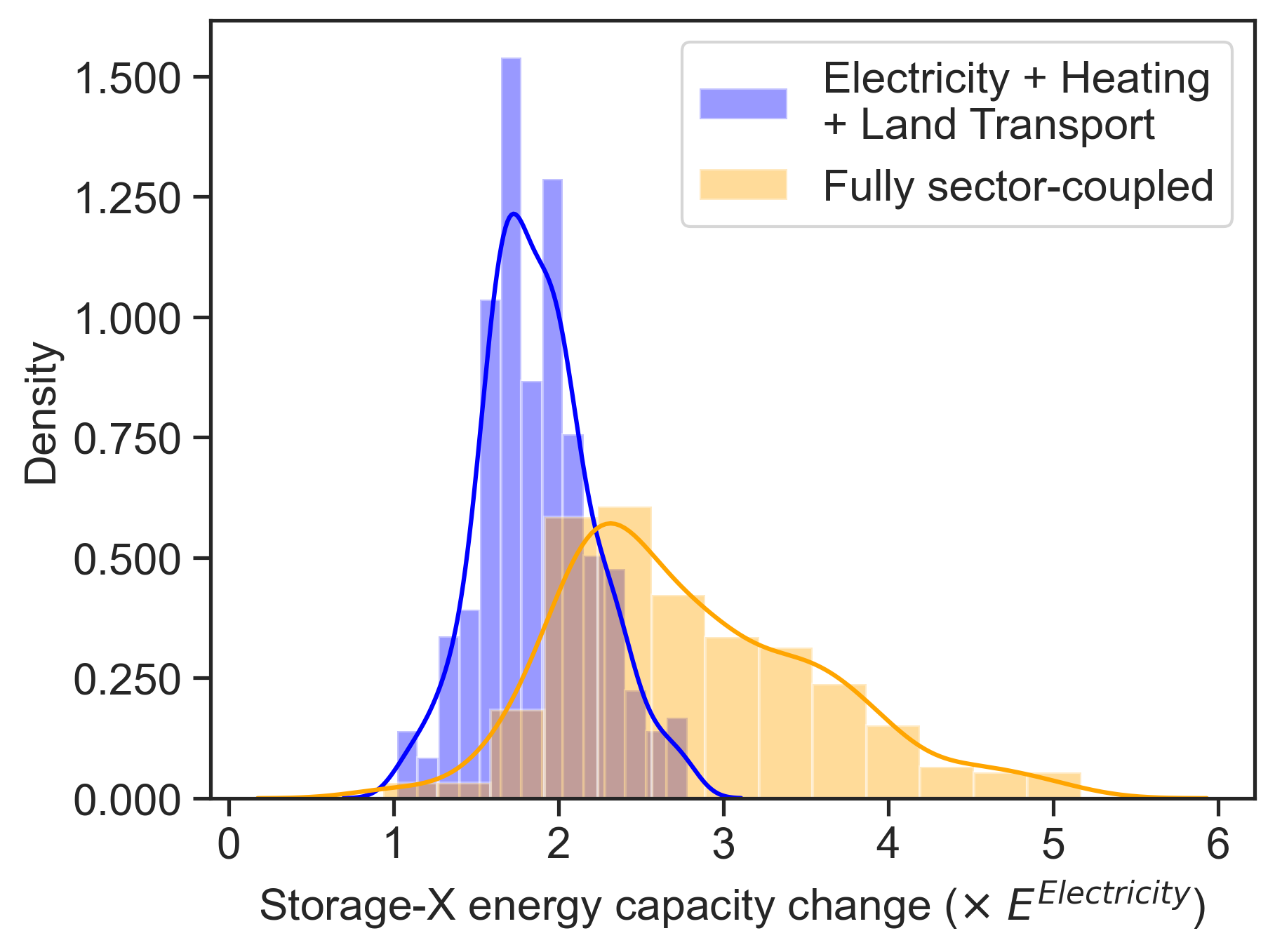}
	\includegraphics[width=0.5\textwidth]{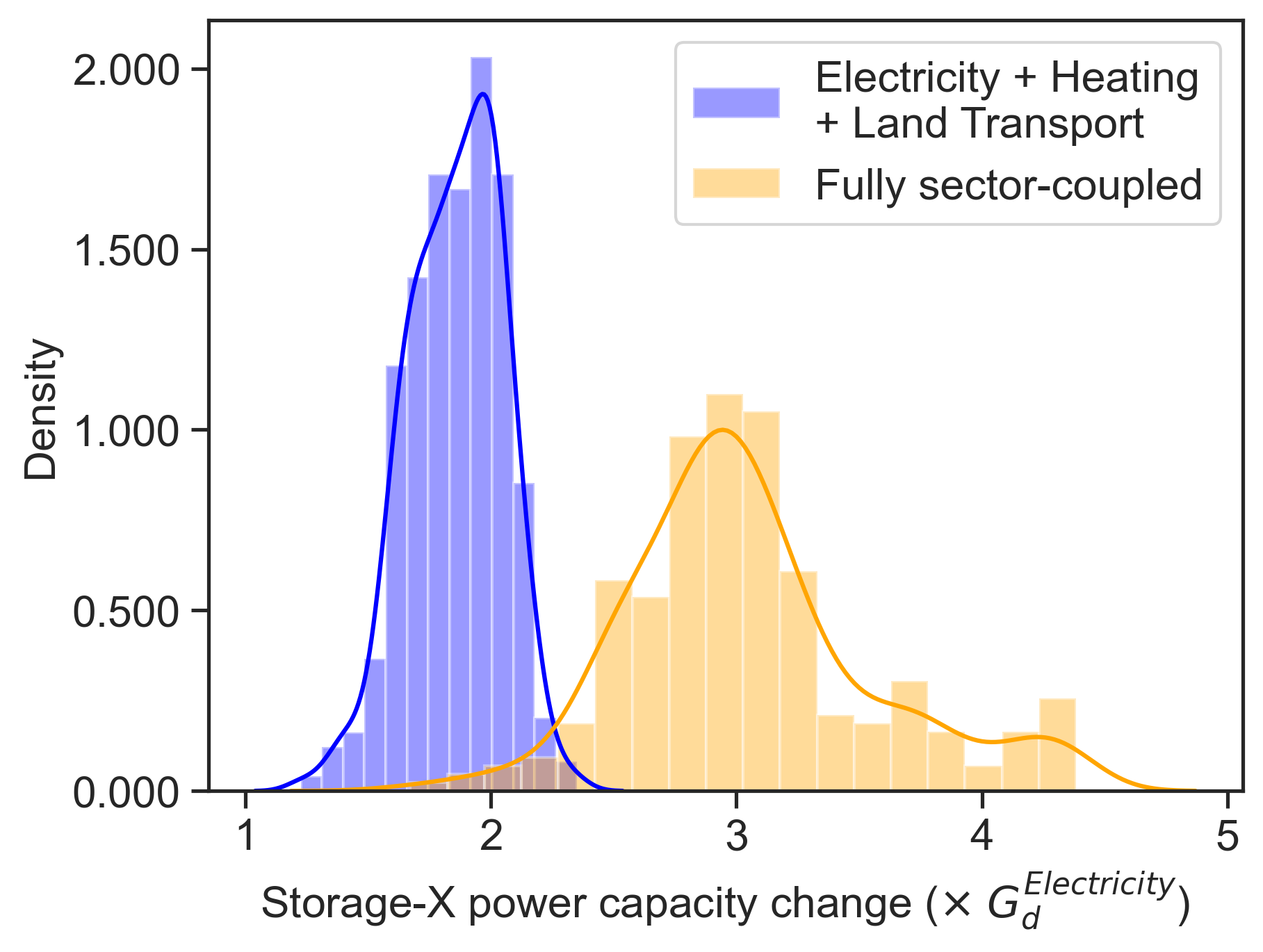}
	\includegraphics[width=0.5\textwidth]{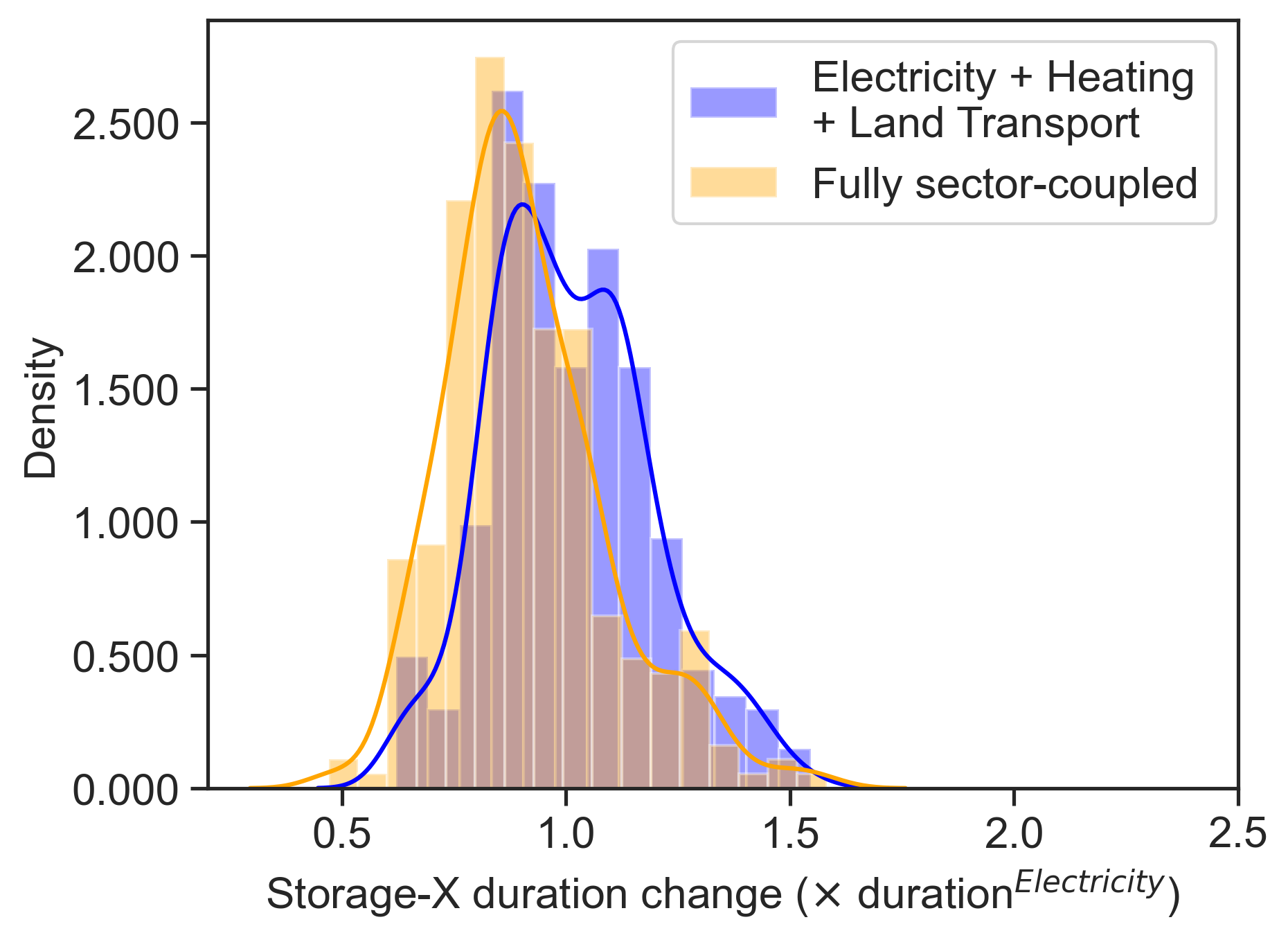}
	\caption{Distribution of the changes in energy capacity (top), power capacity (middle), and duration (bottom) from sector-coupling relative to the Electricity system results. This comparison is obtained by evaluating the same configuration in all three systems while probing the change, which is performed for all configurations within the design space ($E\geq2$~TWh). }
	\label{sfig:dist_changes_from_sectorcoupling}
\end{figure}

\newpage
\begin{figure}[!htbp]
	\centering
	\includegraphics[width=0.9\textwidth]{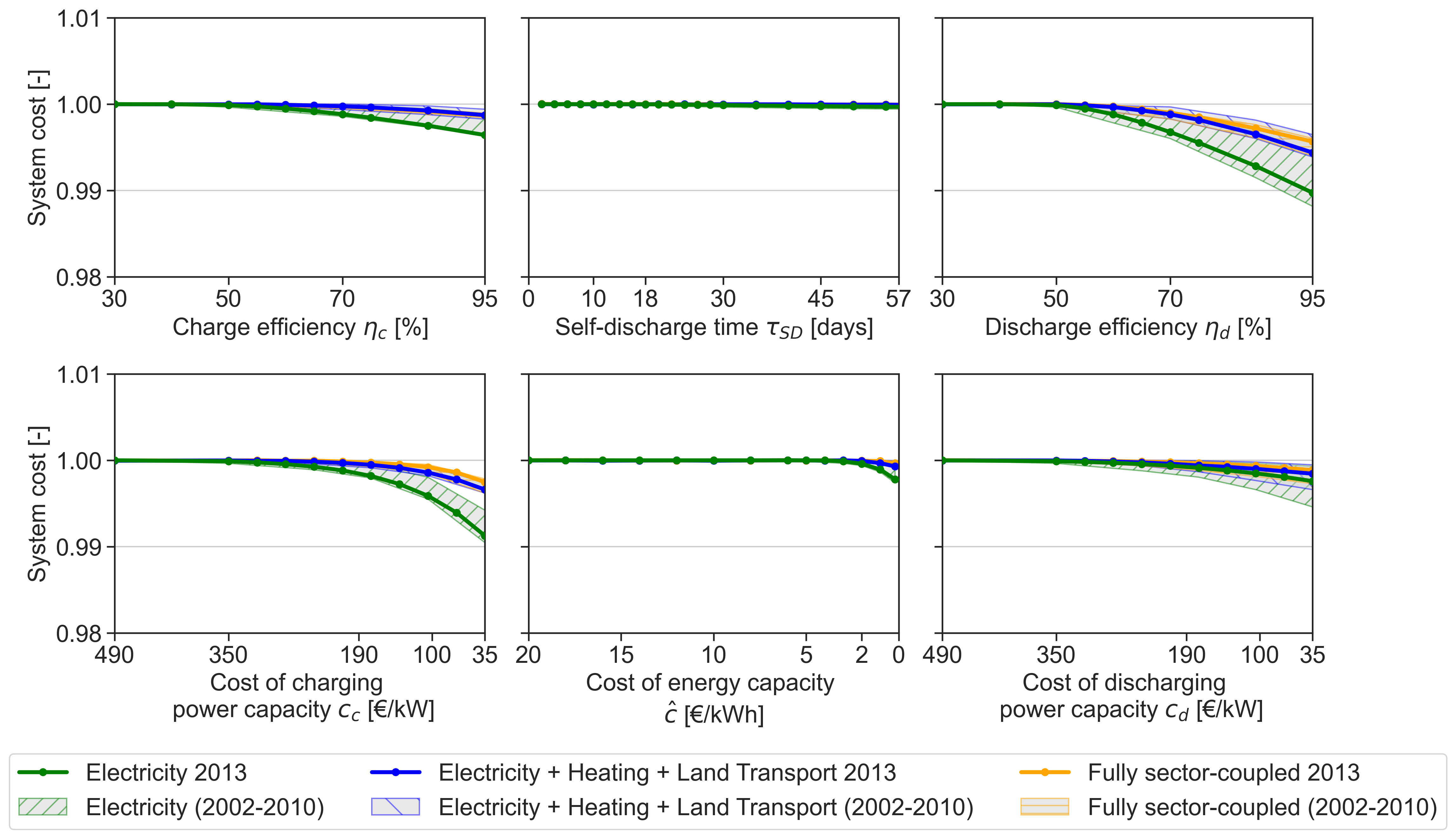}
	\caption{\textbf{System cost reduction obtained with a single-parametric sweep}. The figure shows the results obtained by varying one storage-X parameter at a time for the 2013-weather year (solid lines) while keeping the remainder fixed according to the "Fixed" configuration (Table \ref{tab:storage_reference}). The routine is repeated while varying the weather year in an interval from 2002 to 2010 (hatched area).}
	\label{sfig:single_parameter_sweep_cost_reduction}
\end{figure} 

\end{document}